\documentclass[11pt]{article}

\usepackage{a4wide}
\usepackage{xspace}
\usepackage{multicol}
\usepackage{amsmath}
\usepackage{graphicx}
\usepackage{epsfig}
\usepackage{hyperref} 
\hypersetup{linktocpage} 
\hypersetup{
    colorlinks,
    citecolor=blue,
    filecolor=blue,
    linkcolor=blue,
    urlcolor=blue}
\usepackage{amssymb}

\newcommand{\beq}{\begin{equation}}
\newcommand{\eeq}{\end{equation}}

\newcommand{\bdm}{\begin{displaymath}}
\newcommand{\edm}{\end{displaymath}}

\newcommand{\bea}{\begin{eqnarray}}
\newcommand{\eea}{\end{eqnarray}}

\newcommand{\as}{\alpha_s}

\newcommand{\sanepruning}{\text{Y-pruning}}
\newcommand{\Sanepruning}{\text{Y-pruning}}

\newcommand{\anomalouspruning}{\text{I-pruning}}

\newcommand{\anomalous}{\text{I}}

\def\ord{{\cal O}}

\def \msb{\overline{\textrm{MS}}}
\def\d{\partial}

\def \d{{\rm d} }
\def \d0 {D\O \;}
\def \yc{y_\text{cut}}
\def \zc{z_\text{cut}}
\def \event2{\textsc{Event2} }

\def \CA{C/A }

\def\as{\alpha_s}

\def \CA{C/A }

\title{\textbf{Jet substructure with analytical methods}}
\author{
  Mrinal Dasgupta,\!$^{1}$ Alessandro Fregoso,$^2$ Simone Marzani$^{3}$ and Alexander Powling$^{2}$ \\
\\
{\sl  \small $^1$Consortium for Fundamental Physics, School of Physics \& Astronomy,} \\ {\sl \small University of Manchester, Manchester M13 9PL, United Kingdom }\\[2pt]
{\sl  \small $^2$School of Physics \& Astronomy,}\\ {\sl \small University of Manchester, Manchester M13 9PL, United Kingdom}\\[2pt]
{\sl  \small $^3$Institute for Particle Physics Phenomenology,}\\ {\sl \small 
Durham University, Durham DH1 3LE, United Kingdom}\\[2pt]
}
\date{}

\begin{document}

\maketitle

\vspace{-10.0cm}
 \begin{flushright}
   DCPT/13/88 \\ IPPP/13/44 \\ MAN/HEP/2013/11
 \end{flushright}
\vspace{8cm}

\begin{abstract}
 We consider the mass distribution of QCD jets after the application of jet 
substructure methods, specifically the mass-drop tagger, pruning, trimming and 
their variants. In contrast to most current studies employing Monte Carlo methods, we carry out analytical calculations at the next-to--leading order level, 
which are sufficient to extract the dominant logarithmic behaviour for each technique, and compare our findings to exact fixed-order results. Our results 
should ultimately lead to a better understanding of these jet substructure methods which in turn will influence the development of future substructure tools for LHC phenomenology.

\end{abstract}

\clearpage

\tableofcontents
\newpage

%
\section{Introduction}
In the recent past much attention has been devoted to jet substructure 
techniques as discovery tools for new heavy particles that may be produced with large boosts at the LHC~\cite{boost1,boost2}. Although the earliest discussion in the literature of substructure methods as discovery tools for boosted heavy particles was two decades ago~\cite{mhs}, there have been several rapid advances of late. A large number of techniques have been recently proposed and their theoretical feasibility studied mainly with Monte Carlo methods, although analytical studies exist in some cases~\cite{Rubin:2010fc,Field:2012rw, Larkoski:2013eya, Walsh:2011fz,Feige:2012vc}.

On the experimental front similar progress has been made and many substructure methods have been successfully incorporated into experimental analyses of QCD jets~\cite{ATLAS:2012am, Aad:2012meb, Aad:2013gja, CMS-substructure-studies} and searches for new physics~\cite{Aad:2012raa,Aad:2012dpa, ATLAS:2012dp, ATLAS:2012ds, Chatrchyan:2012ku,Chatrchyan:2012cx, Chatrchyan:2012yxa}. The work that has been performed till date has shown that jet substructure methods will form an important component of the current and future LHC physics program.

While the progress made in the general area of substructure studies and boosted objects is greatly encouraging, it is perhaps also the right time to examine in more detail some of the questions that have cropped up and continue to be raised about the various tools that have been developed and employed in this context. A set of substructure tools that shall particularly concern us in this article go by the collective name of boosted-object methods and examples include the mass-drop tagger~\cite{BDRS}, pruning~\cite{pruning1,pruning2} and trimming~\cite{trimming} techniques. All these methods are specifically designed to enhance signal jets arising from boosted heavy particles and discriminate against QCD background, using the inherently different jet substructure typically obtained in the two cases. Given that several such tools have been put forward it is natural to ask questions about their efficiency and robustness relative to each other and attempt to compare them. For instance, while having a number of tools available implies a desirable element of flexibility, it also gives rise to a danger of duplication and redundancy. Moreover in order to exploit the desirable flexibility aspect, a clear understanding of which substructure method to use in a given study is imperative. Such an understanding can only be obtained by detailed and informed comparisons of the performance of different substructure methods over a wide range of values of jet masses, transverse momenta and other parameters involved in those methods.

To the best of our knowledge while studies do exist which compare the performance of these substructure methods (see for instance~\cite{boost1,boost2,Soper:2010xk,QuirogaArias:2012nj}), these have been mainly Monte Carlo studies and analytical insight into their behaviour is still lacking. Such an understanding would have some clear advantages over reliance purely on information from event generators. At the very 
least there would be no need to worry about the results obtained and conclusions reached being dependent for instance on the parton shower model or event generator tune used, as may be the case with Monte Carlo studies~\cite{Richardson:2012bn}. Moreover, while pure Monte Carlo studies may be sufficient to alert us to similarities as well as different features of substructure methods, they do not offer direct insight into {\it{why}} these features may arise in the first place. Clearly with analytical formulae in hand one is much better placed to explain peculiar features that can emerge in either Monte Carlo or experimental studies. This in turn may also facilitate the removal of any undesirable aspects of substructure methods and lead to the development of better and more robust tools. 

Yet another issue that one can raise, concerns the nature of calculations that 
need to be performed to best describe observables such as jet masses, after the application of the boosted-object techniques. In the case of plain inclusive 
jet mass distributions it is well-known that there are large double logarithms $1/m_j \, \alpha_s^n \ln ^{2n-1} p_t/m_j$ that arise in the jet-mass distribution accompanied by less singular but still logarithmically enhanced terms. In the region of interest for boosted object studies one may have $m_j \ll p_t$ even when electroweak scale jet masses are considered, due to the multi-TeV values of jet $p_t$ which can be attained at the LHC. This implies that the logarithms in question, which arise in the distribution of QCD background jets, can be large even at values of jet masses where one may expect to see a signal peak. 
An accurate description of the QCD jet mass distribution therefore requires resummation of the large logarithms in question. For a detailed discussion of resummation for hadron collider jet masses and issues therein we refer the reader to Refs.~\cite{BanDasKKKMar, DasKKKMarSpannow, jetmass_harvard, jetmass_mit}. In particular in Ref.~\cite{DasKKKMarSpannow} the inclusive jet mass distribution 
was computed at next-to--leading logarithmic (NLL) accuracy for hadron collider jets. However due to the presence of non-global logarithms~\cite{Dassalam1,Dassalam2} the resummation of NLL terms for inclusive jet mass can only be achieved in the large $N_c$ limit, which however should still be well suited for 
comparisons to experimental data. Also due to the presence of soft gluon clustering effects~\cite{AppSey1,BanDas,BanDasDel} the resummation can currently only be performed for jets defined in the anti-$k_t$ algorithm, where such clustering effects are absent to the NLL accuracy that is generally sought~\cite{antikt}. Thus for jets defined in other algorithms such as Cambridge-Aachen (C/A)~\cite{cambridge,aachen} currently only Monte Carlo event generators can be used to provide theoretical estimates for inclusive jet mass distributions.
It is well known however that boosted taggers such as the mass-drop procedure~\cite{BDRS} and similar methods like pruning~\cite{pruning1,pruning2} and trimming~\cite{trimming} all aim to discriminate against QCD background by placing cuts on soft gluon radiation inside the jet. By their very nature therefore these tools should at least partially remove the logarithms of $p_t/m_j$ that 
afflict the plain jet mass distribution. An intriguing possibility therefore arises as to whether pure fixed-order tools such as \textsc{Nlojet}++~\cite{nagy} may in fact suffice to give an accurate description of jet masses with substructure algorithms. It is therefore of importance to investigate what structure emerges when one considers perturbative calculations of jet masses with substructure algorithms. Once this is 
understood it should point to whether fixed-order, all-order resummed or Monte Carlo techniques may provide the best theoretical description for this class of LHC observables.

In our present paper we set out to answer some of these questions. We take as examples the aforementioned techniques of mass-drop, pruning and trimming and aim to compute them through to next-to--leading order (NLO). We do not carry out complete NLO calculations but work using essentially eikonal approximations to treat the QCD radiation within a jet. These approximations are known to correctly reproduce the logarithmic enhancements that we seek to study. Our main aim is to highlight the presence or absence of logarithmic enhancements for various substructure methods and understand their nature and origin. On this basis it should be possible to decide on whether the resulting logarithms if any may be resummable and to comment on the straightforwardness or otherwise of the resummation required, compared for instance to the case of plain jet mass. To this end we determine the leading and in some instances next-to--leading logarithmic behaviour that emerges for each substructure algorithm at the NLO level, and examine the issue of non-global 
logarithms and test our findings by comparing to exact fixed-order estimates.

To understand these aspects of the jet mass distributions it is possible to work explicitly with jets produced in $e^{+}e^{-}$ collisions as for our initial study of plain jet mass~\cite{BanDasKKKMar}. The additional aspect of initial state radiation (ISR) present at hadron colliders is only a relatively 
inessential detail in these studies and will not change the conclusions we arrive at here. In the current article we confine ourselves to establishing the results that emerge at the NLO level while a companion article is devoted to all-order behaviour and comparison with parton showers as well as detailed phenomenological considerations for hadron collider jets~\cite{DasFregMarSal}. We also consider here only pure QCD background jets and a detailed study of the action of substructure algorithms on signal jets will be undertaken in a future article.

The layout of this article is as follows. We devote the next section to some generalities and in order to set the scene for the remainder of the article remind the reader of the NLO structure of the plain jet mass distribution, to which the corresponding structure that emerges for each substructure algorithm can be compared. In the following section we explore the logarithmic structure at leading order (LO) and NLO for the mass-drop tagger, and point to the need for a modified mass-drop procedure~\cite{DasFregMarSal} which we also study. We also explore the question of non-global and clustering logarithms and carry out comparisons to 
fixed-order estimates from \event2~\cite{CataniSeymour}. This is followed in the next section by an investigation of pruning also tested against \event2 and 
lastly we address the question of the trimming technique in a similar manner. Our findings shall show the logarithmic structure of the substructure methods at hand to be very different from one another, in some cases very rich in physics effects and in all cases worthy of further investigation beyond fixed-order which is carried out in the companion article~\cite{DasFregMarSal}. Lastly in our conclusions we summarise our work, discuss its main implications and provide suggestions for future studies which may lead to the development of better jet substructure methods in the near future.

\section{General considerations and plain jet mass}
As we mentioned in our introduction, the features of the substructure methods that we wish to highlight shall all emerge in a simple study where one can take the jets to be produced in a process with a colourless initial state, for instance $e^{+}e^{-}$ annihilation.

In this paper we consider differential distributions in the squared jet-mass 
normalised to the jet energy squared i.e. $\frac{1}{\sigma}\frac{d \sigma}{d v}$, with $v=\frac{m_j^2}{E_j^2}$. When computed in perturbation theory the plain 
jet-mass distribution in the small-$v$ limit behaves as
\bea \label{behaviour}
\frac{v}{\sigma}\frac{d \sigma^\text{(plain)}}{d v} &=& \frac{\as}{\pi} \left(a_{12} L + a_{11} +\ord(v)\right) \nonumber\\ &+& 
\left( \frac{\as}{\pi}\right)^2 \left(a_{24} L^3 + a_{23} L^2 + a_{22}L + \ord\left(v^0\right)\right) +\ord\left(\as^3 \right), \, \, L \equiv \ln \frac{R^2}{v},
\eea
where we have considered for simplicity the approximation of small jet radius, $R \ll 1$ as in Ref.~\cite{BanDasKKKMar}. The counting of the logarithms is usually performed at the level of the so-called integrated distribution:
\beq \label{integrated_distr}
\Sigma(v)= \frac{1}{\sigma} \int^v d v' \frac{d \sigma}{d v'}.
\eeq
Consequently, the contributions with coefficients $a_{i \, 2i}$ are referred to as double logarithms, while the ones with $a_{i\, i}$ are single logarithms. We can take $\alpha_s$ to be defined in the standard $\msb$ scheme unless explicitly specified otherwise and assume its scale to be the jet energy $E_j$.

Then for the case of plain jet-mass one has~\cite{BanDasKKKMar}
\bea
\label{jmcoeffs}
a_{12} &=& C_F, \\ \nonumber
a_{11} &=& -\frac{3 C_F}{4}, \\ \nonumber
a_{24} &=& -\frac{C_F^2}{2}, \\ \nonumber
a_{23} &=& \frac{3}{8} C_F \left(3 C_F + 4 \beta_0 \right),
 \eea
with $\beta_0 = \frac{1}{12} \left(11C_A-2 n_f \right)$. We have not 
reported explicitly the coefficient $a_{22}$ which for the plain jet-mass has numerous sources including multiple emission effects, non-global logarithms, 
clustering logarithms, cross-talk between the resummed exponent and order 
$\alpha_s$ coefficient functions and running coupling effects.

\section{The Mass Drop Tagger}\label{sec:MDT}
\subsection{Definition}
The Mass Drop Tagger (MDT)~\cite{BDRS} involves two parameters $\mu$ and $\yc$, which can be optimised for the study in question. One starts with a hard jet $j$ with radius $R$ defined with the \CA algorithm~\cite{cambridge, aachen} and then one applies the following algorithm:
\begin{enumerate} \label{MDTdef}
\item Break the jet $j$ into two subjets $j_1$ and $j_2$ such that $m_{j_1}>m_{j_2}$.
\item  If a significant mass drop is found $m_{j_1} < \mu m_{j}$ with a splitting which is not too asymmetric, $y= \frac{\mathrm{min}\left(p_{t,j1}^2,p_{t,j2}^2\right)\Delta R^2_{j1,j2}}{m_j^2}>\yc $, then the algorithm tags the jet and exits the loop.
\item Otherwise, redefine $j=j_1$ and go back to step 1.
\end{enumerate}
For our current purpose of checking the structure of large logarithms that emerges in the perturbative calculations for jet masses, we shall use an $e^{+}e^{-}$ adaptation of the above procedure which involves the use of energies and 
angles rather than transverse momenta and distance measures that are invariant under longitudinal boosts, as is the case for hadron collisions. Hence we 
replace the $p_{t,j}$ in the above definitions by energies $E_j$ and define 
$\Delta R^2_{j1,j2}$ as $2\left(1-\cos\theta_{j1,j2} \right)$ where $\theta_{j1,j2}$ is the angle between the jet directions. We then note that the measure $y$  can be expressed as a ratio of energies $y= \frac{\mathrm{min} \left(E_{j1},E_{j2}\right)}{\mathrm{max} \left(E_{j1},E_{j2}\right)}$ . In the limit of a collinear parton splitting $j \to j_1,j_2$, a situation in which we shall be particularly interested below, one can express $y$ in terms of the respective energy fractions $x,1-x$ of the parent parton energy,  $y\simeq \frac{\mathrm{min}\left(x,1-x\right)}{\mathrm{max} \left (x,1-x \right)}$.

The MDT is often used in conjunction with a procedure known as filtering~\cite{BDRS}, an extensive analytical study of which can be found in~\cite{Rubin:2010fc}. In this paper we are going to ignore the effects of filtering on the MDT 
jet mass distribution, because the standard choice $n_\text{filt}=3$ only modifies the mass distribution beyond the $\mathcal{O}\left(\alpha_s^2 \right)$ contributions considered here. Further, it has been argued that at all orders the 
effects of filtering are generally well beyond the logarithmic accuracy we aim for~\cite{DasFregMarSal}.

\subsection{Leading-order results}
\label{sec:MDTLO}
Here we shall carry out the leading-order (LO) calculations relevant to the MDT. We shall examine the distribution $\frac{1}{\sigma} \left(\frac{d\sigma}{d v} \right)$ where $v = m_j^2/E_j^2$ is the squared jet-mass of the measured jet after the application of the mass-drop procedure, normalised to the energy squared of the fat jet. Since we wish to focus on the structure of large logarithms at leading order we can start by examining a configuration where one emits a soft gluon with four-momentum $k$ from a quark-antiquark pair in $e^{+}e^{-}$ annihilation. Moreover this gluon gets recombined with the quark or antiquark to form the massive jet that we focus on and for definiteness let us consider this to be the quark jet. We parametrise the momenta of the partons as 
\bea
\label{lo:kin}
p &=& E_q \left( 1,0,0,1 \right), \nonumber\\
k &=& E_g \left(1,0,\sin \theta, \cos \theta \right),
\eea
where $E_q$ and $E_g$ are the energies of the quark and gluon respectively. In terms of the energy of the overall fat jet $j$ we shall take these energies to be $E_g = x \,E_j$ and $E_q = (1-x)E_j$. Also for the two partons to be recombined by the jet algorithm into a single jet one must have  $\Delta^2_\theta =2 \left(1-\cos \theta \right) < 2 \left( 1-\cos R \right)$, which in the collinear limit is simply $\theta^2 < R^2$. 

Now consider passing this jet through the MDT. On undoing the jet clustering we produce two massless partons $p$ and $k$ so that the mass-drop condition is trivially satisfied. The asymmetry condition is satisfied for values of $x$ such that $1/(1+\yc)> x > \yc/(1+\yc)$. We thus calculate the mass distribution of 
jets that pass the above asymmetry cut. The normalised jet mass can be expressed as $\frac{m_j^2}{E_j^2} = 2x(1-x) \left(1-\cos\theta \right) \approx x\theta^2$ where we employed the soft-collinear approximation.
Treating the emission of the soft gluon in the standard eikonal approximation we obtain in the collinear limit:
\bea \label{eq:LOmdtsetup}
\frac{1}{\sigma}\frac{d \sigma}{d v}^\text{(MDT, LO)}=  \frac{\as C_F}{\pi} \int \frac{d \theta^2}{\theta^2}\int_{\frac{\yc}{1+\yc}}^{\frac{1}{1+\yc}}\frac{d x}{x}\Theta\left(R^2 -\theta^2 \right)\delta \left(v- x \theta^2 \right).
\eea
 
Evaluating the above integral is straightforward and leads to 
\bea \label{LOmdt}
\frac{1}{\sigma}\frac{d \sigma}{d v}^\text{(MDT, LO)}
&=& \frac{\as C_F}{\pi} \frac{1}{v}\ln \left(\frac{1}{\yc}\right )\Theta\left(\frac{\yc}{1+\yc} R^2 -v  \right) \nonumber \\
 &+&\frac{\as C_F}{\pi} \frac{1}{v}\ln \left( \frac{R^2}{v(1+\yc)}\right)\Theta\left(v - \frac{\yc}{1+\yc} R^2  \right).
\eea
\begin{figure}
\begin{center}
\includegraphics[width=0.49 \textwidth]{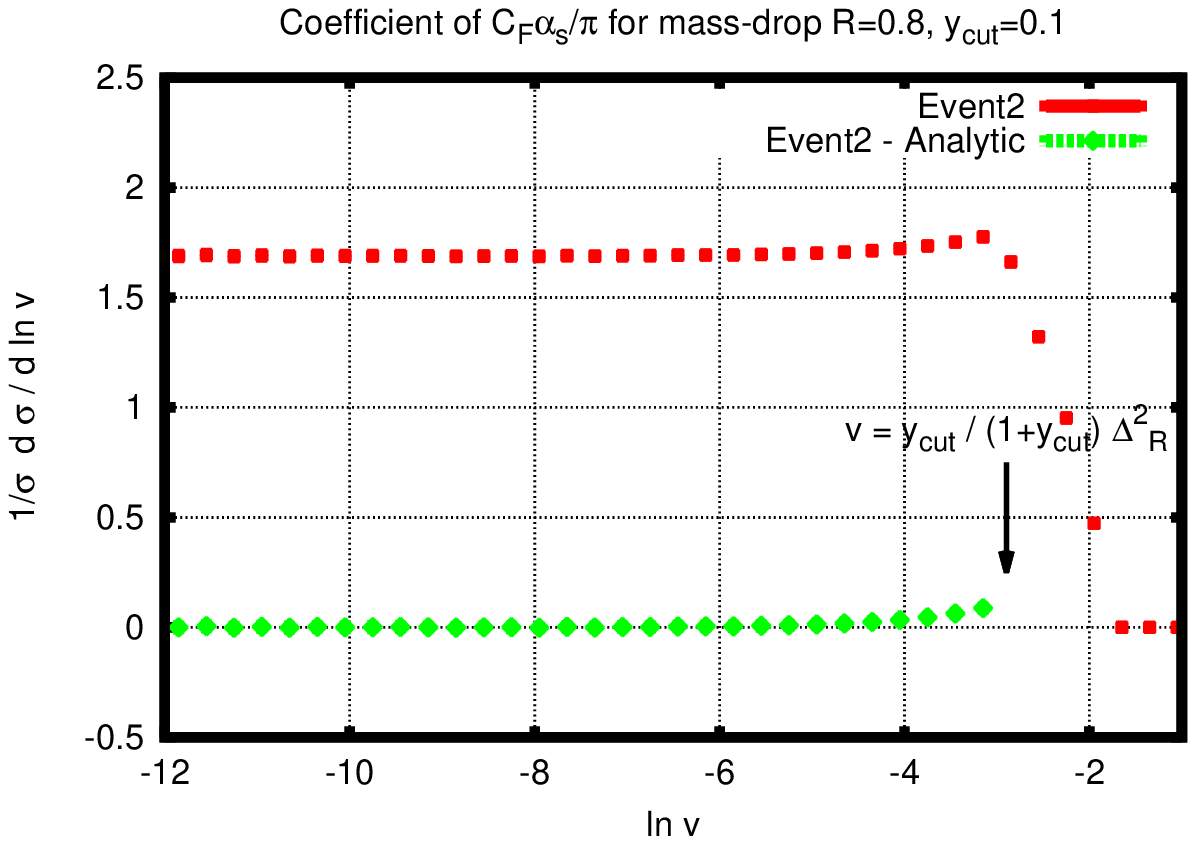}
\includegraphics[width=0.49 \textwidth]{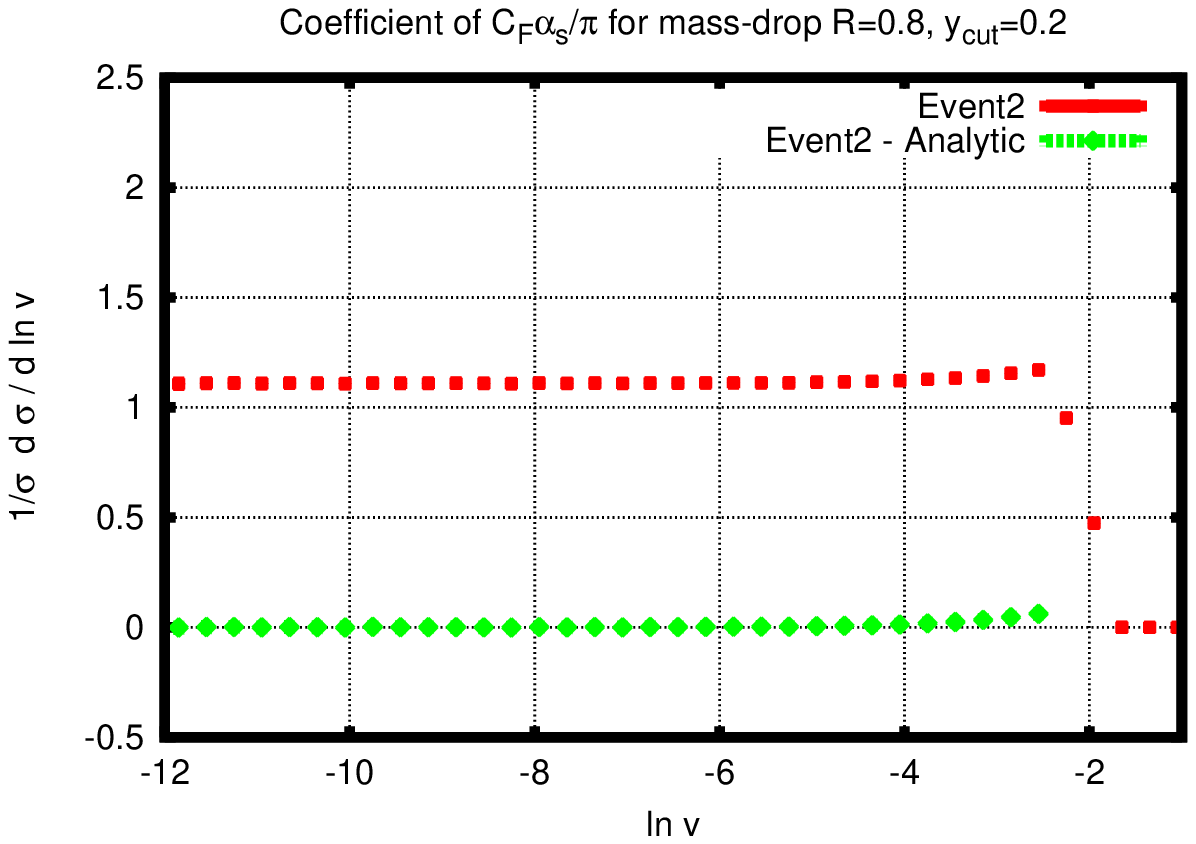}
\includegraphics[width=0.49 \textwidth]{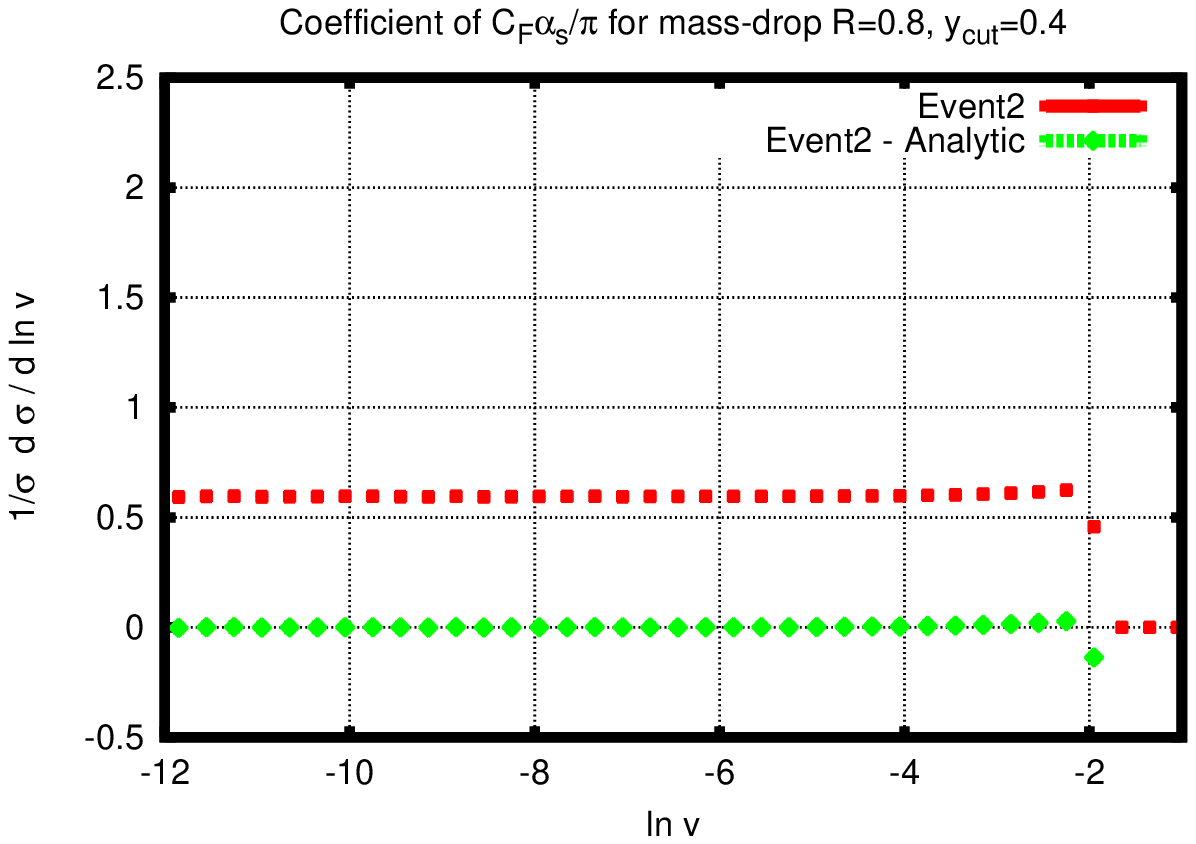}
\caption{Comparison of the analytic calculation Eq.~(\ref{LOmdtcomplete}) with \event2 at LO in the region $v< \frac{\yc}{1+\yc}\Delta_R^2$, for different values of $\yc$. The red curve shows the fixed-order result alone which is flat for small $v$ and hence indicates a single logarithmic behaviour for the integrated distribution. The green curve indicates that, after subtracting our analytical calculation, the result vanishes at small $v$ as expected.}
\label{fig:MDCF}
\end{center}
\end{figure}

Several comments are in order. Firstly we note that at small jet masses $v < R^2 \yc/(1+\yc)$ the result is single-logarithmic in $v$, in contrast to the behaviour of the plain jet mass. The action of the tagger has been to replace a soft logarithm in $v$ by a logarithm in $\yc$. The value of $\yc$ is chosen so that these logarithms are modest and this means that the background will be reduced compared to the plain jet-mass, as intended. For larger masses, $v>R^2 \yc/(1+\yc)$, one returns to the double logarithmic plain jet mass~\footnote{Note that this last statement is only true in Eq.~(\ref{LOmdt}) at the level of  double logarithms, but it can be extended beyond this accuracy with a more careful treatment of the kinematics.}.


One may then expect that the fixed-order perturbative expansion in the case of the MDT jet-mass is more convergent than for plain jet-mass which could be promising from the point of view of using pure fixed-order perturbative estimates to describe the MDT jet-mass. 
Note also that the remaining single logarithm for the MDT is of pure collinear origin, a point that we shall return to subsequently.

It is clearly of interest to verify the result Eq.~(\ref{LOmdt}) 
against fixed-order codes and for this purpose one can use the NLO program \event2 for our present study, involving jets in $e^{+}e^{-}$ annihilation. Before we do so however we note that thus far we have worked in only the soft-collinear limit. In order to perform meaningful comparisons with \event2 it is required to go beyond this limit. It is straightforward to modify our result by considering large-angle emissions as well as improving the collinear region to account for hard emissions. We provide the details of this calculation in the appendix while quoting our full result 
\beq \label{LOmdtcomplete}
\frac{1}{\sigma}\frac{d \sigma}{d v}^\text{(MDT, LO, full)}
= \frac{\as C_F}{\pi} \frac{1}{v}\ln \left(\frac{1}{\yc} e^{-\frac{3}{4}\left(\frac{1-\yc}{1+\yc}\right)} \right )\,, \quad \text{for} \quad v < \frac{\yc}{1+\yc} \Delta_R^2 .
\eeq

From the above result one notes that introducing the full splitting function accounting for hard collinear emissions changes the coefficient of the single logarithmic behaviour obtained at small jet masses relative to the pure soft $\ln 1/\yc$ coefficient obtained before. This is in contrast to the role of soft emissions at large angles which generate only subleading terms in the small jet-mass region. Additionally, fully accounting for large-angle and hard emissions is 
also important to obtain the correct position of the transition point where the behaviour switches from a single-logarithm to the result for plain jet-mass. 
For instance going beyond the collinear limit one notes that relative to the small $R$ result for the transition points $v=R^2 \yc/(1+\yc)$, the finite $R$ 
effects involve replacing $R^2$ by $\Delta_R^2 =2 (1-\cos R)$, as in the above formula. However we do not in this paper concern ourselves with these transitions as our aim is to check the logarithmic structure purely in the small $v$ region.

In order to test our analytic calculation, we compare it to the result obtained with the fixed-order code \event2~\cite{CataniSeymour}. Numerical results are obtained for $e^+e^-$ collisions at the centre-of-mass-energy $E_\text{CM}=1$~TeV. Jets are defined with the \CA algorithm ($R=0.8$) using the FastJet package~\cite{Cacciari:2011ma, Cacciari:2005hq}. We define the \event2 result to be the average of the results from the hardest and second-hardest jets, which should then give a result that can be directly compared to our single-jet calculation. The comparison is shown in Fig.~\ref{fig:MDCF}, which demonstrates that our analytic calculation correctly reproduces the full LO result in the small-$v$ limit, for different values of $\yc$.

Having carried out a leading-order calculation for MDT and obtaining a result 
which is single-logarithmic, it is clearly of interest to explore the structure of logarithms at the NLO level, to which the next sub-section is devoted.

\subsection{Logarithmic behaviour beyond leading-order}\label{sec:MDT_wrong_branch}

\begin{figure}
\begin{center}
\includegraphics[width=0.3 \textwidth]{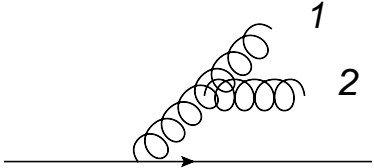}
\caption{NLO configuration that gives rise to an extra logarithm for the MDT.} \label{fig:nonAbelian}
\end{center}
\end{figure}

In the previous section we have shown that the MDT reduces the logarithmic divergence of the LO jet mass distribution. We want to investigate whether this remains true at higher perturbative orders. To be more precise we would like to check whether the MDT mass distribution exhibits only single logarithms beyond LO. 

Unfortunately, as we demonstrate below, a new effect appears at NLO which 
results in an extra-logarithm in the jet-mass distribution, hence spoiling the simple picture of pure single-logarithms encountered before. We call this 
effect the \emph{wrong-branch} issue and after describing it in detail, we explain how to remove it via a {\it{modified Mass Drop Tagger}} (mMDT) suggested in the companion paper~\cite{DasFregMarSal}. The point is simply that in the mass-drop procedure, when the mass-drop or asymmetry condition fails, one proceeds analysing the subjet with the {\it{largest mass}}, rather than the {\it{most energetic}} one, so that we do not necessarily follow the hard parton and we can end up measuring the mass of a (\emph{wrong}) soft branch. This is essentially a flaw in the original mass-drop tagger~\cite{BDRS} which results in consequences for the structure of large logarithms in the perturbative expansion. 

To show this explicitly, let us consider an emission such as the one pictured in Fig.~\ref{fig:nonAbelian}. The figure depicts the branching of a soft gluon $k$ into offspring gluons $k_1$ and $k_2$. We are interested in the {\it{collinear}} regime where the angular distance between the two offspring gluons $\Delta^2 _{\theta_{12}}$ is the smallest distance amongst the various pairs of distances in the C/A algorithm. The C/A algorithm would first cluster the pair of gluons and then cluster the parent to the quark, to form the fat jet. When we undo the last clustering  on the fat jet composed of the hard parton and both gluons, we shall find two subjets: a massive jet $j_1$ composed of the two soft gluons $k_1$ and $k_2$ and a massless jet $j_2$ corresponding to the quark. The mass-drop criterion will be automatically satisfied since in the soft limit we are considering, the subjet $j_1$ will have a much smaller mass than the initial fat jet by virtue of the fact that it is composed of two soft particles rather than one hard parton and additional soft partons~\footnote{To be more precise we compute here the leading-logarithmic behaviour which arises from the region $x,\theta_{12}/\theta \ll 1$, with $\mu$ not too small, so that logarithms of $\mu$ are not parametrically large. In this limit it is straightforward to verify that the mass-drop criterion is always satisfied.}. However, the energy asymmetry condition may be satisfied or otherwise. Assuming it is \emph{not} satisfied means that one next moves on to consider the jet $j_1$ instead of the jet $j$. We thus encounter a situation where we study the jet-mass of $j_1$ instead of the jet-mass of the original jet and the jet-mass distribution of $j_1$ corresponds to a single-logarithmic tagged gluon jet-mass. The failure of the energy asymmetry cut translates into
\begin{equation} \label{MDTgluonconditions}
x< \frac{\yc}{1+\yc} \quad \text{or} \quad x> \frac{1}{1+\yc},
\end{equation}
where $x$ is the energy fraction, relative to the fat jet, of the massive parent gluon jet. At this point we are only interested in computing the most divergent contribution, so we can drop the second condition above, which corresponds to the emission of a hard gluon. 

When the asymmetry cut fails we switch to examining the more massive subjet $j_1$. Here one can consider a collinear branching such that $k_1$ and $k_2$ carry fractions $1-z$ and $z$ of the parent gluon momentum respectively and we denote the angle between them as $\theta_{12}$ where $\theta_{12} \ll 1$. Then we have that $m_{j1}^2/E_j^2 = x^2 z (1-z) \theta_{12}^2$, where we normalised to $E_j^2$, the energy of the fat jet. Moreover to accept the jet $j_1$ the asymmetry condition must also be satisfied in the parent gluon splitting, which implies $\yc/(1+\yc) < z< 1/(1+\yc)$. One can then write 
\begin{multline}
C_F C_A \left(\frac{\alpha_s}{\pi} \right)^2 \int \frac{d x}{x} \frac{d\theta^2}{\theta^2}    \Theta\left(\frac{\yc}{1+\yc}-x\right)
\\  \Theta \left (R^2-\theta^2\right)
\int dz \left(\frac{1}{2} p_{gg}(z)+\frac{T_R n_f}{C_A} p_{qg}(z) \right) \frac{d\theta_{12}^2}{\theta_{12}^2}  \delta \left(v-z(1-z) x^2 \theta_{12}^2 \right) \\  \Theta \left ( z-\frac{\yc}{1+\yc}\right) \Theta \left( \frac{1}{1+\yc}-z\right)\Theta \left(\theta^2-\theta_{12}^2 \right).
\end{multline}

In writing the above result we have used the soft approximation to obtain the probability of producing the parent gluon and then treated the collinear decay to a pair of gluons via the ``reduced'' splitting function
\begin{equation}\label{pgg}
p_{gg}(x) = 2\frac{1-x}{x} + x(1-x),
\end{equation}
and to a quark-antiquark pair via the corresponding splitting function 
\begin{equation}\label{pqg}
p_{qg}(x) = \frac{1}{2}(x^2+(1-x)^2) .
\end{equation}
 Carrying out the integrals is simple and yields the following results in the small $v$ limit, $v\to 0$, where we separate the $C_F C_A$ and $C_F n_f$ contributions:
\bea \label{MDJg1}
\frac{1}{\sigma}\frac{d \sigma}{d v}^\text{(MDT, $C_F C_A$)}&=&
C_F C_A   \left(\frac{\alpha_s}{\pi} \right)^2 \frac{1}{4v} \Bigg\{ \left( \ln\frac{1}{\yc}+\frac{11 \yc^3+9 \yc^2-9 \yc-11}{12 \left(1+\yc \right)^3} \right)\nonumber \\
&&\Bigg[\ln^2\frac{1}{v} +\ord\left( \ln \frac{1}{v} \right)  \Bigg]\Bigg\},
\eea
and
\beq \label{MDJg2}
\frac{1}{\sigma}\frac{d \sigma}{d v}^\text{(MDT, $C_F n_f$)}=
C_F n_f  \left(\frac{\alpha_s}{\pi} \right)^2 \frac{1}{4 v} \Bigg\{ \frac{1-\yc ^3}{6 \left(1+\yc \right)^3} \Bigg[\ln^2\frac{1}{v}
+\ord\left( \ln \frac{1}{v} \right)  \Bigg]\Bigg\}.
\eeq

Note the domain of validity of the above results is for rather small values of $v$ below $v \approx R^2 \yc^3/(1+\yc)^3$, where as previously our expression for the transition point is approximate due to missing finite $R$ effects and the neglect of hard parton recoil. For larger values of $v$ up to a maximum of $v \approx R^2 \yc^2$ one obtains a $1/v \ln^3 v$ behaviour which we do not explicitly compute as we are interested only in the asymptotic small jet-mass limit. Our point is simply that in the small $v$ limit the flaw in the mass-drop tagger that causes us to follow the soft branch, leads to a change in the logarithmic behaviour from that observed at leading order. While the leading order result was purely single-logarithmic, which looked promising in terms of reducing the background, at NLO one encounters $\alpha_s^2 \ln^2v/v$ terms. While these are still less singular than the double logarithms one meets in the plain jet-mass, such behaviour would evidently still require resummed calculations to address the issue of large logarithms in the perturbative expansion at all orders. It is not in fact clear that a compact resummed formula can be written down for the mass-drop tagger which also incorporates the above wrong-branch effects. Hence it is desirable to eliminate these terms via possibly modifying the tagger. 

\begin{figure}
\begin{center}
\includegraphics[width=0.49 \textwidth]{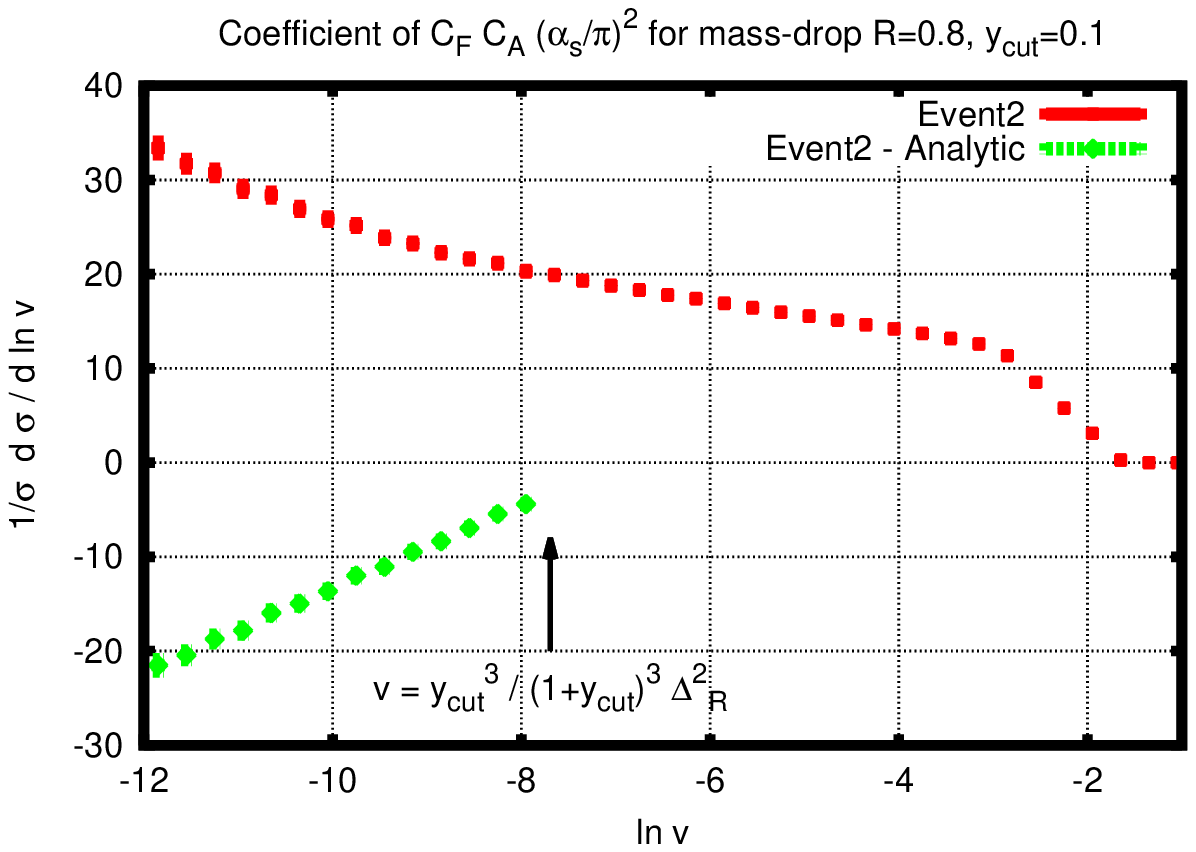}
\includegraphics[width=0.49 \textwidth]{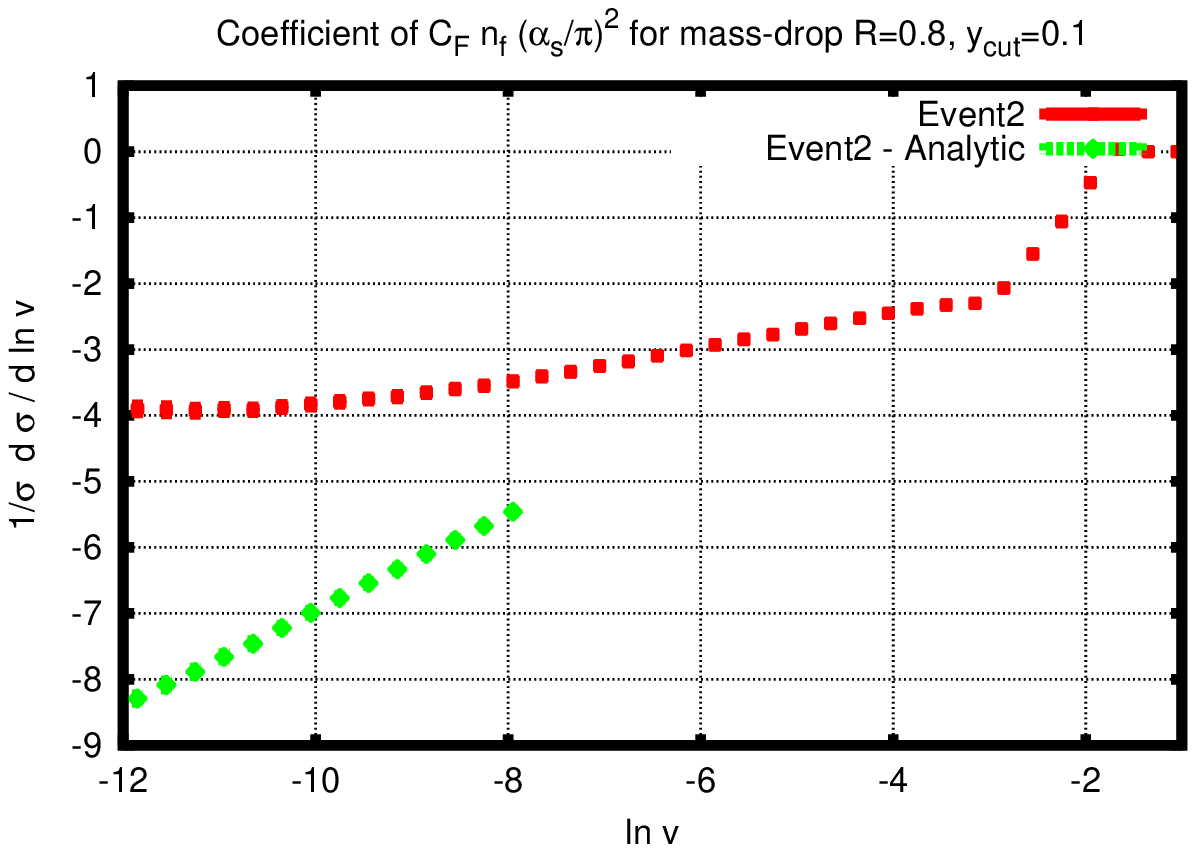}
\includegraphics[width=0.49 \textwidth]{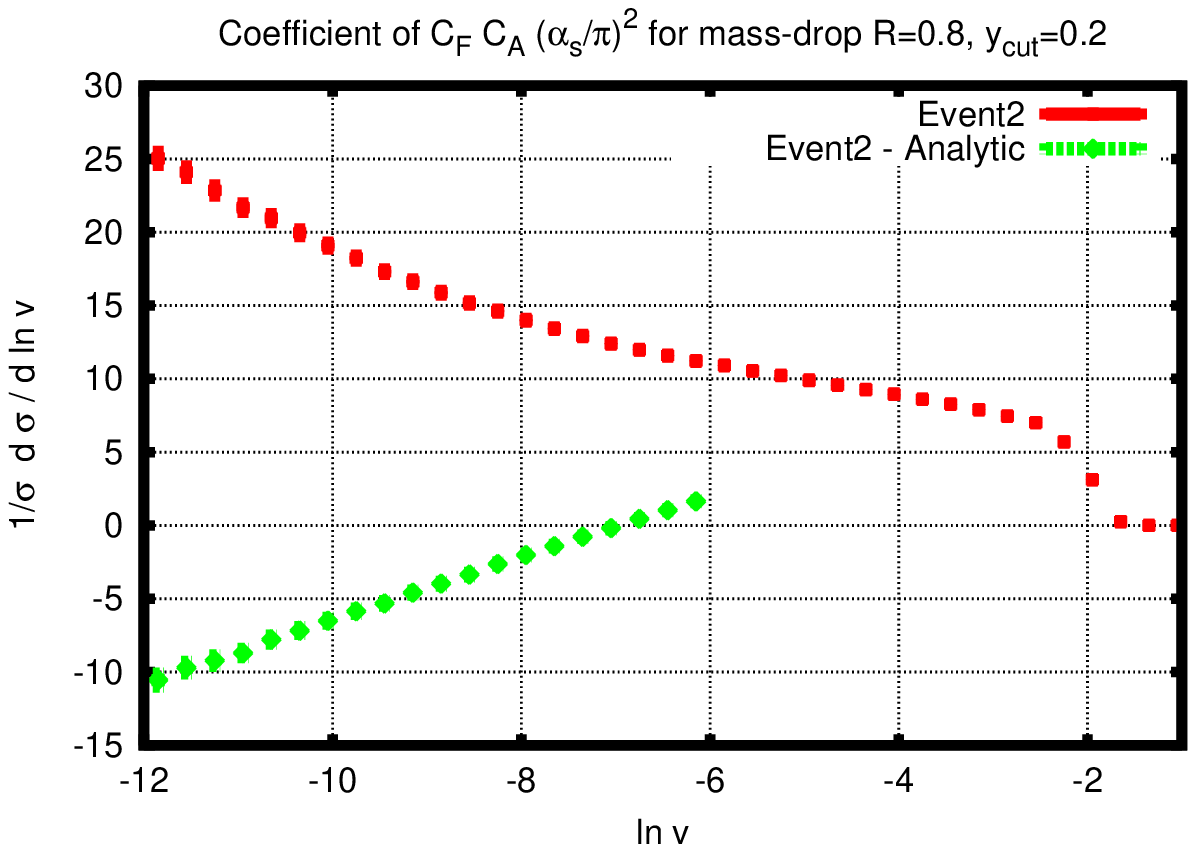}
\includegraphics[width=0.49 \textwidth]{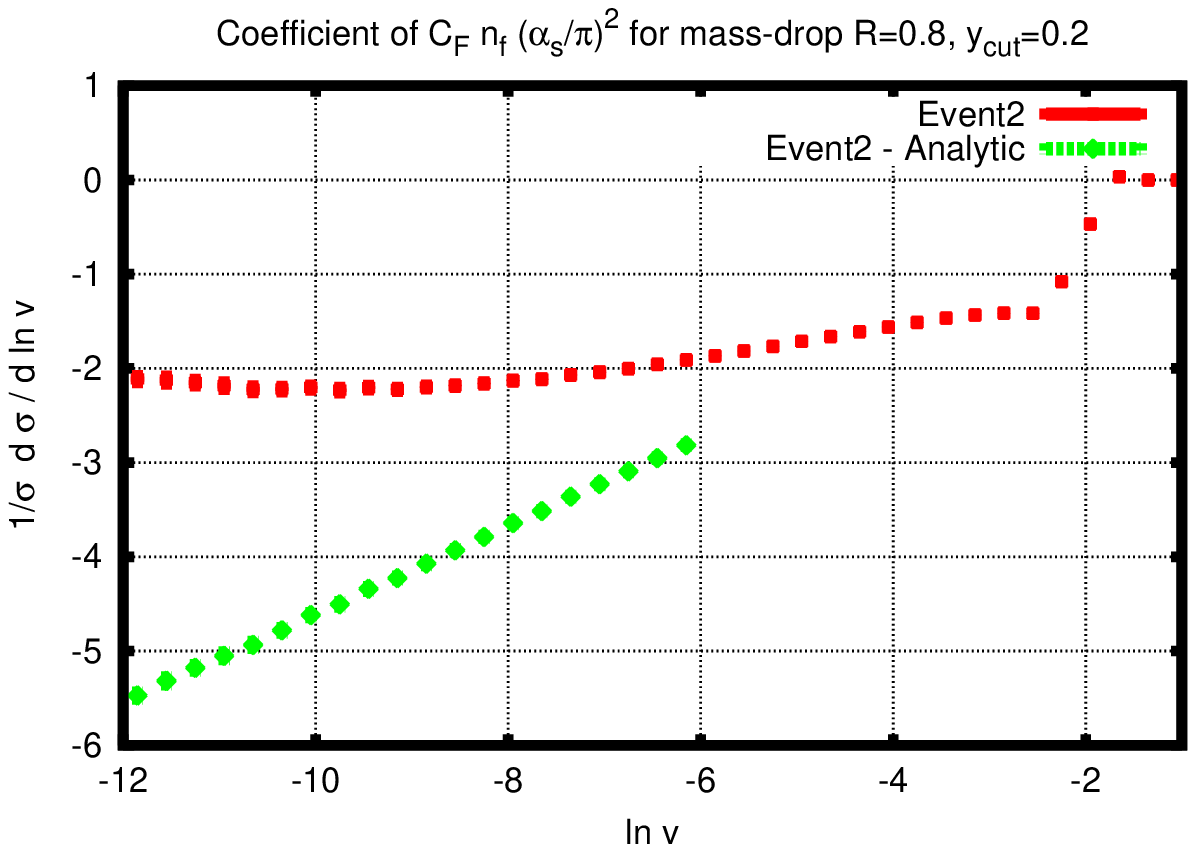}
\includegraphics[width=0.49 \textwidth]{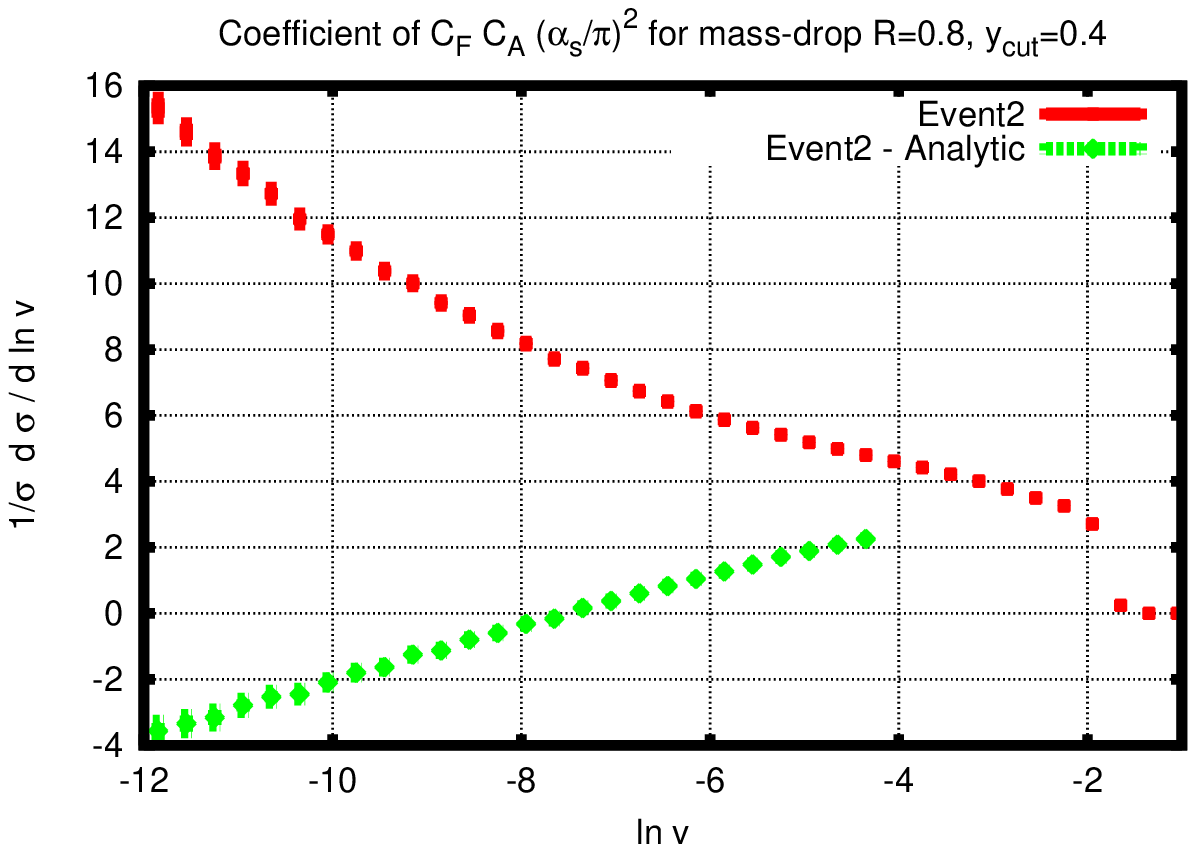}
\includegraphics[width=0.49\textwidth]{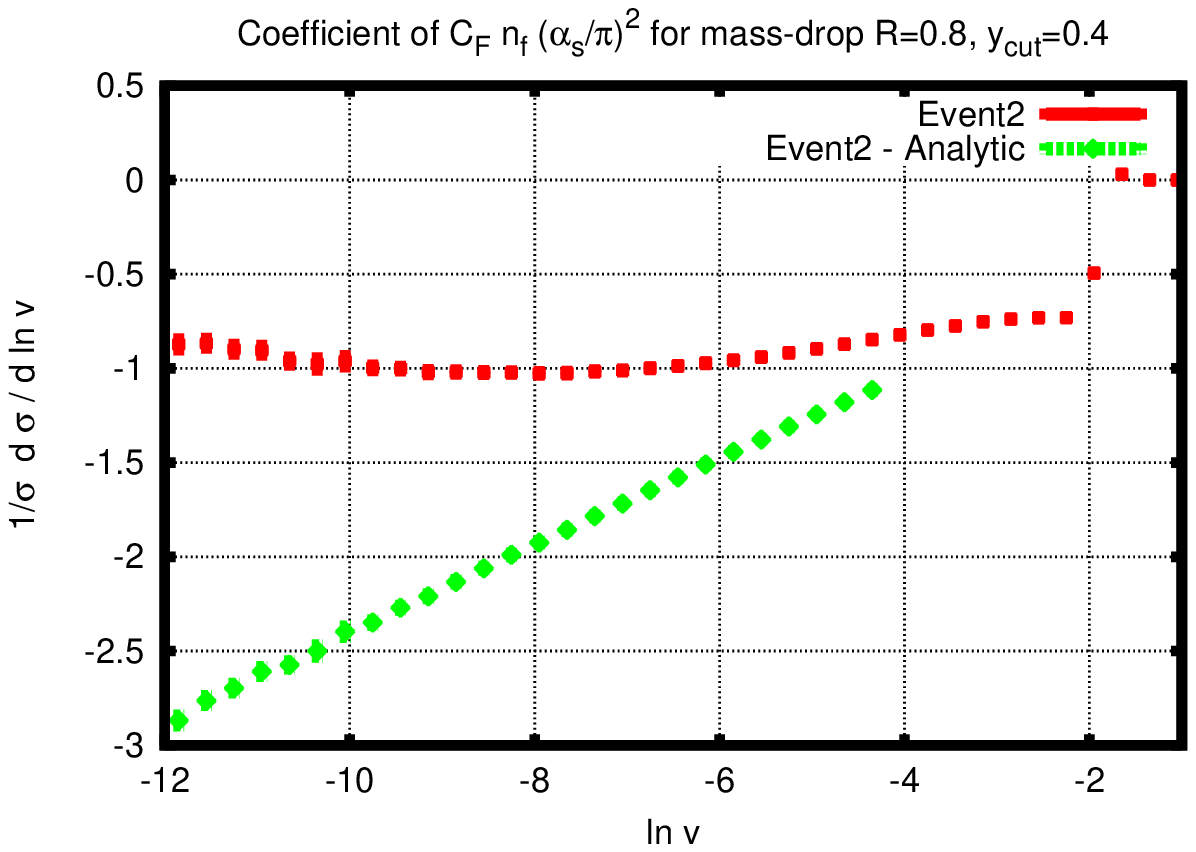}
\caption{
Comparison of the NLO analytic calculations Eq.~(\ref{MDJg1}), on the left, and Eq.~(\ref{MDJg2}), on the right, with \event2, in the region $v< \frac{\yc^3}{(1+\yc)^3}\Delta_R^2$, for different values of $\yc$. The plots demonstrate that the extra logarithm for MDT in the $C_F C_A$ and $C_F n_f$ channels, is correctly captured by the our calculation: the difference between analytical and \event2 results (in green) is consistent with a linear behaviour at small $v$ corresponding to a single-logarithmic leftover, $\alpha_s^2 \ln^2 1/v$, in the integrated distribution.}
\label{fig:MDTwrongbranch}
\end{center}
\end{figure}
Prior to suggesting any modification of the mass-drop we check our calculation Eqs.~(\ref{MDJg1}) and (\ref{MDJg2}) against \event2 for different values of $\yc$. The results are reported in Fig.~\ref{fig:MDTwrongbranch}: when we subtract our calculation of the extra-logarithm from \event2 we obtain a straight line for $d\sigma/d\ln v$ plotted against $\ln v$, which implies a single logarithmic behaviour. This indicates that we control the more divergent behaviour we have subtracted. 

In practice it turns out, as argued and demonstrated in more detail in the companion paper~\cite{DasFregMarSal}, that the numerical effect of following the wrong-branch is small for a variety of reasons. However given that the role of the MDT was to identify hard substructure within a jet it is clearly an unintended anomaly that a soft jet is returned. Also it is of interest to see if removing the wrong-branch problem will lead to a tagger where the jet-mass is purely single-logarithmic at all orders which may turn out to be simpler 
to compute via, for example, resummation. To this end we now consider the modified Mass Drop Tagger (mMDT)~\cite{DasFregMarSal} and its logarithmic structure at NLO.

\section{The modified Mass Drop Tagger}
\subsection{Definition and leading-order calculation}
The modification of the mass-drop tagger that we have proposed in Ref.~\cite{DasFregMarSal} is to replace step 3 of the definition of MDT, with
\begin{enumerate}
\item[3.] Otherwise, redefine $j$ to be the {\it{harder}} between $j_{1}$ and $j_2$ and go back to step 1 (harder means higher transverse mass, $m^2+p_t^2$, or $p_t$ at hadron colliders or more energetic at $e^+e^-$ colliders).
\end{enumerate}
It is fairly obvious that at LO there is no difference between MDT and mMDT, so we refer to section~\ref{sec:MDTLO} for discussions and results.
\subsection{Next-to--leading order calculation: independent emission contribution}
\label{sec:MMDT_NLO}
\begin{figure}
\begin{center}
\includegraphics[width=0.3 \textwidth]{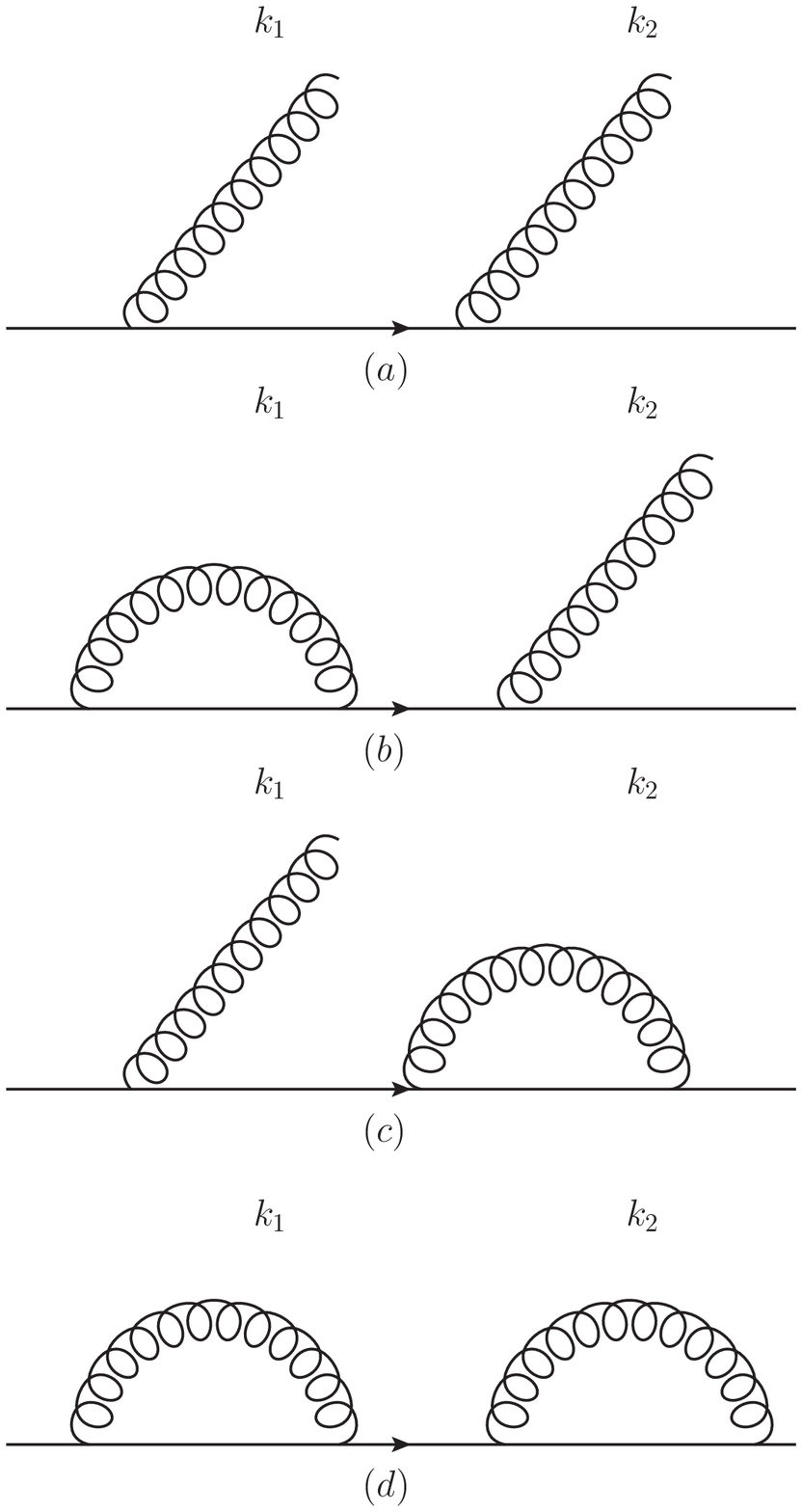}
\caption{Real and virtual contributions to the NLO jet-mass distribution in the $C_F^2$ channel.}
\label{fig:indep}
\end{center}
\end{figure}
Here we shall carry out an approximate NLO calculation for the mMDT jet-mass, exploring all the configurations that give rise to large logarithms in the traditional jet-mass. We shall concentrate on the $v \to 0$ limit, dropping all contributions that are not enhanced.

We start by addressing the independent double-soft and collinear emission of 
two real gluons from a quark (or antiquark), and the corresponding virtual 
corrections. These configurations are well known to be the source of the 
leading double logarithms in the plain jet-mass. We shall see that the modified mass drop procedure reduces this to a single-logarithmic (pure collinear) 
dependence, which is consistent with an exponentiation of the leading-order result Eq.~(\ref{LOmdt}). Details of the resummed calculation formally deriving this exponentiation, using standard techniques can be found in the accompanying article~\cite{DasFregMarSal}. 

Let us consider the independent emission, from a quark, of soft gluons $k_1$ and $k_2$, as depicted in Fig.~\ref{fig:indep}, such that all partons combine 
into a fat \CA jet with radius $R$. One can consider all cases that lead to a non-vanishing jet mass involving double-real emission as well as one-real -- one-virtual contributions. Also shown in the figure is the double-virtual configuration which does not lead to a finite jet-mass and can thus be ignored for our calculation of the differential distribution. In the soft approximation we shall ignore the recoil of the quark against the soft gluons and hence the jet-axis will be given by the quark direction. 

We can write the momenta of the emitted gluons as 
\bea
k_1 &=&  x_1 E_j \left(1,0,\sin \theta_1, \cos \theta_1 \right) \\ \nonumber
k_2 &=& x_2 E_j \left(1, \sin \theta_2 \sin \phi, \sin \theta_2 \cos \phi, \cos \theta_2 \right)
\eea
where $x_i$ are the energy fractions of the fat-jet energy $E_j$ carried by parton $k_i$ while that carried by the hard parton $p_1$ is $1-x_1-x_2$. The angle 
between $k_1$ and $k_2$ shall be denoted by $\theta_{12}$. We start by 
examining the region where $\theta_1^2, \theta_2^2 < R^2$ and $\theta_1^2$ is 
the smallest angular distance, $\theta_1^2 < \theta_2^2, \theta_{12}^2$ and 
hence in the \CA algorithm $k_1$ is clustered to the hard parton, $p_1$, first and $k_2$ is clustered next. When one undoes the algorithm, $k_2$ emerges first and the jet breaks into two subjets consisting of $k_2$ and the massive jet $j_1$ generated by the $p_1$ and $k_1$ recombination. Now one has to take into account that for the overall jet $j$ to be accepted there has to be a mass-drop and the splitting involving emission of $k_2$ should not be too asymmetric.

The mass-drop condition implies $m_{j_1} < \mu m_j$, 
which translates into
\begin{equation}
(p_1+k_1)^2 < \mu (p_1+k_1+k_2)^2,
\end{equation}
which in terms of energy fractions and angles of soft partons gives rise to the mass-drop constraint 
\begin{equation}
x_2 \theta_2^2 > f x_1 \theta_1^2,
\end{equation}
where $f =\frac{1-\mu}{\mu}$ and we have dropped terms bilinear in the soft parton 
momenta, which do not contribute to the large logarithms we aim to compute.
Moreover to satisfy the fact that the splitting should not be too asymmetric, in the soft limit we require the energy fraction $\frac{\yc}{1+\yc}<x_2 <\frac{1}{1+\yc}$. For the composite jet consisting of all three partons to be accepted and to have a given mass $v$ we thus have the constraint 
\begin{equation}
\Theta \left ( \frac{1}{1+\yc} -x_2 \right)\Theta \left ( x_2 -\frac{\yc}{1+\yc }\right) \Theta\left ( x_2 \theta_2^2 -f x_1 \theta_1^2 \right) \delta \left(v- x_1 \theta_1^2 -x_2 \theta_2^2 \right).
\end{equation}

Next we consider the situation that the mass-drop is satisfied but the energy 
asymmetry cut fails due to $k_2$ being too soft, i.e $x_2 < \yc/(1+\yc)$. 
The mMDT then moves on to consider the hardest subjet which is the one made by the gluon $k_1$ and the quark. For this jet, one then obtains essentially the leading order (single gluon) situation with the jet being accepted if the asymmetry cut is satisfied by the emission $k_1$ and rejected otherwise. 
Lastly, there is the possibility that the mass-drop fails in which case one has to impose the constraint on the hardest subjet  again as before. To obtain a non-zero jet mass the hardest jet must be the one involving $p_1$ and $k_1$ 
which implies $1-x_2 >x_2$ or $x_2 < 1/2$.
The contribution of the double-real emission can then be summed up as 
\bea \label{MDTdoublerealtheta}
\Theta^\text{double real}&=& \Theta \left ( \frac{1}{1+\yc} -x_2 \right)\Theta \left ( x_2 -\frac{\yc}{1+\yc }\right) \Theta \left(x_2 \theta_2^2 - f x_1 \theta_1^2 \right) \delta \left ( v -x_1 \theta_1^2-x_2 \theta_2^2 \right) \nonumber \\
&+&\Bigg[ \Theta \left(x_2 \theta_2^2 -f x_1 \theta_1^2 \right)  \Theta \left ( \frac{\yc}{1+\yc}-x_2\right)  
 +\Theta \left (f x_1 \theta_1^2 - x_2 \theta_2^2 \right) \Theta\left(\frac{1}{2}-x_2 \right) \Bigg] \nonumber \\ && \Theta \left(x_1-\frac{\yc}{1+\yc} \right)
 \Theta \left(\frac{1}{1+\yc}-x_1\right) \delta \left ( v-x_1 \theta_1^2 \right). \nonumber\\
\eea

Next considering the case of one real and one virtual emission in precisely the same way (and noting that the real emission has to pass the asymmetry cut to 
yield a non-zero jet-mass value) we can write the summed contribution of the 
independent emission terms as 
\bea
\label{MDTrealvirttheta}
\Theta^\text{1 real, 1 virt} &= &
-\Theta \left(x_1-\frac{\yc}{1+\yc} \right)
 \Theta \left(\frac{1}{1+\yc}-x_1\right) \delta \left (v - x_1 \theta_1^2 \right) \nonumber\\ &&- \Theta \left(x_2-\frac{\yc}{1+\yc} \right)
 \Theta \left(\frac{1}{1+\yc}-x_2\right) \delta \left(v -x_2 \theta_2^2 \right),
\eea
where we also included a minus sign in each term above, relevant to the 
emission probability for a virtual gluon, which is otherwise identical in the eikonal approximation to the corresponding real emission probability.

The complete set of constraints is given by $\Theta^\text{nlo}=\Theta^\text{double real}+\Theta^\text{1 real, 1 virt}$. We first point out that the combination of the first contribution in Eq.~(\ref{MDTdoublerealtheta}) and of the last 
in Eq.~(\ref{MDTrealvirttheta}) results in cancellation of the divergences in the limit where $x_1$ or $\theta_1$ tend to zero. As a consequence, these 
two terms combined do not produce large logarithms in the jet mass.
 
We can combine the remaining terms into the expression
\bea \label{MDTnlotheta}
\Theta^\text{nlo}&=& -\Theta \left( x_1-\frac{\yc}{1+\yc} \right) \Theta \left( \frac{1}{1+\yc}-x_1 \right) \delta \left (v-x_1 \theta_1^2 \right) \nonumber\\ &\times&
\left[ \Theta \left( x_2-\frac{\yc}{1+\yc} \right) \Theta \left ( x_2 \theta_2^2 -f x_1 \theta_1^2 \right)
+   \Theta\left(x_2- \frac{1}{2} \right) \Theta \left (f x_1 \theta_1^2 - x_2 \theta_2^2 \right)  
\right].\eea
It is straightforward to show that the second term in the square brackets will not lead to a logarithmic enhancement, so we shall drop it in the rest of the 
calculation. The above constraints then have to be considered together with the two-gluon independent emission in the required phase space region:
\bea
\label{eq:secord}
\frac{1}{\sigma}\frac{d \sigma}{d v}^\text{(mMDT, $C_F^2$)} &=& \left ( \frac{ C_F \alpha_s}{\pi} \right)^2 \int \frac{d x_1}{x_1} \frac{d x_2}{x_2} \frac{d \theta_1^2}{\theta_1^2} \frac{d  \theta_2^2}{\theta_2^2} \frac{d\phi}{2\pi}\, \Theta^\text{nlo} \, \Theta \left (\theta_2^2- \theta_1^2 \right) \Theta \left(\theta_{12}^2- \theta_1^2 \right).
\eea
The angular constraints in the above result originate from the fact that as we mentioned before we are studying the region $\theta_1^2 < \theta_2^2, \theta_{12}^2$ where $ \theta_{12}^2 \approx \theta_1^2+\theta_2^2 -2 \theta_1 \theta_2 \cos \phi $ and $\phi$ is an azimuthal angle. We have also inserted a factor of two in order to take care of the situation where $\theta_2$ is the smallest 
angle, which by symmetry gives an identical result to the case considered above. 

Further we note that for the purpose of extracting only the logarithmic behaviour we seek here, it is possible to make additional simplifications. To be specific one can ignore the constraint $\Theta \left(\theta_{12}^2- \theta_1^2 \right)$, since we have checked that to our accuracy it suffices to consider only 
the region 
$\theta_2^2>\theta_1^2$
over the full range of values of $\phi$ and hence the $\phi$ integral above is trivial. Moreover, the angular configuration for which $\theta_{12}$ is the smallest distance, and 
therefore the gluons are recombined together, does not contribute to mMDT distribution, to the accuracy considered here. This is not the case for the MDT, where one obtains a single-logarithmic contribution due to following the soft massive branch that results. We also do not explicitly indicate above the constraint that the angles $\theta_{i}^2$ are of course less than the fat jet radius $R^2$.

The angular integrals in Eq.~(\ref{eq:secord}) are thus easily done and one arrives at
\bea
\frac{1}{\sigma}\frac{d \sigma}{d v}^\text{(mMDT, $C_F^2$)} &=&
-  \left ( \frac{C_F \alpha_s}{\pi} \right)^2\frac{1}{v}  \int_{ \frac{\yc}{1+\yc}}^{ \frac{1}{1+\yc}} \frac{d x_1}{x_1} \int_{ \frac{\yc}{1+\yc}}^{1} \frac{d x_2}{x_2} 
\left[ \Theta(x_2-f x_1)\Theta \left(R^2-\frac{v}{x_1} \right)\right.  \nonumber \\&& \left. \times \ln \frac{x_1 R^2}{v} +
\Theta(f x_1-x_2)\Theta \left(R^2-\frac{v f}{x_2} \right) \ln \frac{R^2 x_2}{v f}
\right]. \nonumber \\
\eea

We have now to compute the integrals over the energy fractions $x_1$ and $x_2$. In order to capture also hard-collinear contributions one needs to 
additionally account for energetic emissions in the collinear domain. In the present case as for the plain jet-mass it suffices to replace the factors $1/x_i$ with the full LO splitting function $p_{gq}(x_i)$. Although the algebra is rather cumbersome because we have to consider many different integration regions, the final result we obtain is remarkably simple. In particular, if we focus on the region $v<\frac{\yc}{1+\yc} R^2$ identified already in the leading-order result derived before, we find:
\beq \label{MDTCF2final}
\frac{1}{\sigma}\frac{d \sigma}{d v}^\text{(mMDT, $C_F^2$)}  =  - \left ( \frac{C_F \alpha_s}{\pi} \right)^2\frac{1}{v}  
\ln f_q(\yc)   \left[\ln \left((1+\yc)f_q(\yc)\right)-\frac{\yc(3 \yc+2)}{4(1+\yc)^2}\right] 
\ln \frac{R^2}{v},
\eeq
where we have introduced $f_q(\yc)=\frac{1}{\yc} e^{-\frac{3}{4}\left(\frac{1-\yc}{1+\yc}\right)}$.
The NLO result Eq.~(\ref{MDTCF2final}) exhibits a single logarithmic behaviour.  The analytic calculation gives a clear picture of the underlying physics. The logarithms in the mMDT distribution stem from the collinear region $\theta_i^2 \ll 1$. The integrals over the energy fractions $x_i$ are bounded so that 
logarithms in the jet mass of soft origin are replaced by logarithms of the cut-off $\yc$. Soft emissions at large angles do not make a contribution to single-logarithmic accuracy. A feature of our results is that the argument of the logarithms we compute is $R^2/v$, where the $R$ dependence we obtain is not exact due to the use of the collinear approximation. We note that the $R$ dependence only modifies our answer at next-to--leading logarithmic level i.e below the single logarithmic accuracy we aim to control here, and hence we do not extend our calculations to include finite $R$ corrections but continue to work purely in the collinear (small-$R$) approximation.

\begin{figure}
\begin{center}
\includegraphics[width=0.49 \textwidth]{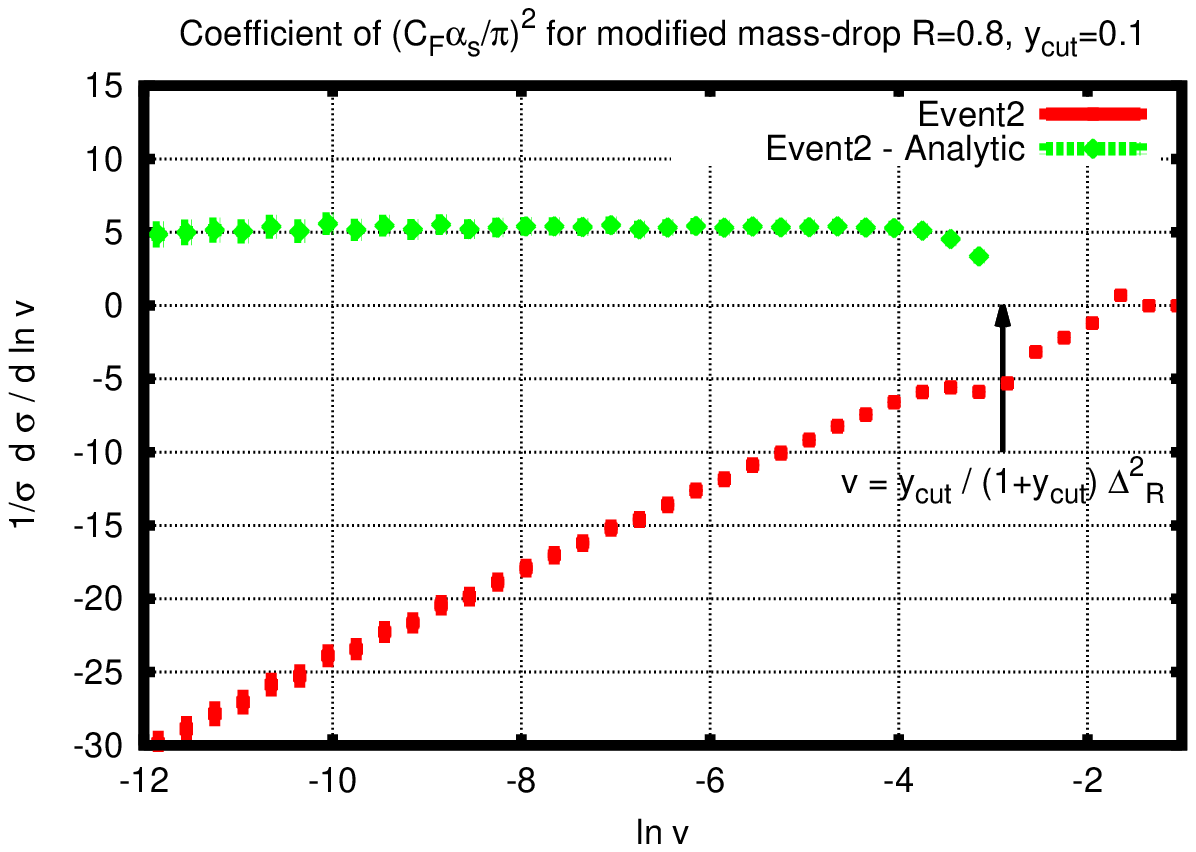}
\includegraphics[width=0.49 \textwidth]{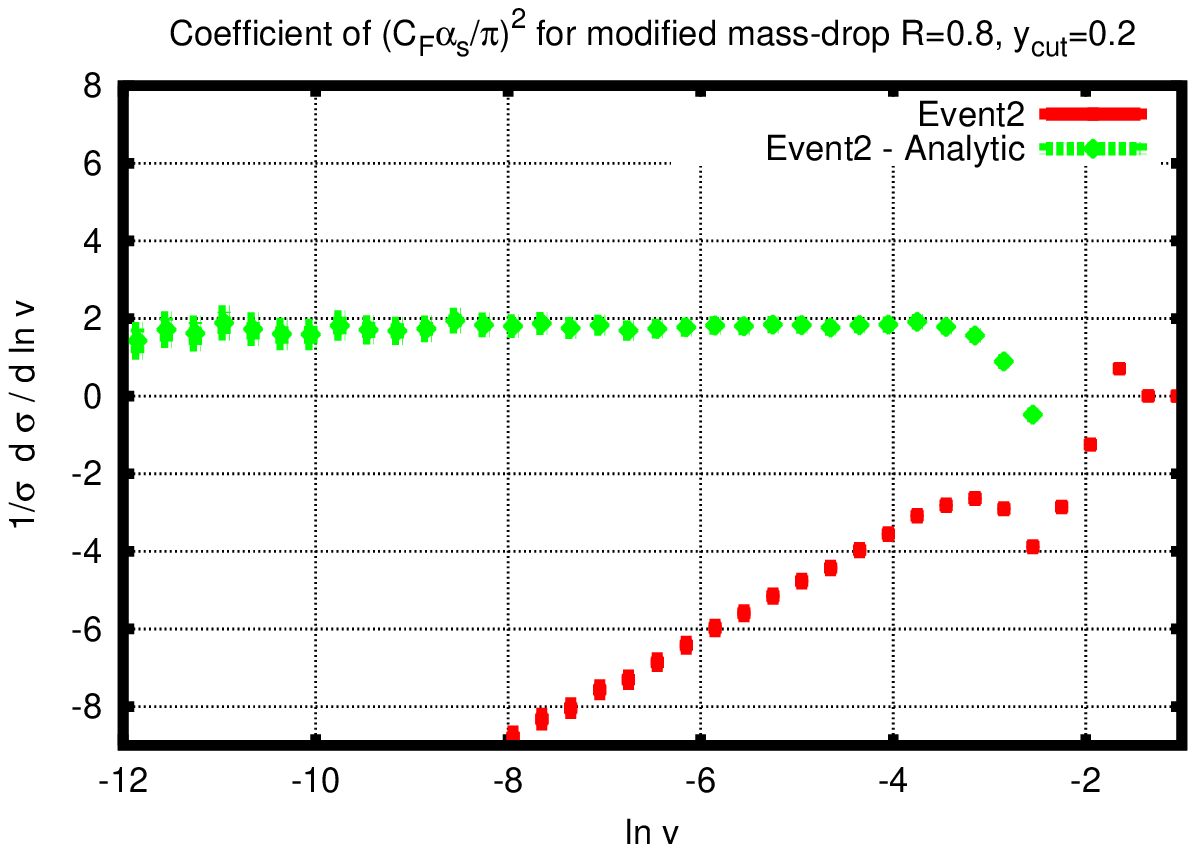}
\includegraphics[width=0.49 \textwidth]{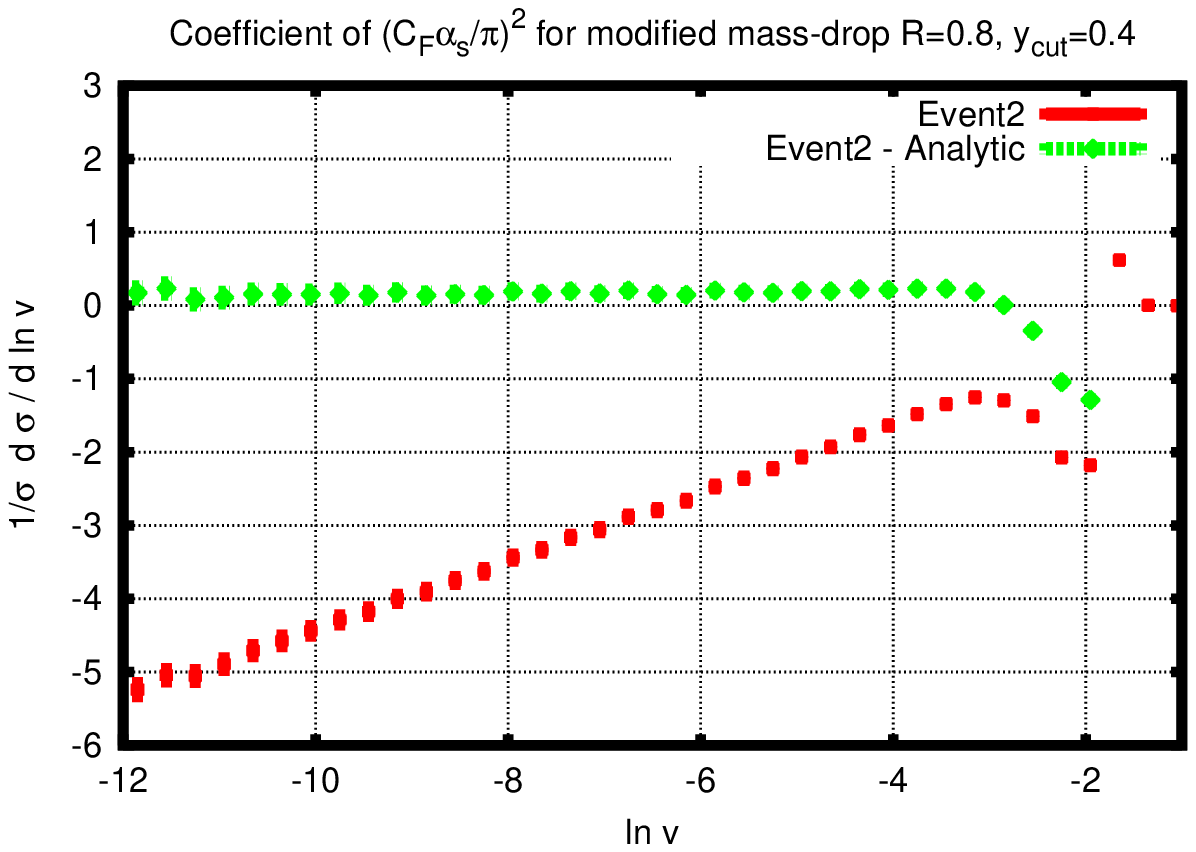}
\caption{
Comparison of the analytic calculation Eq.~(\ref{MDTCF2final}) with \event2 for the coefficient of $C_F^2$, in the region $v< \frac{\yc}{1+\yc}\Delta_R^2$, for different values of $\yc$. The red curve shows the fixed-order result alone which behaves like a straight line at small $v$ and hence indicates a single logarithmic behaviour for the integrated distribution. The green curve indicates that, after subtracting our analytical calculation, the result is flat at small $v$, as expected.}
\label{fig:MDCF2}
\end{center}
\end{figure}

The comparison of our analytical calculation to \event2, for the $C_F^2$ channel, is shown in Fig.~\ref{fig:MDCF2}. It indicates that on subtracting our 
results from those of \event2 we obtain a constant at small $v$, for $d\sigma/d \ln v$. This implies that we have eliminated the $\alpha_s^2 \ln v$ term that dominates $d\sigma/d \ln v$ at this order, indicating the correctness of our result.

Additionally if we take the small-$\yc$ limit of the LO and NLO results obtained so far, Eq.~(\ref{LOmdt}) and Eq.~(\ref{MDTCF2final}) respectively, we have
\bea
\frac{1}{\sigma}\frac{d \sigma}{d v}^\text{(mMDT, LO)}&=&   \frac{C_F \alpha_s}{\pi} \frac{1}{v}  \ln \frac{e^{-\frac{3}{4}}}{\yc}+\dots, \nonumber \\
\frac{1}{\sigma}\frac{d \sigma}{d v}^\text{(mMDT, $C_F^2$)} &=&  -\left ( \frac{C_F \alpha_s}{\pi} \right)^2\frac{1}{v}  \ln^2 \frac{e^{-\frac{3}{4}}}{\yc} \ln \frac{R^2}{v}+ \dots
\eea
which is consistent with the exponentiation of the LO term i.e. with an all-order structure whose small $\yc$ limit reads:
\begin{equation}
\frac{1}{\sigma}\frac{d \sigma}{d v}^\text{(mMDT, all-orders)} = \frac{d}{dv} \exp \left(-C_F \frac{\alpha_s(E_j R)}{\pi} \ln \frac{e^{-3/4}}{\yc} \ln \frac{R^2}{v}\right),
\end{equation}
where we have set the scale of the coupling in the exponent to be $E_j R$~\cite{BanDasKKKMar}.

Beyond the small $\yc$ limit, i.e including finite $\yc$ terms, the mMDT jet-mass still admits exponentiation. In fact we shall encounter more finite $\yc$ terms after having considered the $C_F C_A$ and $C_F n_f$ contributions in the 
next sub-section. An all-order proof of exponentiation for the mMDT, including a matrix structure to treat finite $\yc$ effects, is presented in~\cite{DasFregMarSal}.

\subsection{Next-to-leading order calculation: flavour changing contributions}
Now we shall turn to the $C_F C_A$  and $C_F n_f$ channels. 
There are several aspects to be considered there. First of all, we need to consider one-loop running coupling corrections to the LO result, which can be 
obtained by redoing the leading-order calculation with proper account of the 
fact that the argument of the running coupling is the transverse momentum $k_t$ of the soft emission with respect to the emitting parton, and yields (see also~\cite{DasFregMarSal}):
\begin{equation}
\frac{1}{\sigma} \frac{d\sigma}{d v}^\text{(mMDT, r.c.)} = C_F \frac{\alpha_s{(E_j R)}}{\pi} \, \frac{1}{v} \, \frac{1}{1-\lambda} \ln \left(\frac{1}{\yc} e^{-\frac{3}{4}\left(\frac{1-\yc}{1+\yc}\right)} \right), \, \lambda =\beta_0 \frac{\alpha_s(E_j R)}{\pi} \ln \frac{R^2}{v}.
\end{equation}
The above result gives an NLO contribution of the form
\begin{equation} \label{CFCArc}
\frac{1}{\sigma} \frac{d\sigma}{d v}^\text{(mMDT, NLO r.c.)} =C_F  \beta_0 \left( \frac{ \alpha_s}{\pi } \right)^2  \frac{1}{v} \ln \left(f_q(\yc)\right)  \ln \frac{R^2}{v}.
\end{equation}

The second effect to be considered, which yields relevant single logarithms, 
corresponds to a splitting of the type depicted in Fig.~\ref{fig:nonAbelian}, 
where the offspring gluons $k_1$ and $k_2$ are the closest pair in angle, in a region of phase-space where the asymmetry condition fails because the parent gluon $k$ is {\it{too hard}}, i.e. $x> \frac{1}{1+\yc}$. The mMDT will then follow the branch corresponding to the massive gluon jet, rather than the quark jet leading to a {\it{flavour changing}} contribution of the form:
\begin{multline}
C_F C_A \left(\frac{\alpha_s}{\pi} \right)^2 \int \frac{d x}{x} \frac{d \theta^2}{\theta^2}   
\Theta\left(x- \frac{1}{1+\yc}\right)  \\  \Theta \left (R^2-\theta^2\right)
\int d z \left(\frac{1}{2}p_{gg}(z)+\frac{T_R n_f}{C_A}p_{qg}(z) \right)  \frac{d\theta_{12}^2}{\theta_{12}^2}  \delta \left(v-z(1-z) x^2 \theta_{12}^2 \right) \\  \Theta \left ( z-\frac{\yc}{1+\yc}\right) \Theta \left( \frac{1}{1+\yc}-z\right)\Theta \left(\theta^2-\theta_{12}^2 \right).
\end{multline}

In the above we have made the collinear approximation for the gluon splitting $\theta_{12} \ll 1$ as that is the relevant region which produces the single-logarithmic term we seek. Also, in the region of integration where the parent gluon $k$ becomes collinear to the quark, we make the usual substitution $\frac{1}{x}\to p_{gq}(x)$, to account for hard collinear corrections. Separating $C_F C_A$ and $C_F n_f$ contributions, in the region $v< \frac{\yc}{1+\yc} R^2$, we obtain:
\bea \label{MMDJg1}
\frac{1}{\sigma}\frac{d \sigma}{d v}^\text{(mMDT, $C_F C_A$)}&=&
C_F C_A   \left(\frac{\alpha_s}{\pi} \right)^2 \frac{1}{v}  \left( \ln\frac{1}{\yc}+\frac{11 \yc^3+9 \yc^2-9 \yc-11}{12 \left(1+\yc \right)^3} \right) \nonumber \\&\times&
 \left( \ln (1+\yc) -\frac{\yc (2+3\yc)}{4(1+\yc)^2}\right)
  \ln \frac{R^2}{v},
\eea
and
\beq \label{MMDJg2}
\frac{1}{\sigma}\frac{d \sigma}{d v}^\text{(mMDT, $C_F n_f$)}=
C_F n_f  \left(\frac{\alpha_s}{\pi} \right)^2 \frac{1}{v} \frac{1-\yc ^3}{6 \left(1+\yc \right)^3} \left( \ln (1+\yc) -\frac{\yc (2+3\yc)}{4(1+\yc)^2}\right)
 \ln \frac{R^2}{v}.
\eeq

Thus, we conclude that the behaviour of mMDT in the $C_F C_A$ and $C_F n_f$ channels  is {\it{single-logarithmic}}, while the MDT exhibits an extra logarithm as in Eqs.~(\ref{MDJg1}, \ref{MDJg2}). The reason for this difference is that the flavour changing contribution in the mMDT arises from an {\it{energetic}} rather than a soft parent gluon while for the MDT contribution we considered the 
splitting of a soft parent gluon which resulted essentially in an extra 
logarithm in $v$. Moreover due to the limited phase space available for the 
failure of the asymmetry condition and the fact the we have a hard parent gluon, the above results vanish as $\yc \to 0$. As values of $\yc \sim 0.1$ are commonly used in phenomenology we can expect that the impact of the terms above, on the behaviour of the mMDT jet-mass distribution will be at best modest. In any case their resummation is simple and contributes to the flavour changing matrix structure of the resummed answer, stated in the companion article~\cite{DasFregMarSal}.

\subsection{Non-global logarithms}
Finally, we turn our attention to the issue of non-global logarithms~\cite{Dassalam1,Dassalam2}. We have already noted that the leading (single) logarithms 
are collinear in origin. When calculating the non-global terms in the usual jet mass distribution~\cite{BanDasKKKMar} one considers soft large-angle contributions with similar opening angles and the non-global logarithms arise from integrating over energy fractions. In the case of the mMDT, the integrals over the energy fractions are cut off by $\yc$  and hence one may anticipate that the non-global logarithms in $v$ are absent. To be more explicit, to study the leading-order non-global contribution, we take account of the situation when a gluon with momentum $k_1$ is emitted outside the jet and it emits a softer gluon $k_2$ inside the jet, which gives rise to non-global logarithms in the plain jet mass. Further we shall ignore the effect of soft gluon clustering in the \CA algorithm, which has been shown to reduce non-global logarithms significantly~\cite{AppSey1,BanDasDel}, which amounts to computing an upper bound on the non-global contribution.
Thus one has the constraint\footnote{Since non-global logarithms arise away 
from the collinear region, in this calculation we shall not make a collinear approximation and hence use the distance measure $\Delta_\alpha^2=2(1-\cos\alpha)$.}
\begin{equation}
\Theta^{\mathrm{NG}} = \Theta \left (x_1-x_2 \right) \Theta \left( \Delta_{\theta_1}^2-\Delta_R^2 \right) \Theta \left(\Delta_R^2-\Delta_{\theta_2}^2 \right) \Theta \left(\frac{1}{1+\yc}-x_2 \right )
\Theta \left(x_2-\frac{\yc}{1+\yc} \right ),
\end{equation}
where we have imposed that $x_1 >x_2$ and the condition that $k_2$ be inside 
the jet, $\Delta_R^2>\Delta_{\theta_2}^2$  while $k_1$ is outside $\Delta_R^2<\Delta_{\theta_1}^2$. We have also applied the asymmetry condition on the gluon energy fraction $x_2$, as is required to obtain a finite jet mass with the mMDT.

Considering the $C_F C_A$ correlated emission term of the squared matrix element for two gluon emission, along with the above constraint we are led to
\begin{equation}
\frac{1}{\sigma}\frac{d \sigma}{d v}^\text{(mMDT, NG)} = 4 C_F C_A \left ( \frac{\alpha_s}{2 \pi} \right)^2 \int \frac{dx_1}{x_1} \frac{d x_2}{x_2} \int  d\cos \theta_2 \int d \cos \theta_1 \, \Omega_2 \, \Theta^{\mathrm{NG}}
\delta \left(v-x_2 \Delta_{\theta_2}^2 \right).
\end{equation}
where $\Omega_2$ is the angular function~\cite{Dassalam1}
\begin{equation} \label{omega2}
\Omega_2 = \frac{2}{\left(1-\cos \theta_1\right)\left(1+\cos \theta_2\right) |\cos \theta_1-\cos \theta_2|}.
\end{equation}
We obtain
\beq
\frac{1}{\sigma}\frac{d \sigma}{d v}^\text{(mMDT, NG)} = C_F C_A \left ( \frac{\alpha_s}{2 \pi} \right)^2  \cot^2 \left( \frac{R}{2}\right)
\frac{1+\yc}{\yc}\left(\ln \frac{1}{\yc}-(1-\yc) \left(1-\ln(1+\yc) \right) \right),
\eeq
valid for $v<\frac{\yc}{1+\yc} \Delta_R^2$. Thus, 
the mMDT distribution has no non-global logarithms at $\ord\left(\as^2\right)$. For the same reason one may expect Abelian clustering logarithms discussed in 
Ref.~\cite{BanDasKKKMar,BanDas,BanDasDel} to also be absent here. 

Although we have only carried out an order $\alpha_s^2$ calculation we believe that the mMDT is free of non-global logarithms at any order, due to the application of the asymmetry cut. While the calculation carried out here applies also to the original MDT, which is also free of non-global logarithms at order $\alpha_s^2$, the all-order statement does not apply to that case. To see this one can consider a soft emission that emits a softer gluon into the jet as in the present case. The soft and collinear branching of the soft emission inside the 
jet (which occurs at the order $\alpha_s^3$ level) generates a massive gluon jet, which the tagger follows due to the wrong-branch issue. This leads to an $\alpha_s^3 \ln^3 v$ non-global contribution which then generalises to higher orders. This is yet another good reason to abandon the original MDT.

\begin{figure}
\begin{center}
\includegraphics[width=0.49 \textwidth]{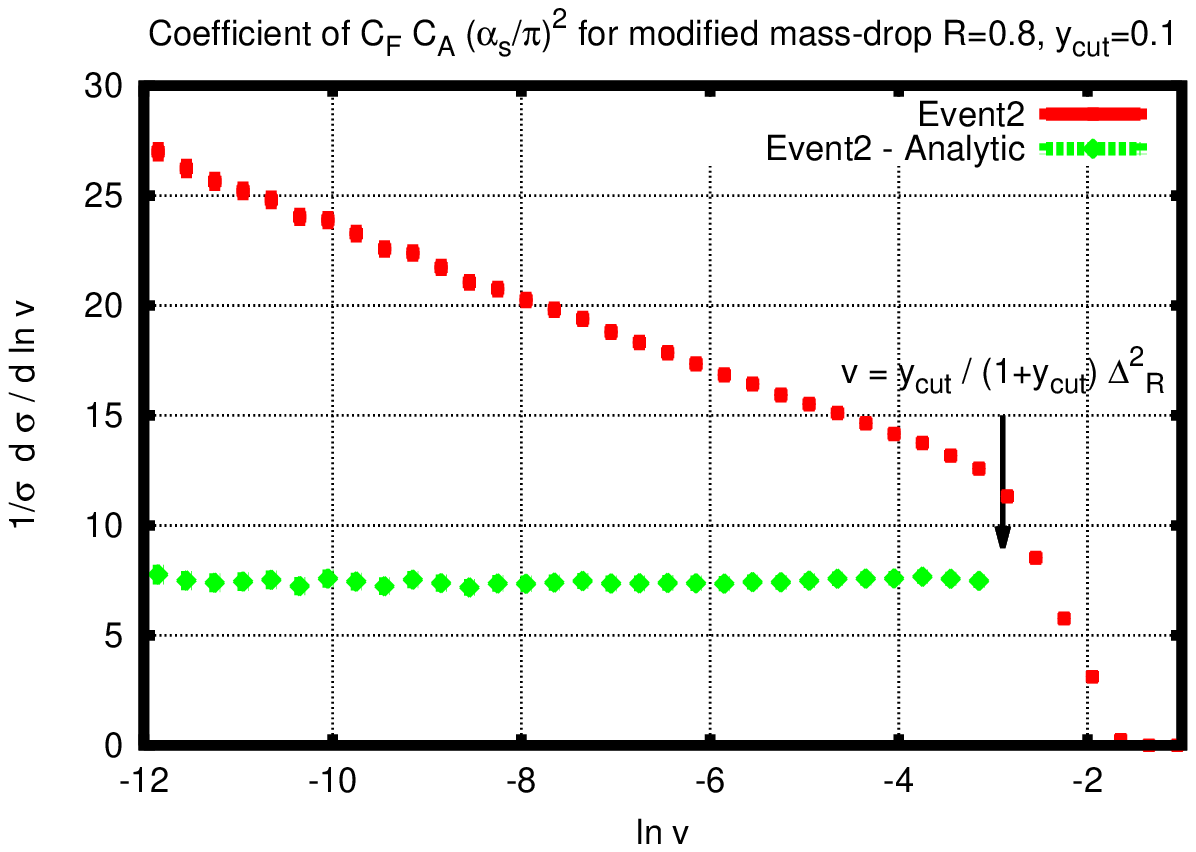}
\includegraphics[width=0.49 \textwidth]{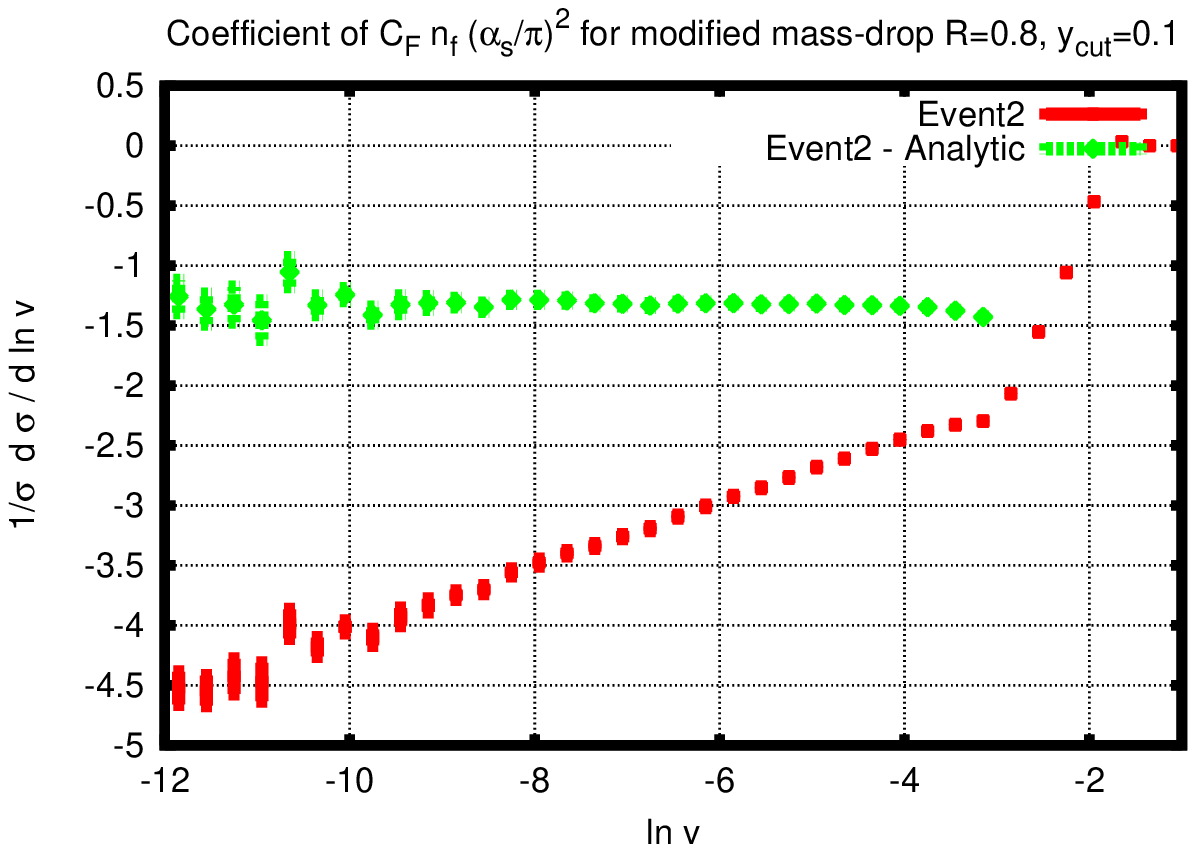}
\includegraphics[width=0.49 \textwidth]{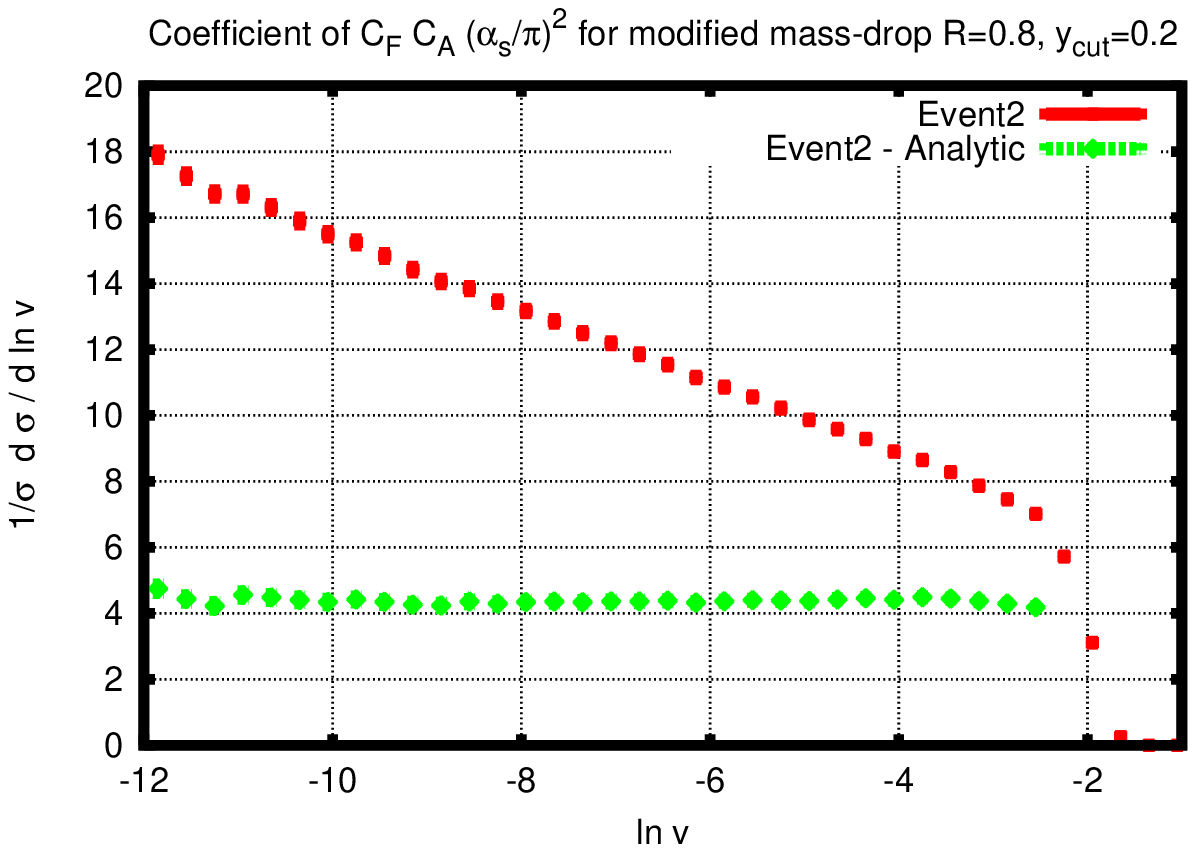}
\includegraphics[width=0.49 \textwidth]{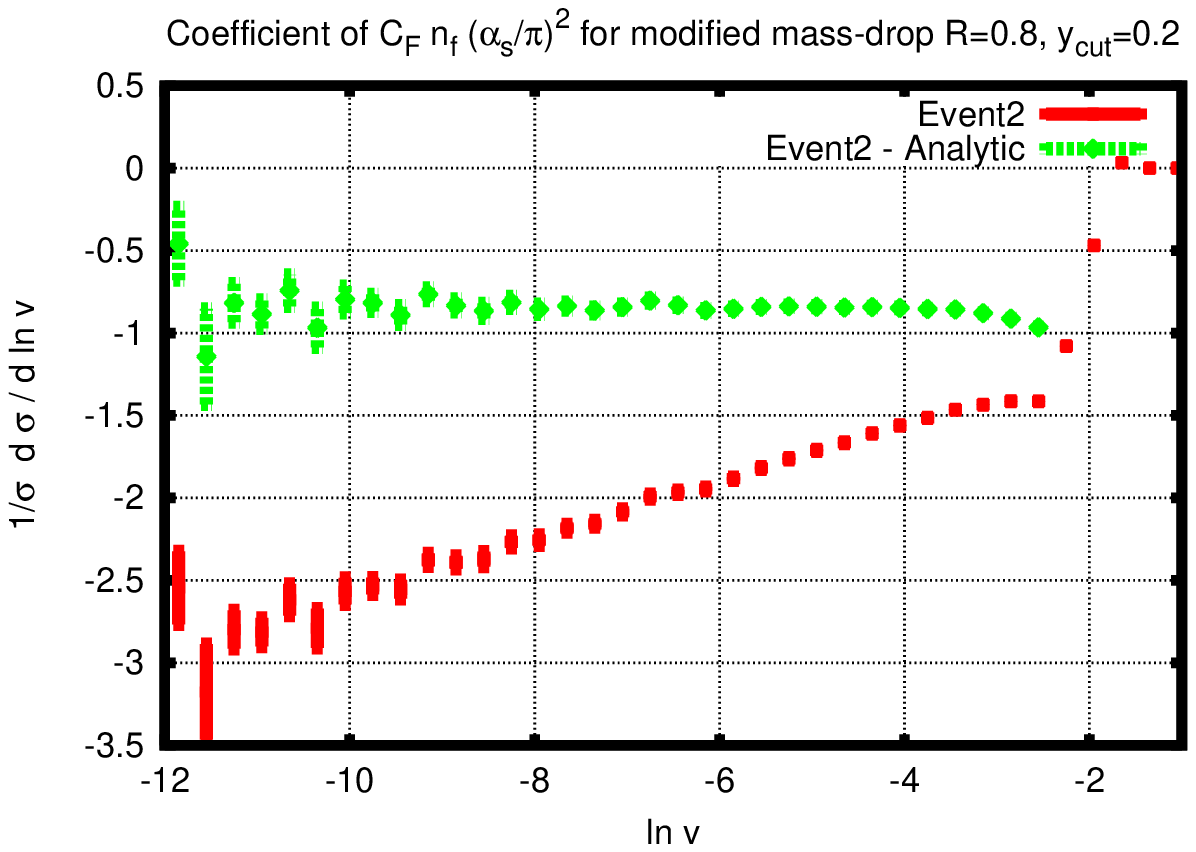}
\includegraphics[width=0.49 \textwidth]{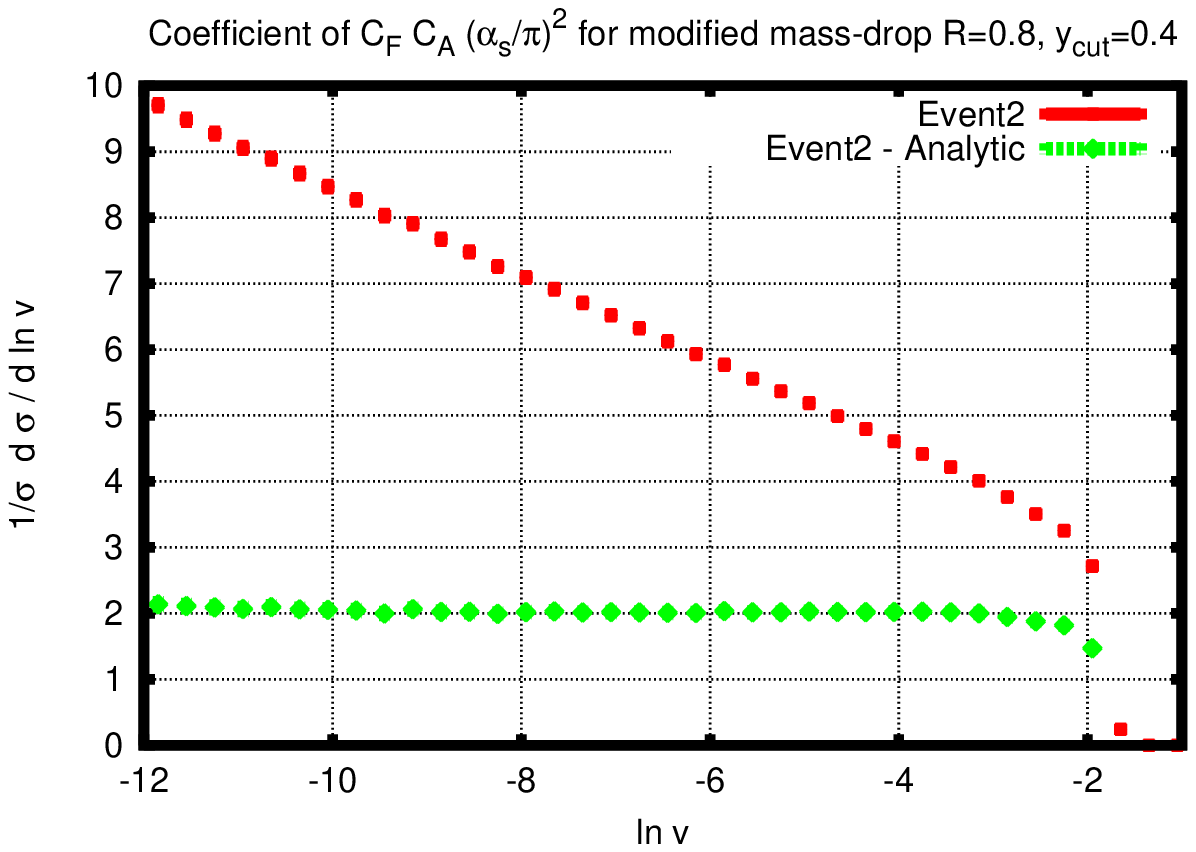}
\includegraphics[width=0.49 \textwidth]{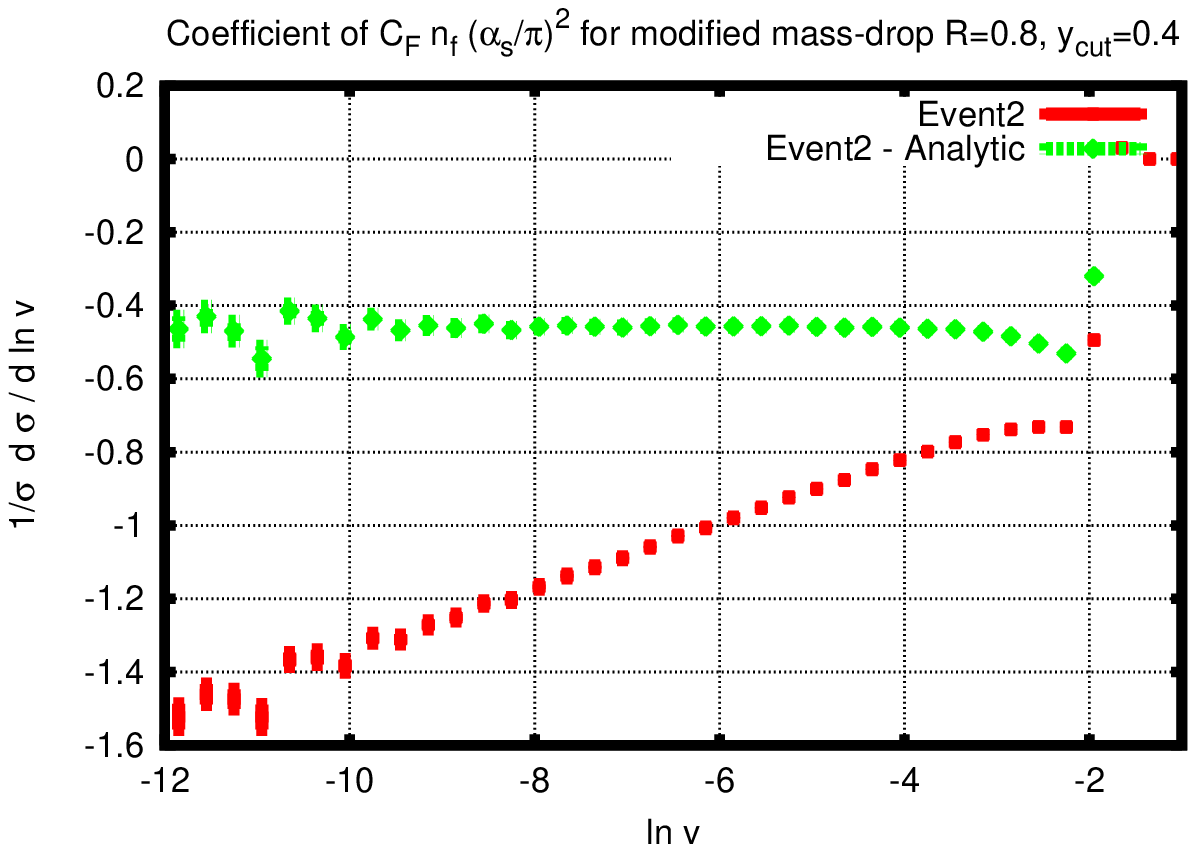}
\caption{
Comparison of the analytic calculations Eq.~(\ref{MMDJg1}) and Eq.~(\ref{MMDJg2}) with \event2 for the coefficients of $C_F C_A$, on the left, and of $C_F n_f$, on the right, in the region $v< \frac{\yc}{1+\yc}\Delta_R^2$, for different values of $\yc$. The red curve shows the fixed-order result alone which behaves like a straight line at small $v$ and hence indicates a single logarithmic behaviour for the integrated distribution. The green curve indicates that, after subtracting our analytical calculation, the result is flat at small $v$, as expected.}
\label{fig:mMDTnonAbelian}
\end{center}
\end{figure}

We are now in a position to compare our results for the $C_F C_A$ and $C_F n_f$ contributions, Eq.~(\ref{CFCArc}) together with Eq.~(\ref{MMDJg1}) and Eq.~(\ref{MMDJg2}), to the full result from \event2. The results are shown in Fig.~\ref{fig:mMDTnonAbelian}, which shows that we have full control on the single logarithms.

\subsection{Summary}
Given the simple but somewhat lengthy calculations that have been carried out thus far, we feel a summary encapsulating our findings regarding the mMDT is in order. We have found that the mMDT jet-mass distribution contains only single 
logarithmic enhancements in contrast to the plain jet-mass. These logarithms can be resummed and in the limit of small $\yc$ (which is a good practical approximation in any case), the result is a straightforward exponentiation of the leading-order effect, with running coupling effects included. Beyond small $\yc$ 
there is a matrix structure to the exponentiation arising due to the flavour changing feature discussed above. Non-global and clustering logarithms are absent. We can consider our result in terms of the values of the coefficients $a_{i,2i}$ and $a_{i,i}$ for the mMDT jet-mass as for the case of plain jet-mass, Eq.~(\ref{jmcoeffs}) and we have found that for mMDT one has the result that $a_{12}$, $a_{24}$ and $a_{23}$ all vanish. Hence we are left with the coefficients of single logarithmic terms which in the limit of small $\yc$ are given by 
\bea
a_{11}^{\mathrm{mMDT}} &=& C_F \ln \frac{e^{-3/4}}{\yc}+\ord(\yc), \\ \nonumber
a_{22}^{\mathrm{mMDT}} &=& - C_F^2 \ln^2 \frac{e^{-3/4}}{\yc} +C_F \beta_0 \ln \frac{e^{-3/4}}{\yc}+\ord(\yc). 
\eea

We conclude this section by pointing out that the mMDT has {\it{remarkable}} properties in that it results in a jet-mass distribution for a single-jet which can be studied in any jet algorithm without non-global and clustering effects. The pure collinear nature of the resulting logarithms makes them very straightforward to resum at hadron colliders, with no non-trivial soft large-angle colour structure involved. A resummed and matched calculation for the mMDT should 
thus be a straightforward exercise and this augurs well for direct comparisons to LHC data. Lastly the removal of double logarithms leads to the absence of undesirable Sudakov peaks in the background as is discussed in some detail in Ref.~\cite{DasFregMarSal}. In particular the form of $a_{22}$ reported above suggests that one can choose the value of $\yc$ such that $a_{22}$ vanishes which 
practically speaking would mean that the result for mMDT would be well approximated by just its leading order form. This in turn would mean that $d\sigma/d\ln v$ would be essentially constant, implying a {\it{flat}} background distribution.
In the next section we shall turn our attention to the case of pruning.

\section{Pruning}

\subsection{Definition}
We shall now consider the calculation of the jet mass with pruning~\cite{pruning1}. Starting with a fat jet (here we consider \CA with radius $R$) one reruns the jet algorithm over the constituents of the fat jet with the additional conditions:
\begin{enumerate}
\item  For each pair of objects $ij$ considered for recombination, compute the distance $\Delta_{\theta_{ij}}^2$ and momentum (energy) fraction
$z=\frac{\mathrm{min}\left(p_{ti},p_{tj}\right)}{|\underline{p}_{ti}+\underline{p}_{tj}|}$.
\item If $\Delta_{\theta_{ij}}^2>R_\text{prune}^2$ and $z<\zc$ do not recombine $i$ and $j$ and discard the softer one.  Continue with the algorithm. The resulting jet is the pruned jet.
\end{enumerate}
The pruning radius is chosen in relation to the mass and the transverse momentum of the fat jet: $R_\text{prune}^2= R_\text{fact}\frac{2 \, m_\text{fat jet}^2}{p_t^2}$. In this study we choose $R_\text{fact}=\frac{1}{2}$.
We shall adopt a definition of pruning where one uses the distance measure as 
in $\Delta_{\alpha}^2 =2 (1-\cos \alpha)$ in both the \CA algorithm and in the definition of $R_\text{prune}^2$, and because we are looking at $e^+e^-$ collisions, we use the energies of the jets rather than their $p_t$.
\subsection{Leading-order calculation}
At LO the calculation for pruning is straightforward. At this order we can 
consider the fat jet to be made up of a quark-gluon pair that arises from the 
splitting of an initial quark such that the quark and gluon carry respectively a fraction $1-x$ and $x$, of the energy of the fat jet. We note that the quantity $R_\text{prune}^2$ at this order is just $x(1-x)\Delta_{\theta_{ij}}^2$ which is always less that $\Delta_{\theta_{ij}}^2$. Thus the energy fractions of both constituents have to be larger than $\zc$ in order to survive pruning and hence to obtain a finite jet mass. 

Imposing this condition and inserting the gluon emission probability in the soft limit, we are led to consider the following integral, where we work in the soft and collinear limit:
\begin{equation} 
\label{LOpruningsetupsc}
\frac{1}{\sigma}\frac{d \sigma}{d v}^\text{(pruned, LO)} = \frac{\as C_F}{\pi}\int \frac{d\theta^2}{\theta^2} \int_{\zc}^{1-\zc}\frac{d x}{x}\Theta\left(R^2 -\theta^2 \right) \delta \left(v- x \theta^2 \right),
\end{equation}
which gives us the result:
\bea 
\label{LOpruningsc}
\frac{1}{\sigma}\frac{d \sigma}{d v}^\text{(pruned, LO)}
&=& \frac{\as C_F}{\pi} \frac{1}{v}\ln \left( \frac{1-\zc}{\zc} \right)\Theta\left(\zc R^2 -v  \right) \nonumber \\
 &+&\frac{\as C_F}{\pi} \frac{1}{v}\ln \left( (1-\zc)\frac{R^2}{v} \right)\Theta\left(v - \zc R^2  \right).
\eea
The result obtained is like the case of the MDT and mMDT at this order in that it is single logarithmic for small jet masses, with the logarithm being of collinear origin. The soft logarithm in $v$ has been removed and replaced with essentially a logarithm of $\zc$, which we do not consider large. Values of $\zc$ may be considered similar to those of $\yc$ in the mMDT and MDT i.e $\zc \sim 0.1$. 

In order to obtain the complete answer including hard collinear emission and 
terms varying as powers of $\zc$ we can easily go beyond the soft approximation in the collinear region. Moreover one can also treat soft large-angle radiation, which does not contribute relevant logarithms at this order. The calculation as for the MDT and mMDT can be found in the appendix and the result reads:
\beq
\label{LOpruningfull}
\frac{1}{\sigma}\frac{d \sigma}{d v}^\text{(pruned, LO, full)}
= \frac{\as C_F}{\pi} \frac{1}{v}\ln \left( \frac{1-\zc}{\zc}e^{-\frac{3}{4}\left( 1- 2 \zc \right)} \right)\,, \quad \text{for} \quad v<\zc \Delta_R^2,
\eeq
which is the same as the LO result for (m)MDT, Eq.~(\ref{LOmdt}), if we replace $\zc \leftrightarrow \frac{\yc}{1+\yc}$.

\begin{figure}
 \begin{center}
\includegraphics[width=0.49 \textwidth]{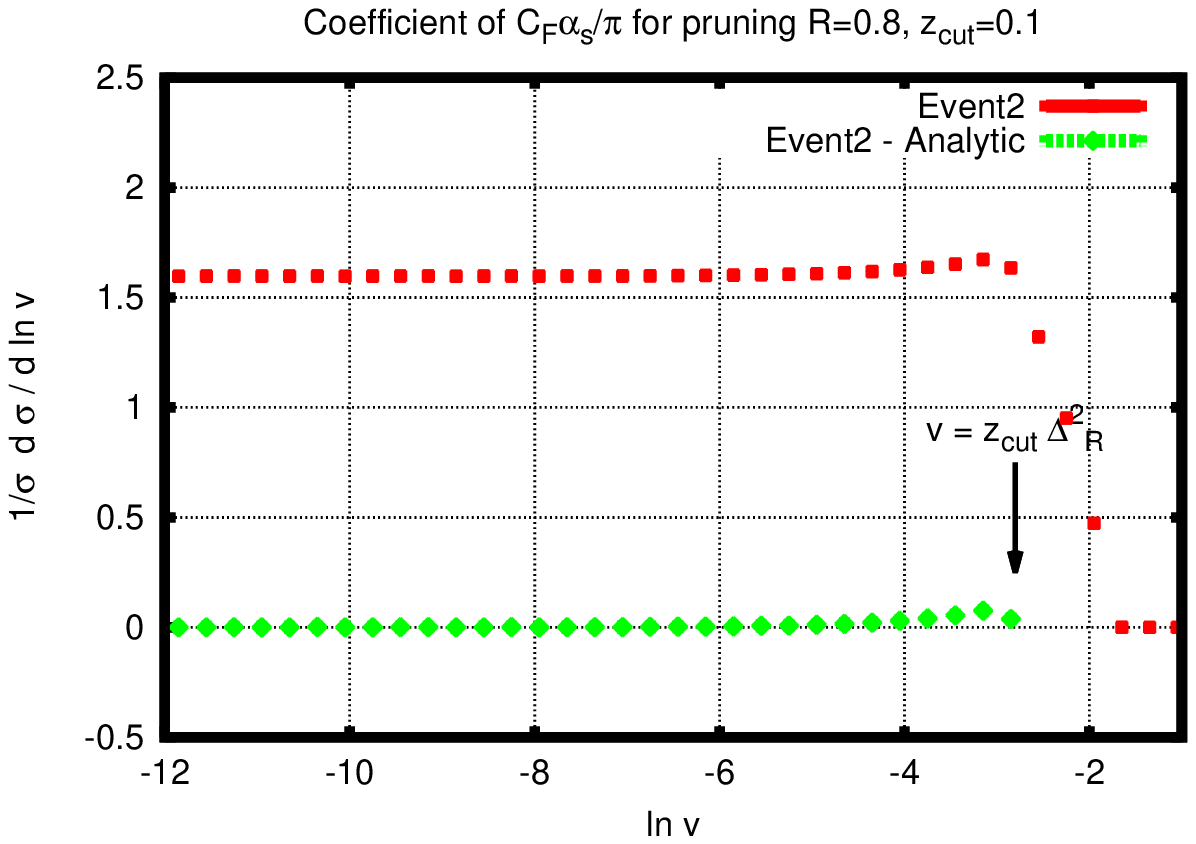}
\includegraphics[width=0.49 \textwidth]{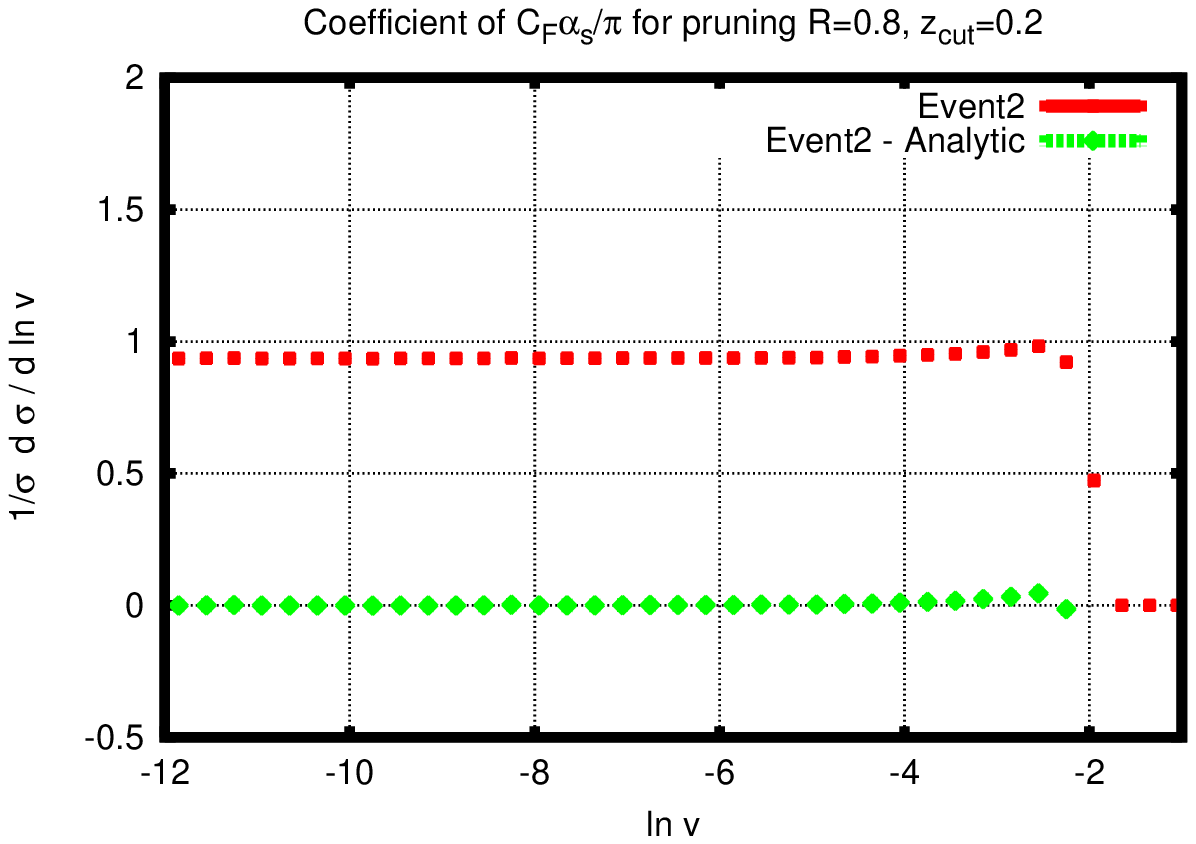}
\includegraphics[width=0.49 \textwidth]{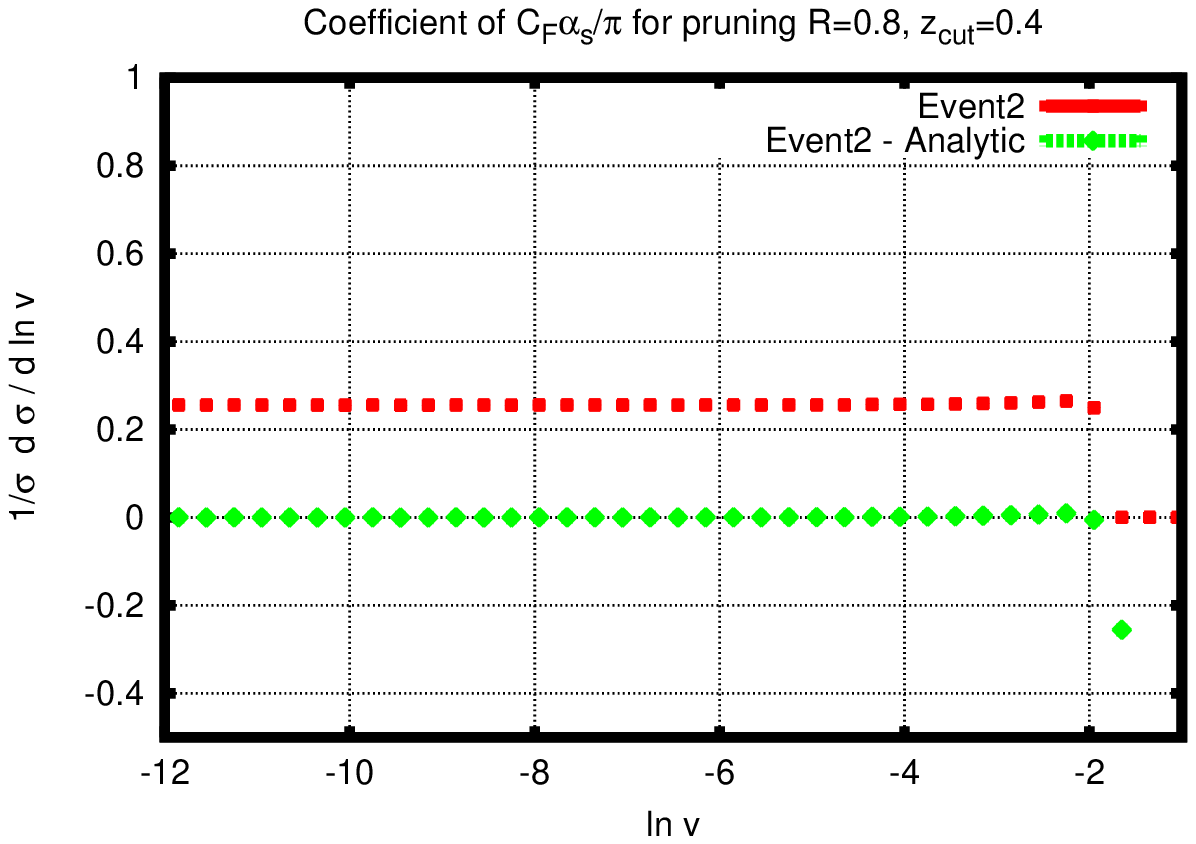}
\caption{Comparison of the analytic calculation Eq.~(\ref{LOpruningfull}) with \event2 at LO in the region $v< \zc \Delta_R^2$, for different values of $\zc$. The red curve shows the fixed-order result alone which is flat for small $v$ and hence indicates a single logarithmic behaviour for the integrated distribution. The green curve indicates that, after subtracting our analytical calculation, the result vanishes at small $v$ as expected.}
\label{fig:pruningCF}
\end{center}
\end{figure}

In Fig.~\ref{fig:pruningCF} we show the comparison of our analytical calculation Eq.~\eqref{LOpruningfull} with leading-order results from \event2. Plotting the difference between \event2 and the analytical result for the differential 
distributions, versus $\ln v$, yields a result which is consistent with zero for small $v$, for all values of $\zc$ studied. This indicates that we have correctly computed the coefficient of the single logarithm we anticipated. 

\subsection{Next-to-leading order calculation: independent emission contribution} 
\label{sec:pruningCF2}

In this section  we shall carry out the NLO calculations for pruning and 
investigate its logarithmic structure. As for the mMDT, we first consider the 
Abelian $C_F^2$ channel and we concentrate on the $v \to 0$ limit. We shall show below that considering soft and collinear emissions one obtains at this order {\it{double logarithmic behaviour}} absent in the MDT and mMDT and in contrast to what we observed in the LO calculation of the preceding section.

Also the presence of double logarithms indicates that pruning is as singular 
as the plain jet-mass itself and hence from the NLO level onwards {\it{does not remove soft gluon effects}} in the manner it was possibly intended to. We will also demonstrate that pruning suffers additionally from the presence of non-global logarithms although the size of the non-global effects can be expected to be reduced compared to the plain jet-mass case. Additionally further complications will arise via the presence of Abelian clustering logarithms~\cite{BanDasKKKMar,BanDas} which complicate the calculation of pruning to single-logarithmic accuracy. For this reason, in this section we only compute the coefficients $a_{24}$ and $a_{23}$ in the expansion of Eq.~(\ref{behaviour}) and leave the determination of single logarithms to future work. Nevertheless, as mentioned we will discuss some single-logarithmic effects, specifically those due to non-global logarithms, in the next sub-section.

We consider contributions as in the mass-drop case from emissions $k_1$ and $k_2$ which may be real or virtual (see Fig.~\ref{fig:indep}), in the independent emission approximation, expected to yield the leading singular behaviour in the $v \to 0$ limit. We start by working in the soft and collinear limit and, for brevity, we will consider here also the small-$\zc$ limit. A more complete derivation incorporating finite-$R$ and finite-$\zc$ effects  is reported in appendix~\ref{prune:nlo}.
We organise the calculation according to the phase-space constraints on the double-real emission contribution. To this end we introduce the quantity $R_\text{prune}^2 = m_\text{fat jet}^2/E_j^2$ which is given by $R_\text{prune}^2 \approx x_1 \theta_1^2+ x_2 \theta_2^2$ where $x_1$ and $x_2$ are the fractions of the jet energy carried by $k_1$ and $k_2$ respectively and $x_i, \theta_i \ll 1$.

Now one can consider the following regions separately: the region where both emissions are at an angular distance from the emitting quark which is greater than $R_\text{prune}$, both emissions are within an angular distance $R_\text{prune}$ or one emission is at a distance larger than $R_\text{prune}$ and the other at a distance smaller than $R_\text{prune}$. In the case of the one-real one-virtual contributions, the situation is similar to the single emission case considered in the leading order calculation in that the angular distance of the emission is {\it{always}} greater than the normalised squared jet-mass. Hence these contributions only survive pruning if the energy fraction of the real emission is greater than $\zc$ and there is {\it{no constraint on angle}} other than the requirement that the real gluon is within the angular distance $R$, the fat jet radius.  One can however divide the integration region for the virtual corrections in precisely the same way as for the real emission i.e introduce $R_\text{prune}^2 = x_1 \theta_1^2+ x_2 \theta_2^2$ and then consider the integration regions  such that the emission angles (more precisely the $\theta_i^2$) are both greater than $R_\text{prune}^2$, both less than $R_\text{prune}^2$ or one is greater and the other less than $R_\text{prune}^2$. Doing so lets us combine the real and virtual corrections together so as to conveniently 
cancel divergences.

We find after considering real and virtual terms together, that in the regions 
where both $\theta_1^2,\theta_2^2>R_\text{prune}^2$ or $\theta_1^2, \theta_2^2 < R_\text{prune}^2$, real-virtual cancellations ensue, which result in no large logarithmic terms in the jet-mass, for small jet masses. In particular, we note that the absence of large logarithms in the $\theta_1^2,\theta_2^2< R^2_\text{prune}$ region is a consequence of the dynamical choice of the pruning radius. We shall see that this is not the case for trimming and the dominant logarithms will arise precisely from this region.

Finally, we consider the region of phase space where one gluon (say $k_2$) is emitted in the core of the jet, i.e.\ within the pruning radius, and the other is emitted at an angular distance larger than $R_\text{prune}$. Adding real 
and virtual corrections in this angular region leads to
\bea \label{thetapruningCF2}
\Theta^\text{nlo}&=& \Big[ \Theta(x_1-\zc) \delta \left(v-x_1 \theta_1^2-x_2 \theta_2^2\right)+\Theta(\zc-x_1) \delta \left(v-x_2 \theta_2^2\right)\nonumber\\
&-&\Theta(x_2-\zc)  \delta \left(v-x_2 \theta_2^2\right)-\Theta(x_1-\zc)  \delta \left(v-x_1 \theta_1^2\right)\Big] \Theta \left(R_\text{prune}^2-\theta_2^2 \right) \Theta \left(\theta_1^2 -R_\text{prune}^2\right), \nonumber\\
\eea
where the first line comes from double real emissions and the second one 
contains real-virtual contributions. To clarify this point further we are considering the situation where $k_2$ is within the core of the jet while $k_1$ is beyond $R_\text{prune}$. If $k_1$ has an energy fraction $x_1$, defined 
with respect to the fat jet energy, which is greater than $\zc$, it survives pruning while if the energy fraction is below $\zc$ it gets removed. When $k_1$ survives pruning (the first term above) then both gluons contribute to the jet-mass $v$. On the other hand when $k_1$ is pruned away only $k_2$ makes a contribution to $v$, indicated by the second term within square brackets above. The last two terms simply denote the case where there is only one real gluon which must survive pruning to obtain a finite jet mass and hence the condition is $x_{1,2} > \zc$. These terms acquire a minus sign due to the additional presence of a virtual emission.

 The combination of the first and last terms above gives only subleading terms due to the cancellation that occurs in the limit where $x_2$ or $\theta_2$ 
vanishes. Thus we have two relevant integrals to compute:
\begin{equation}
\label{eq:intsc1}
I_1 =  \left(\frac{\alpha_s C_F}{\pi} \right)^2 \int  \frac{d x_1}{x_1}\frac{d x_2}{x_2} \frac{d \theta_1^2}{\theta_1^2} \frac{d \theta_2^2}{\theta_2^2} \\ 
\delta \left(v- x_2 \theta_2^2 \right) \Theta(\zc-x_1) \Theta(\theta_1^2-R_\text{prune}^2) \Theta(R_\text{prune}^2-\theta_2^2) 
\end{equation}
and 
\begin{equation}
\label{eq:intsc2}
I_2=  -\left(\frac{\alpha_s C_F}{\pi} \right)^2 \int  \frac{d x_1}{x_1}\frac{d x_2}{x_2} \frac{d \theta_1^2}{\theta_1^2} \frac{d \theta_2^2}{\theta_2^2}\delta \left(v- x_2 \theta_2^2 \right) \Theta(x_2-\zc) \Theta(\theta_1^2-R_\text{prune}^2) \Theta(R_\text{prune}^2-\theta_2^2) 
\end{equation}
where as mentioned before we have $R_\text{prune}^2 = x_1 \theta_1^2+ x_2 \theta_2^2$. In writing these results we remind the reader that they make the assumption that $x_i, \theta_i \ll 1$. In particular we have not been careful about upper limits on the $x$ integrals, for instance in $I_2$ the maximum value of $x_2$ should be $1-\zc$ rather than unity. To correct for such effects, which will generate finite $\zc$ corrections as well as to incorporate soft emissions at large angles will require us to extend the calculation above, which we do in 
appendix \ref{prune:nlo}. The main features of pruning already emerge within the approximations made above and hence in the main text here we shall focus on the results that emerge from carrying out the integrations above.

The results for $I_1$ and $I_2$ are relatively easy to obtain and are mentioned below:
\bea
\label{eq:I1}
I_1 =  \left(\frac{\alpha_s C_F}{\pi} \right)^2 \frac{1}{v} \left(\frac{1}{6} \ln^3 \frac{R^2 \zc}{v} \right)\Theta \left(\zc R^2 -v \right)
\eea
\bea
\label{eq:I2}
I_2  = -  \left(\frac{\alpha_s C_F}{\pi} \right)^2 \left[\frac{1}{2 v} \left(\ln \frac{1}{\zc} \ln^2 \frac{R^2}{v} -\ln^2 \frac{1}{\zc} \ln \frac{R^2}{v}\right)+
\frac{1}{6v} \ln^3 \frac{1}{\zc} \right]\Theta \left(\zc R^2-v \right) \\
- \left (\frac{\alpha_s C_F}{\pi} \right)^2 \frac{1}{6v}\ln^3 \frac{R^2}{v}\Theta \left(v- \zc R^2 \right).
\eea

Thus, as we claimed the differential distribution for pruning at this order has a leading term varying as $\ln^3 v/v$, due to $I_1$, which is the same order of divergence as the plain jet-mass itself. Physically this term represents the 
contribution from a soft gluon that dominates the fat jet-mass and is pruned 
away due to failing the $\zc$ criterion. The final pruned jet which is returned has no hard substructure and we declare it to belong to the ``$\anomalous$-pruned'' class, i.e.\ it is made of only one hard prong~\cite{DasFregMarSal}.

\begin{figure}
\begin{center}
\includegraphics[width=0.49\textwidth]{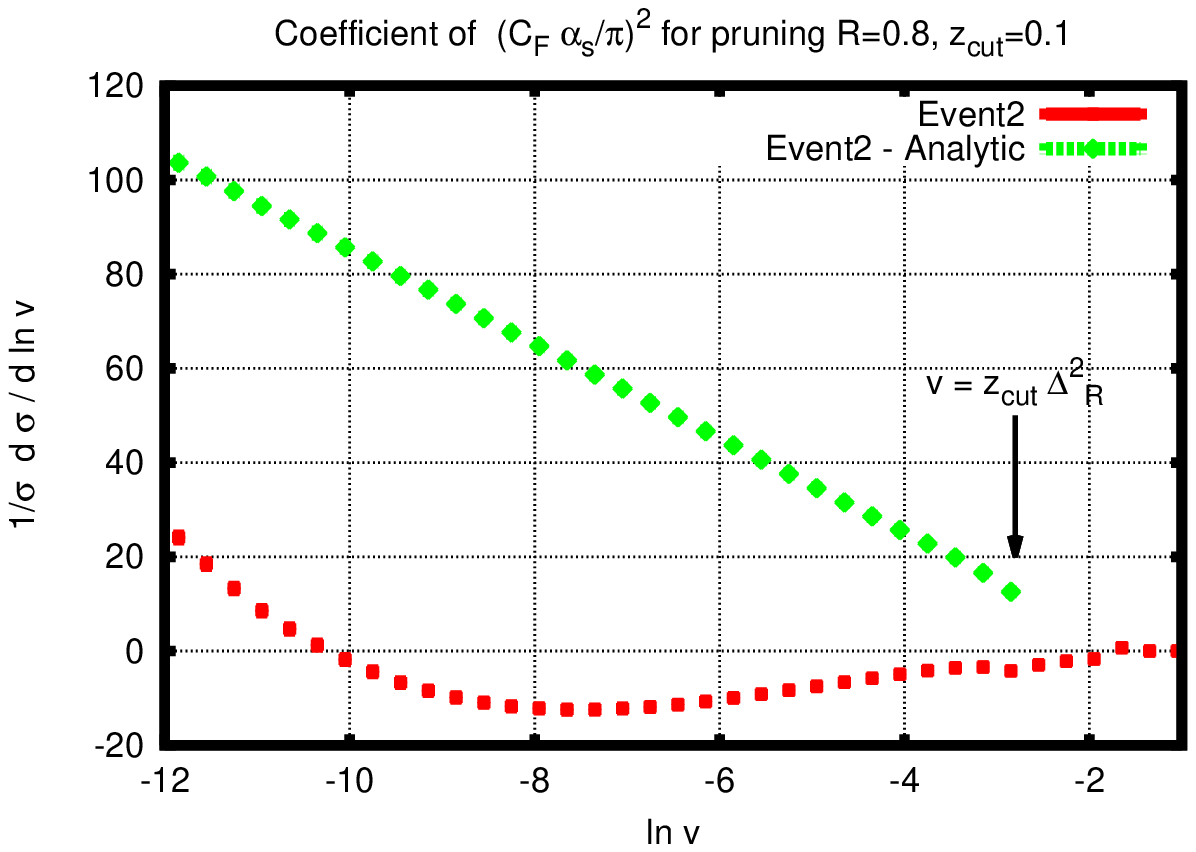}
\includegraphics[width=0.49\textwidth]{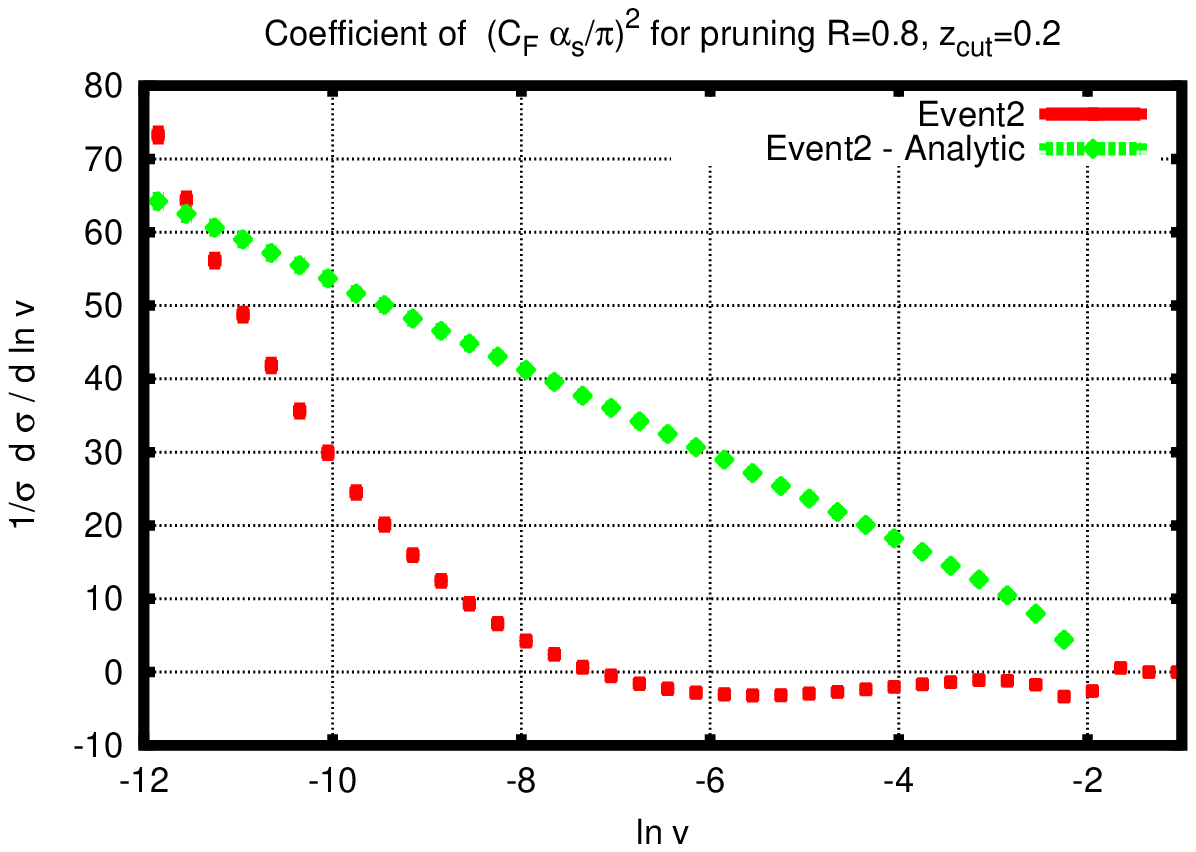}
\includegraphics[width=0.49\textwidth]{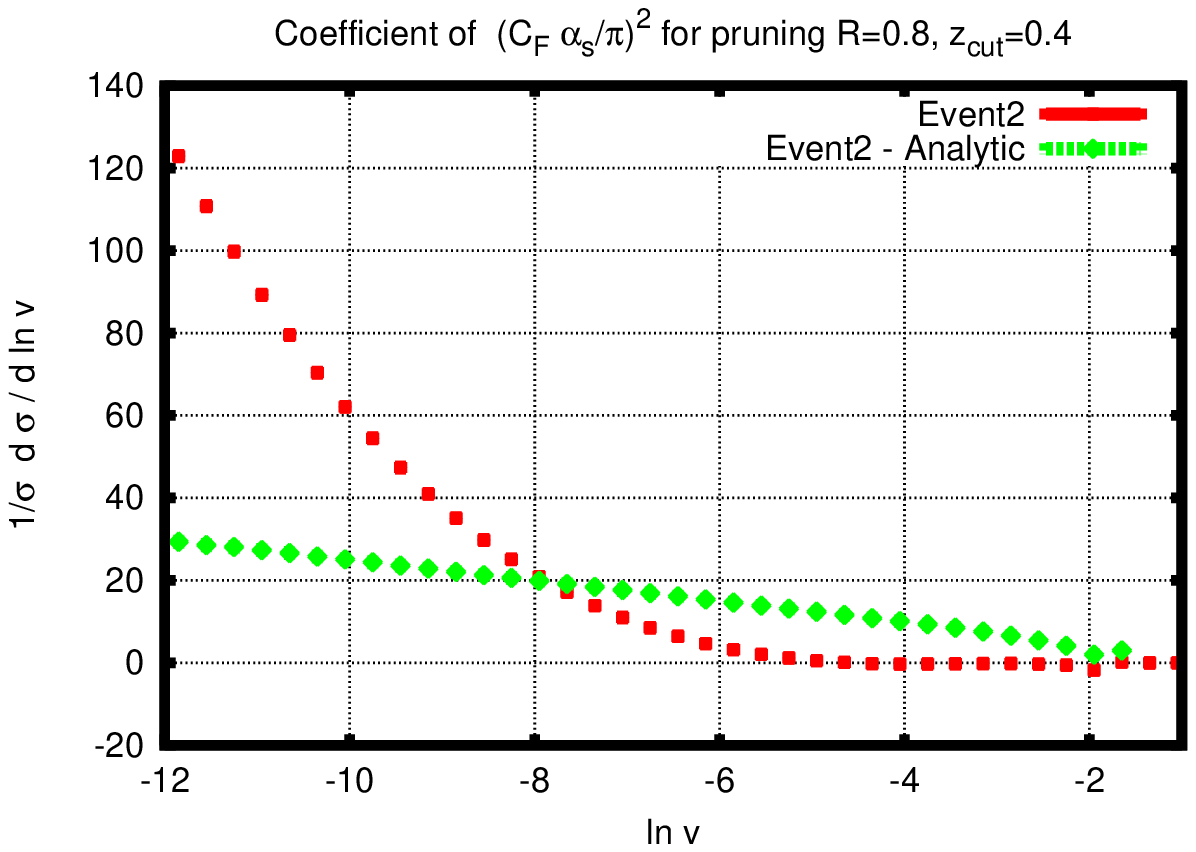}
\caption{
Comparison of the analytic calculation Eq.~(\ref{pruningNLOCF2_final}) with \event2 for the coefficient of $C_F^2$, in the region $v< \zc \Delta_R^2$, for different values of $\zc$. The green curve indicates that, after subtracting our analytical calculation, the result is a straight line at small $v$, as expected for a single-logarithmic leftover ($\alpha_s^2 \ln^2 1/v$, in the integrated distribution).}
\label{fig:pruningCF2}
\end{center}
\end{figure}

Our results thus far can be compared to results from \event2 for the $C_F^2$ 
channel. In order to have clean comparisons however, as for the mMDT it is 
first desirable to extend our calculations beyond the soft-collinear limit. We note that pruning is subject to a large number of physics effects at single-logarithmic level including the presence of {\it{clustering}} logarithms and for this reason we do not carry out a calculation of the single-logarithmic terms 
but confine ourselves to obtaining the coefficients $a_{24}$ and $a_{23}$. 
To this end we can compute the integrals $I_1$ and $I_2$ lifting the simultaneous soft and collinear restrictions where appropriate so as to include hard-collinear and soft wide-angle effects. Doing so we obtain the result:
\bea \label{pruningNLOCF2_final}
\frac{1}{\sigma}\frac{d \sigma}{d v}^\text{(pruned, $C_F^2$)} & =& \left(\frac{\alpha_s C_F}{\pi} \right)^2\frac{1}{v}   \Bigg\{ \frac{1}{6} \ln^3 \frac{1}{v}
+\left[\frac{1}{2} \ln \frac{\zc^2}{1-\zc}-\frac{5}{4}\zc+\frac{\zc^2}{8}+\ln \left(2\tan \frac{R}{2} \right) \right]  \ln^2\frac{1}{v} \Bigg\}\nonumber \\
&+& \ord\left(\as^2\frac{1}{v} \ln \frac{1}{v}\right),
\eea
which is valid in the $v\to 0$ limit.
Details of the calculation can be found in the appendix \ref{prune:nlo}. The 
comparison with \event2 is shown in Fig.~\ref{fig:pruningCF2}, where we can see that the difference between the full NLO calculation and our analytic result is a straight line, i.e.\ a single-logarithmic contribution. In the next section we shall discuss the $C_F C_A$ colour channel, showing in particular that the 
pruned jet-mass is affected by non-global logarithms.

\subsection{Next-to-leading order calculation: non-Abelian terms}
Starting from order $\alpha_s^2$, we also have to consider the contribution 
from gluon splitting. In turn, depending on the precise kinematical details of the gluon emission and decay, this contribution gives rise to running coupling 
effects (which dress the leading-order gluon emission), non-global logarithms 
and extra contributions that arise when the parent gluon is so energetic that the recoiling quark energy is below $\zc$. This last contribution causes the quark to be pruned away leaving us to examine the mass of the resultant gluon jet. The leading contribution here varies as $\as^2 L^3$, in the integrated cross-section, as we shall demonstrate below and hence falls within our aimed accuracy. Running coupling and non-global effects on the other hand matter at $\as^2 L^2$ level in the integrated cross-section. The former are simple to compute (as for the mass-drop case) while the latter acquire complications due to clustering effects in the \CA algorithm. We shall therefore not compute the non-global logarithms precisely but shall compute an upper bound for them as for the mMDT and demonstrate that they are in fact formally present unlike for the mMDT.

Let us begin by considering the most divergent $C_F C_A \, \as^2 L^3$ term. Consider again the configuration that is depicted in Fig.~\ref{fig:nonAbelian}, where a parent gluon $k$ branches into almost collinear offspring $k_1,k_2$. 
Assuming that the parent gluon jet carries an energy fraction $x > 1-\zc$ of the fat jet energy $E_j$, the quark carries an energy fraction that is less than $\zc$. In the recombination with application of pruning, the final \CA merging, which combines the gluon jet with the massless quark, now fails as the quark is too soft and is discarded. We are left to study the mass distribution of the gluon jet:
\begin{multline}
\frac{1}{\sigma}\frac{d \sigma}{d v}^\text{(pruned, $C_F C_A$)} 
= C_F C_A \left(\frac{\as}{\pi}\right)^2 \int_{1-\zc}^1 d x \,p_{gq}(x)\frac{d\theta^2}{\theta^2} \Theta \left(R^2-\theta^2 \right)\\ \times \int_{\zc}^{1-\zc} \frac{dz}{z} \frac{d\theta_{12}^2}{\theta_{12}^2} \left(\theta^2-\theta_{12}^2 \right) \delta \left(v-z x^2 \theta_{12}^2 \right).
\end{multline}
In the above we considered the soft ($z \ll 1$)and collinear branching of an energetic parent gluon and have incorporated the fact that the angle between $k_1$ and $k_2$, $\theta_{12}$, must be smaller than $\theta$ (which one can take to essentially be the angle between the quark and the parent gluon or equivalently the harder off-spring gluon). The integral is simple and, in the $v\to 0$ limit, it gives us:
\begin{equation}
\label{pruncfca}
 \frac{1}{\sigma}\frac{d \sigma}{d v}^\text{(pruned, $C_F C_A$)} = \frac{1}{2v} C_F C_A \left(\frac{\alpha_s}{\pi} \right)^2 \ln^2 \frac{R^2}{v} \left (\ln \frac{1}{1-\zc}-\frac{\zc}{4} \left(\zc+2 \right) \right)+\mathcal{O} \left (\as^2 \frac{1}{v} \ln \frac{1}{v} \right).
\end{equation}

It is clear that due to the limited phase space available for the above effect  it vanishes as $\zc \to 0$. We can test our calculation Eq.~(\ref{pruncfca}) by subtracting it from order $\alpha_s^2$ results from \event2 in the $C_F C_A$ channel. After subtracting the analytical result from the differential distribution obtained with \event 2 (see Fig.~\ref{fig:pruningothers}), we find a 
linear behaviour consistent with an $\as^2 L^2$ single-logarithmic leftover in the integrated cross-section, which is what we would expect and implies that we control the most divergent $\as^2 L^3$ effect we computed above.
\begin{figure}
\begin{center}
\includegraphics[width=0.49\textwidth]{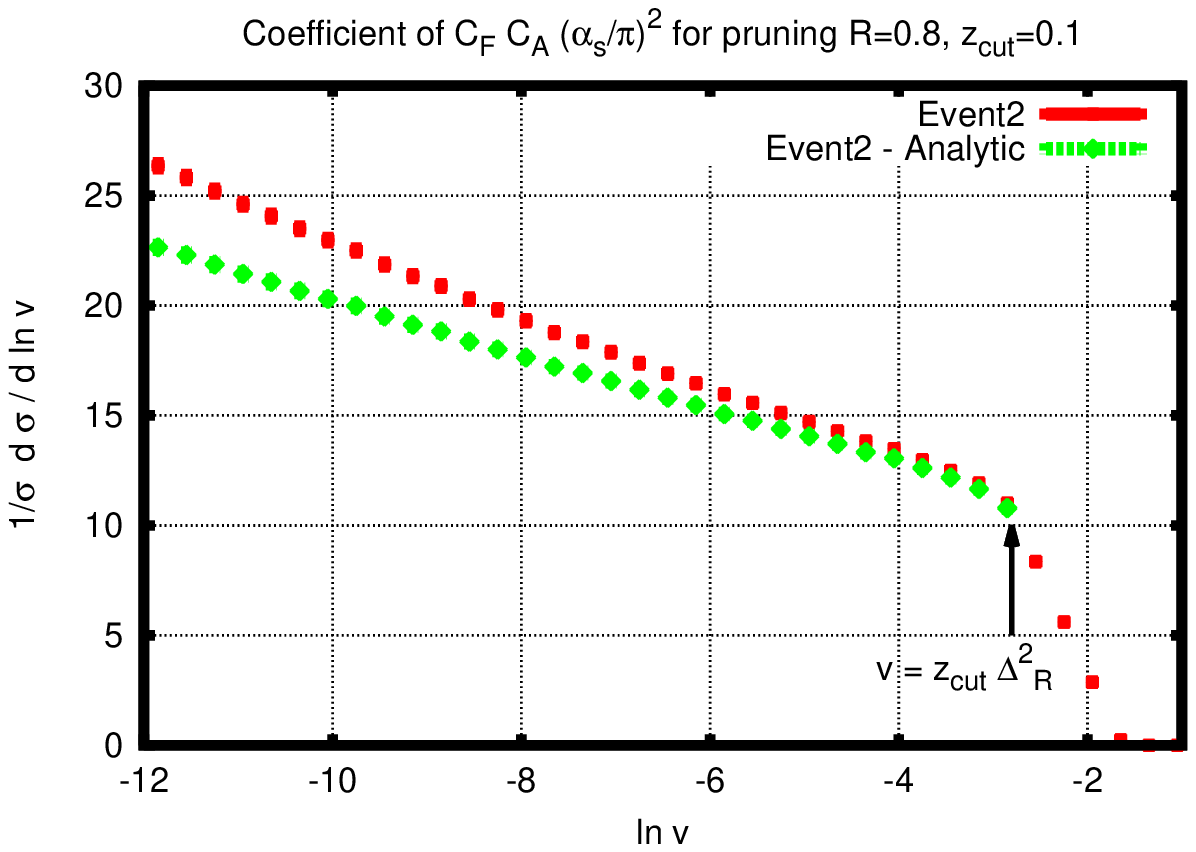}
\includegraphics[width=0.49\textwidth]{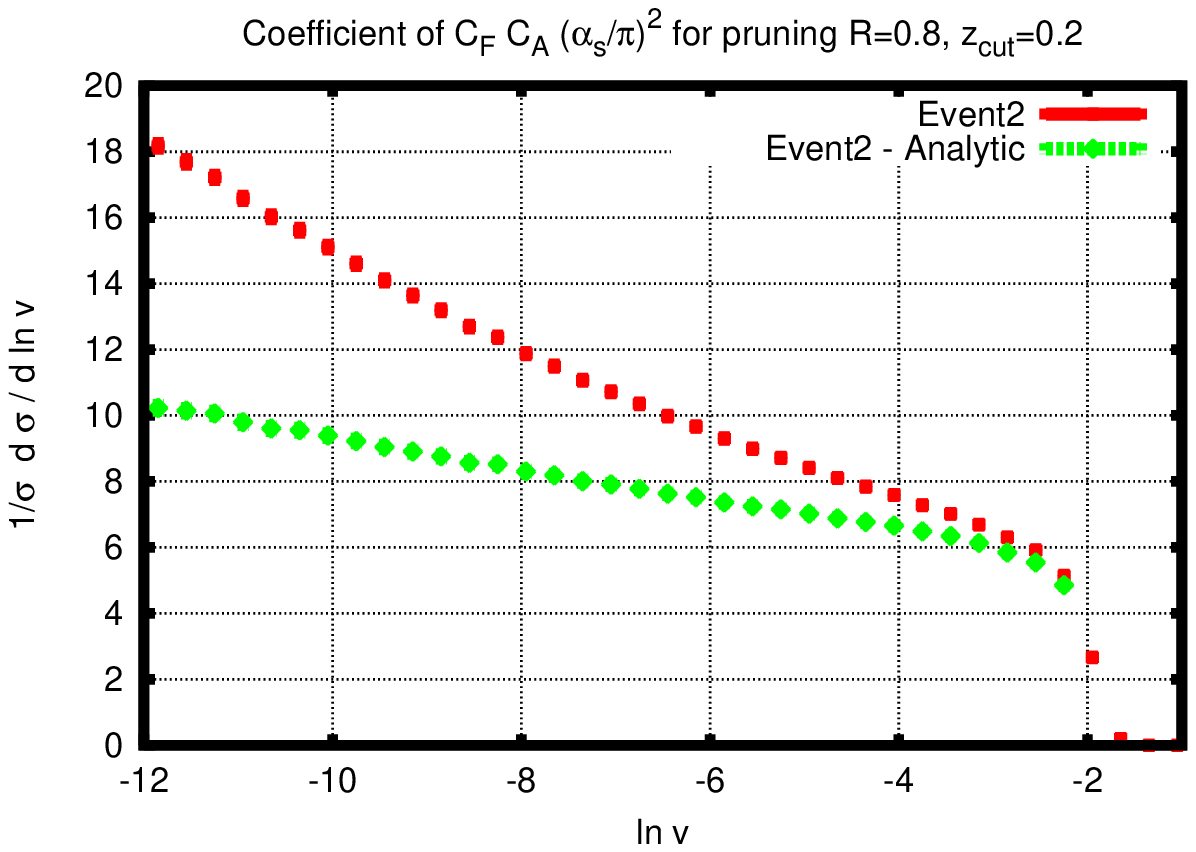}
\includegraphics[width=0.49\textwidth]{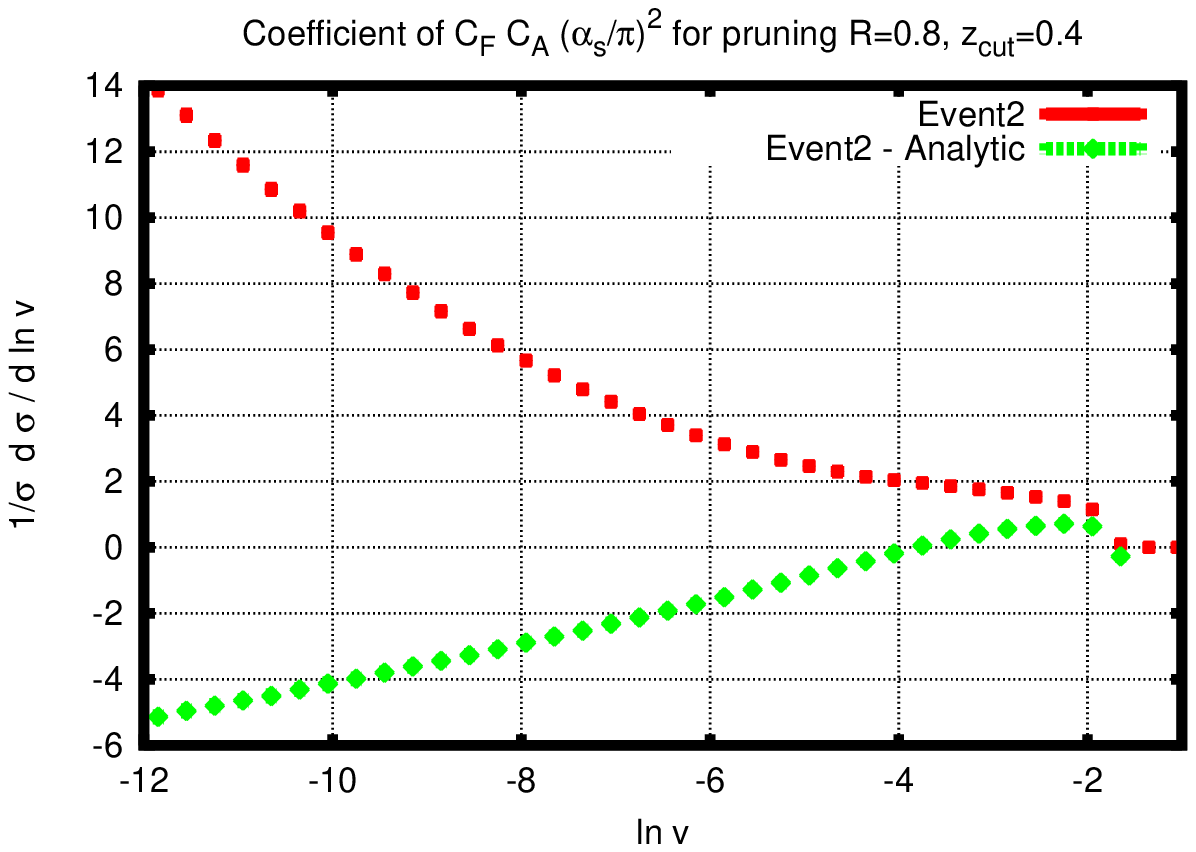}
\caption{
Comparison of the analytic calculation Eq.~(\ref{pruncfca}) with \event2 for the coefficient of $C_F C_A$, in the region $v< \zc \Delta_R^2$, for different values of $\zc$. The green curve indicates that, after subtracting our analytical calculation, the result is a straight line at small $v$, as expected for a single-logarithmic leftover ($\alpha_s^2 \ln^2 1/v$, in the integrated distribution).}
\label{fig:pruningothers}
\label{fig:pruningothers}
\end{center}
\end{figure}

Lastly, we discuss the role of non-global logarithms, absent for the case of 
the mMDT. In the mMDT the non-global logarithms do not arise since the $\yc$ cut-off eliminates soft radiation. For pruning the corresponding $\zc$ only 
applies to objects separated by an angle larger than $R_{\text{prune}}$. In 
particular, let us consider a jet made up of a hard quark and two soft gluons $k_1$ and $k_2$. If $k_1$ is separated by an angle greater than $R_{\text{prune}}$ from the hard quark but has an energy fraction (defined with respect to the fat jet's energy) $x_1$ below $\zc$ it is removed and does not contribute directly to the pruned jet-mass. 
However it can emit a much softer gluon $k_2$ into the core of the jet i.e. 
within an angle $R_{\text{prune}}$ of the hard quark. Thus this softer gluon, 
which cannot be pruned away, makes an essential contribution to the pruned jet-mass distribution: this is a classic non-global configuration. In the C/A 
algorithm employed here (and working for convenience in the small-angle approximation) 
one must additionally have the requirement that the angle $\theta_{12}$ between $k_1$ and $k_2$ must not be the smallest angle when one considers angular separations between pairs of partons. In situations where $\theta_{12}$ is the smallest angle, the non-global contributions are eliminated 
by clustering of the soft gluons~\cite{AppSey1}. However this clustering region leads to the appearance of additional single-logarithms (clustering logarithms) in the $C_F^2$ channel~\cite{BanDas}, which we do not explicitly evaluate here.  
The constraint on emitted gluon energies and angles described here can be 
summarised as 
\begin{equation}
\Theta^{\mathrm{NG}} = \Theta \left (x_1-x_2 \right) \Theta \left (\zc -x_1 \right)\Theta\left(\theta_{12}^2- \theta_2^2 \right)  \Theta \left( \theta_1^2-R_\text{prune}^2 \right) \Theta \left(R_\text{prune}^2- \theta_2^2 \right),
\end{equation}
where we have $R_{\text{prune}} = x_1 \theta_1^2+x_2\theta_2^2$.

We note that since we have $\theta_1>\theta_2$ soft gluon clustering is avoided if $\theta_{12}>\theta_2$. We further note that using $\theta_{12}^2 = \theta_1^2+\theta_2^2-2 \theta_1 \theta_2 \cos \phi$, where $\phi$ is an azimuthal angle, and the fact that $x_1 < \zc$, the angular and energy constraints together imply that clustering is absent for $x_1 < 1/\left (4 \cos^2 \phi \right)$, which is always satisfied if $\zc <1/4$. In practice this value of $\zc$, obtained in the small-angle approximation, will be corrected by finite angle effects, so that one may expect the true value of $\zc$ where clustering switches on, to deviate from $1/4$ by terms of order $R^2$, with $R$ the fat jet radius, which sets the overall angular scale of the problem. In what follows below we shall assume that the clustering is absent, by focusing on the region of small $\zc$. In this case one can ignore the $\phi$ dependence and integrate the squared matrix element for correlated gluon emission (see e.g. Ref.~\cite{Dassalam1}, freely over azimuth $\phi$), to obtain
\bea
\frac{1}{\sigma}\frac{d \sigma}{d v}^\text{(pruned, NG)} &=& 4 C_F C_A \left ( \frac{\alpha_s}{2 \pi} \right)^2 \int \frac{dx_1}{x_1} \frac{d x_2}{x_2} \int  d\cos \theta_2 \int d \cos \theta_1 \, \Omega_2 \, \Theta^{\mathrm{NG}}
\delta \left(v-x_2 \Delta_{\theta_2}^2 \right) \nonumber\\
\eea
where $\Omega_2$ was defined in Eq.~(\ref{omega2}) and we can take its small-angle limit here and we have ignored the $\theta_{12} > \theta_2$ constraint in  
$\Theta^{\mathrm{NG}}$ above. The angular integrations give rise to a single-logarithmic behaviour the coefficient of which is determined by the energy integrals:
\beq \label{pruningNG}
\frac{1}{\sigma}\frac{d \sigma}{d v}^\text{(pruned, NG)}= C_F C_A \left(\frac{\as}{2\pi}\right)^2 4 {\rm Li}_2(\zc) \frac{1}{v} \ln \frac{1}{v},
\eeq
which is valid for $ v< \zc^2 R^2$.
Thus, the pruned mass distribution {\it{does exhibit non-global logarithms at NLO}}. The coefficient ${\rm Li}_2(\zc)$ vanishes linearly as $\zc \to 0$ and given the relatively small $\zc$ values used in phenomenology, one may expect the non-global logarithms not to have a sizeable impact compared for instance to their role in the plain jet mass calculations presented in Ref.~\cite{BanDasKKKMar}. In fact for the plain jet-mass in the anti-$k_t$ algorithm considered in Ref.~\cite{BanDasKKKMar}, one obtains in the limit of small jet-radius, the 
coefficient $\pi^2/3$, rather than $4 {\rm Li}_2(z)$. 
Taking $\zc=0.1$ one finds $4 {\rm Li}_2(\zc)$ is roughly twelve percent of the value for plain jet-mass reported in Ref.~~\cite{BanDasKKKMar}. Of course our considerations here, which imply a small role for non-global logarithms in pruning, are only confined to the leading order $\alpha_s^2$ calculation. The role of non-global logarithms and their impact beyond this order should also be 
considered before one turns to detailed phenomenology for pruning.

Moreover as we explained the result in Eq.~(\ref{pruningNG}) is correct up to 
terms coming from clustering of the two soft gluons. For small 
enough value of $\zc$, i.e.\ $\zc< 1/4 +\ord(R^2)$, here corrections do not produce relevant logarithms of the jet mass. As one increases $\zc$ one may also 
expect clustering effects to play a role in reducing the size of the non-global contribution.

\subsection{$\Sanepruning$} \label{sec:sanepruning}

In this section we explore a variant of pruning~\cite{DasFregMarSal} that eliminates the double-logarithmic structure discussed in section~\ref{sec:pruningCF2}. The modification is as follows:
If at any stage of the re-clustering procedure there was at least one  merging for which $\Delta_{\theta_{12}}^2>R_\text{prune}^2$ and $z>\zc$, the jet is deemed to pass the ``$\sanepruning$'', i.e.\ two-prong, requirement. In this case the jet mass was dominated by (semi)-hard radiation and it is likely that the pruning radius was set appropriately for that radiation. Otherwise discard the jet~\footnote{In preliminary presentations given about this work, the working names that had been used for $\sanepruning$ and $\anomalouspruning$ were, respectively, ``sane" and ``anomalous pruning".}.

\begin{figure}
\begin{center}
\includegraphics[width=0.49\textwidth]{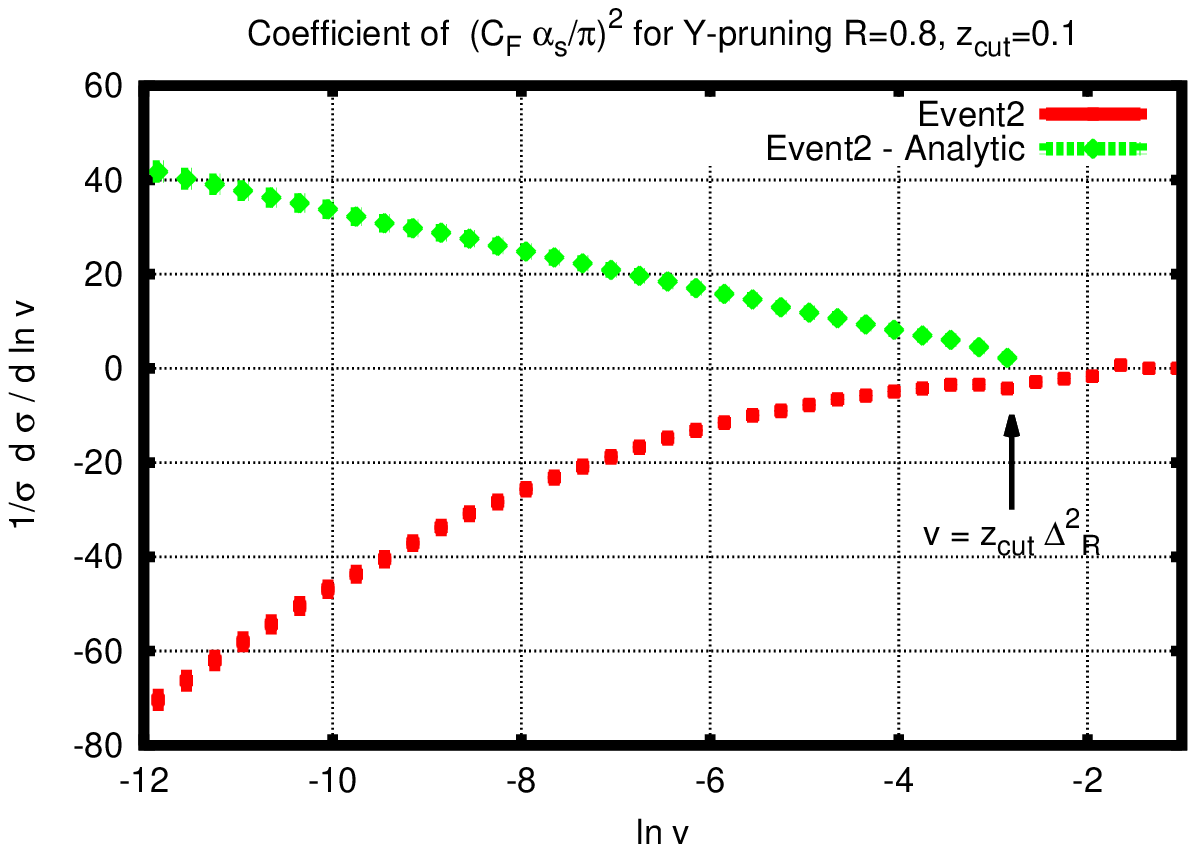}
\includegraphics[width=0.49\textwidth]{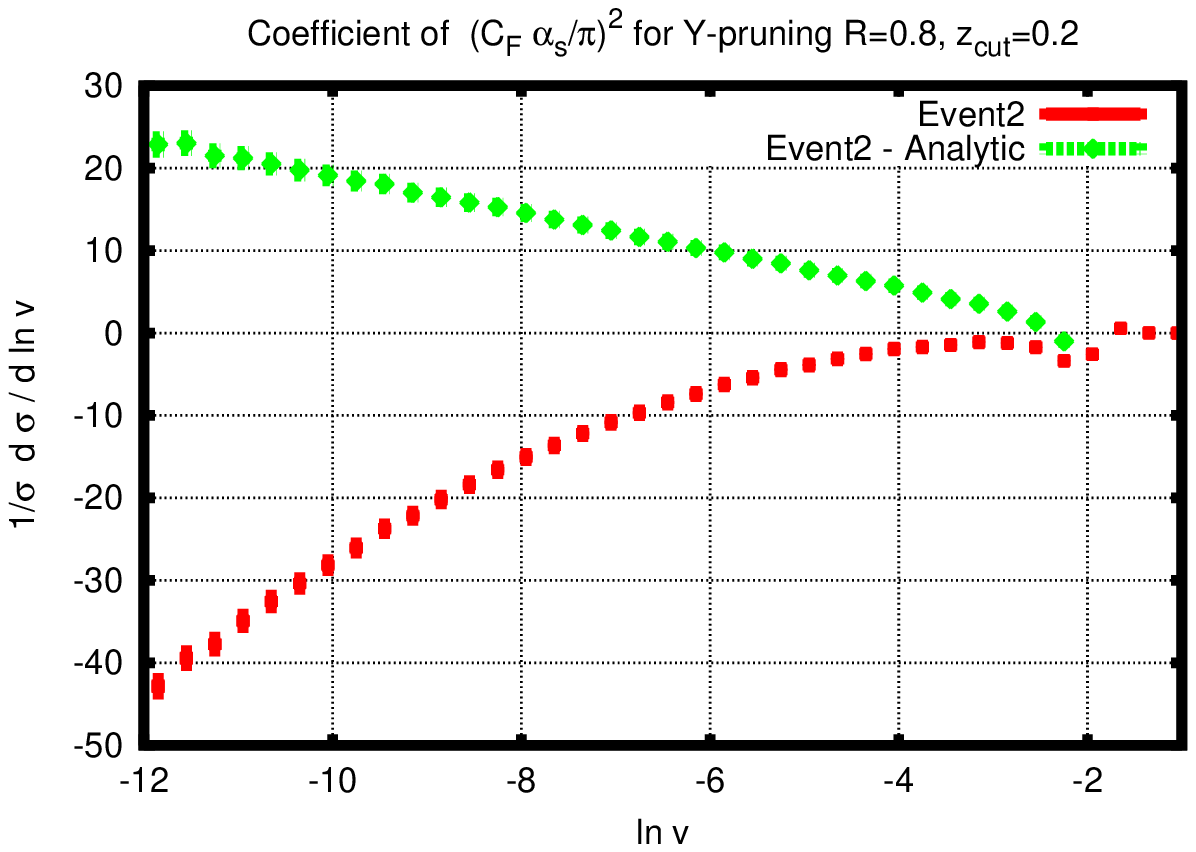}
\includegraphics[width=0.49\textwidth]{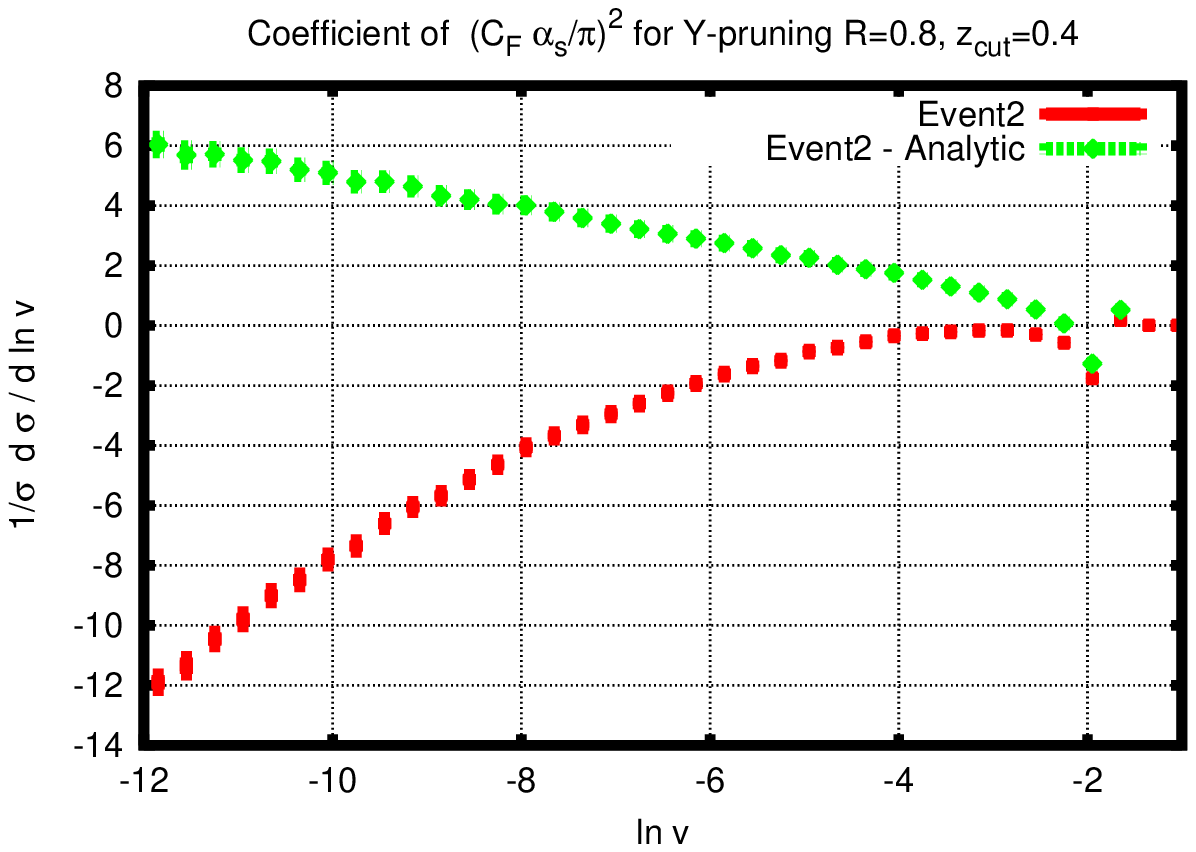}
\caption{
Comparison of the analytic calculation for $\sanepruning$ with \event2 for the coefficient of $C_F^2$, in the region $v< \zc \Delta_R^2$, for different values of $\zc$. The green curve indicates that, after subtracting our analytical calculation, the result is a straight line at small $v$, as expected for a single-logarithmic leftover ($\alpha_s^2 \ln^2 1/v$, in the integrated distribution).}
\label{fig:goodpruningCF2}
\end{center}
\end{figure}

It is obvious that the double-logarithmic contribution that arose in $\anomalouspruning$ (introduced earlier) will be eliminated by this additional requirement. When the emission 
that dominates the jet-mass gets pruned away, we are left with a jet where the 
mass arises from an emission with $\Delta_\theta^2 < R_{\text{prune}}^2$. Hence no emission satisfies the extra requirement above and we discard the jet. On the other hand the contribution from the integral $I_2$ survives and contributes to $\sanepruning$. Therefore $\sanepruning$ has one logarithm less than pruning and plain jet mass, in that the leading divergence is $\as^2 L^3$ at the NLO level. We shall discuss further implications of this point in our summary for pruning. We can compare $\sanepruning$ results from \event2 and we do not need a new analytical calculation but use the result for $I_2$ for the check. We find the difference between our answers and those of \event2 is linear at small $v$, for the distribution in $\ln v$, which indicates that we control the leading divergence and are left with single logarithms $\as^2 L^2$ in the integrated cross-section (see Fig.~\ref{fig:goodpruningCF2}).

\subsection{Comment on the structure of the result for pruning for $v > \zc^2 R^2$}
Thus far we have observed that for $v < \zc R^2$ the pruning result contains double logarithms arising from $\anomalouspruning$ where an emission that 
dominates the mass of the fat jet is too soft to survive the $\zc$ cut-off and is pruned away. Strictly speaking the double logarithmic behaviour has a more 
restricted range of validity than that observed so far. While in our present paper we are mainly interested in the very small $v$ region, for phenomenological purposes one would generally wish to examine a broader range of $v$ values. In doing so one finds that there is an interesting behaviour that emerges in the region $\zc^2 R^2 < v< \zc R^2$. Here the double logarithms cancel away against similar terms that arise from the region where in the double real-emission terms {\it{both}} gluons are beyond the pruning radius. Since this region does not contribute relevant terms when $v \to 0$ (and is irrelevant for our fixed-order checks) but only when $v >\zc^2 R^2$, we do not explicitly compute it in the main text here but provide details of the calculation in appendix \ref{sec:prunemdt}. Here we simply note the result that emerges in the soft-collinear limit which reads 
\bea 
\label{pruningNLOCF2_mmdt}
\frac{1}{\sigma}\frac{d \sigma}{d v}^\text{(pruned, $C_F^2$)} = -\left(\frac{C_F \alpha_s}{\pi} \right)^2 \frac{1}{v} \ln^2 \frac{1}{\zc} \ln \frac{R^2}{v}, \quad \zc^2 R^2 < v < \zc R^2,
\eea
where we have not explicitly mentioned hard-collinear correction terms. We 
note that this result coincides with the soft-collinear result for the mMDT and is consistent with the observation in our companion paper~\cite{DasFregMarSal} that in the range of values of $v$ indicated above, the mMDT and pruning are essentially identical (this is true beyond the small $\zc$ limit if one makes the translation $\zc \leftrightarrow \yc/(1+\yc)$). 

\subsection{Summary}
In contrast to the purely single logarithmic behaviour we witnessed for the mMDT, pruning reveals a much richer structure. At leading order it is purely 
single logarithmic and resembles the mMDT but the situation changes dramatically at the NLO level. One obtains a leading $\alpha_s^2 L^4$ double logarithmic 
behaviour for the integrated cross-section which arises from the situation when a gluon that dominates the original jet mass gets removed by pruning. We refer to this situation as $\anomalouspruning$, since the final jet is one-pronged and it comprises of no emission that gets examined for and passes the pruning criterion. We thus have the following results for the coefficients in pruning, where we report below the results in the small $\zc$, small $R$ limit \footnote{Recall that these coefficients are obtained by defining $L \equiv \ln R^2/v$.}:
\bea
a_{12}^{\mathrm{pruned}} &=& a_{12}^{\mathrm{I-pruned}} +a_{12}^{\mathrm{Y-pruned}} = 0+0=0,  \nonumber\\
a_{11}^{\mathrm{pruned}} &=& a_{11}^{\mathrm{I-pruned}}+a_{11}^{\mathrm{Y-pruned}}=0+C_F \ln \frac{e^{-3/4}}{\zc}=C_F \ln \frac{e^{-3/4}}{\zc},  \nonumber \\
a_{24}^{\mathrm{pruned}} &=& a_{24}^{\mathrm{I-pruned}}+a_{24}^{\mathrm{Y-pruned}}=\frac{1}{6} C_F^2+0=\frac{1}{6} C_F^2,  \nonumber \\
a_{23}^{\mathrm{pruned}} &=&a_{23}^{\mathrm{I-pruned}}+a_{23}^{\mathrm{Y-pruned}}=
 \left( - \frac{1}{2}C_F^2 \ln \frac{1}{\zc}-\frac{3}{8}\right)+ \left( - \frac{1}{2}C_F^2 \ln \frac{1}{\zc}+\frac{3}{8}\right)
\nonumber \\ &=& - C_F^2 \ln \frac{1}{\zc}.
\eea
We do not report the result for $a_{22}$ as a variety of terms contribute at this level, including the role of running coupling, non-global logarithms and clustering logarithms. These effects are of course calculable and we have in fact estimated the leading non-global contribution in this article. We leave more 
complete calculations to future work on pruning. 

We further note that resummation of the large logarithms in pruning is possible and is carried out in detail in Ref.~\cite{DasFregMarSal}. The leading order result Eq.~(\ref{LOpruningsc}) can be combined with the leading logarithmic term from the integral $I_2$, Eq.~(\ref{eq:I2}) to yield a resummed structure of the form (ignoring finite $\zc$ terms for simplicity)
\begin{equation}
C_F \frac{\alpha_s}{\pi} \frac{1}{v} \ln \frac{1}{\zc} \times \exp \left[- C_F \frac{\alpha_s}{2\pi} \ln^2 \frac{R^2}{v} \right ].
\end{equation}
This corresponds to the basic resummation structure of what we have chosen to 
label as $\sanepruning$. In practice the form above is a fairly crude representation of the full result for $\sanepruning$ reported in Ref.~\cite{DasFregMarSal}, but sufficient for our purpose here. We note that the resummed result  for $\sanepruning$ involves a double-logarithmic Sudakov form factor for the jet mass $v$. This form factor can be corrected for single logarithmic effects including non-global logarithms. These shall arise at order $\alpha_s^2$ in the form factor and hence shall first be seen at order $\alpha_s^3$ in the expansion for $\sanepruning$ i.e.\ beyond the NLO fixed-order calculations of this article.
The resummation of the $\anomalous$-contribution, corresponding to the integral $I_1$ Eq.~(\ref{eq:I1}), is also possible and has a more complex structure. 
The resummed answer involves a product of the Sudakov form factors in the fat jet mass and the pruned jet mass, with an integral over the fat jet mass~\cite{DasFregMarSal}. 

One notes that the crucial point for pruning is the appearance of double logarithmic form factors that give rise to Sudakov peaks. The transition between single-logarithmic and double-logarithmic regime happens at values of $v\sim\zc^2 R^2$, which for high $p_t$ jets, say $3$~TeV, appear in the vicinity of the electroweak scale. This is a potentially undesirable feature especially for data driven background estimates, in the context of phenomenology. It is certainly obvious that in any case an accurate calculation of pruning is even more involved than the calculation of plain jet mass 
\cite{DasKKKMarSpannow} and far more difficult than for the case of the modified mass drop tagger. In the following section we shall consider the case of trimming.


\section{Trimming}\label{sec:trimming}

\subsection{Definition}
We now turn our attention to the calculation of the jet mass where we use the 
procedure of trimming~\cite{trimming}, to obtain the final massive jet. To obtain a trimmed jet one considers the constituents of a fat jet and  reclusters them in subjets of definite radius $R_\text{sub}<R$, with $R$ the radius of the fat jet. We then eliminate the subjets with transverse momentum $k_t$ softer than a specified fraction of the $p_t$ of the original fat jet. The list of subjets with $k_t> \zc p_t$ 
constitutes the trimmed jet \footnote{The parameter $\zc$ was referred to as $f_\text{cut}$ 
in the original reference Ref.~\cite{trimming} and we have relabelled it for 
ease of comparison with the other substructure methods.}. In the following we consider that the original fat jets as well as its subjets are defined with the \CA algorithm, although other choices are possible. 

\subsection{Leading-order results}
In principle, for our purposes, the trimming method is similar to pruning 
except that it uses a fixed radius $R_\text{sub}$, rather than one chosen dynamically according to the mass of the fat jet.
The leading order calculation is straightforward. Below we consider the soft-collinear approximation and the emission of a single soft gluon which is recombined with a quark to form the fat jet. For convenience we also adopt the small-$\zc$ limit in the following derivation. As usual, a more complete calculation is left to the appendix. Then for $\theta^2 < R_\text{sub}^2$ one 
always has a contribution to the jet mass distribution irrespective of the value of the gluon energy, while for $\theta^2 > R_\text{sub}^2$ the quark and gluon form distinct subjets and if the soft gluon has a fraction $x$ of the fat jet's energy that is below $\zc$ it is removed and there is no contribution to the jet mass distribution. Therefore we are led to the following expression for the jet mass distribution:

\bea \label{LOtrimmingsetup}
\frac{1}{\sigma}\frac{d \sigma}{d v}^\text{(trimmed, LO)}&=& \frac{\as C_F}{\pi}\int \frac{d \theta^2}{\theta^2}\int \frac{d x}{x}\Theta\left(R^2-\theta^2 \right)\Big[\Theta\left(R_\text{sub}^2-\theta^2 \right)+ \Theta\left(\theta^2-R_\text{sub}^2 \right)\Theta(x-\zc)    \Big] \nonumber \\ &\times& \delta \left(v- x \theta^2 \right). \nonumber \\
\eea

Computing the integrals leads to:
\bea \label{LOtrimming}
\frac{1}{\sigma}\frac{d \sigma}{d v}^\text{(trimmed, LO)}
&=&C_F \frac{\alpha_s}{\pi} \frac{1}{v}\Bigg[ \ln \frac{R_\text{sub}^2}{ v} \Theta \left(\zc R_\text{sub}^2-v \right) + \ln \frac{1}{\zc}  \Theta \left(v-\zc R_\text{sub}^2 \right)\Theta(\zc R^2-v)  \nonumber\\&+& \ln \frac{R^2}{v} \Theta \left (v-\zc R^2 \right) \Bigg]. \nonumber\\
\eea

The above result has an interesting structure. It suggests that at the smallest values of $v$, i.e. for the region $v< \zc R_\text{sub}^2$, the trimmed jet 
mass distribution is {\it{double-logarithmic}} just like the case of the plain 
jet mass and in contrast to the leading order results for mMDT and pruning where we obtained only a single-logarithmic behaviour. The result for trimming for $v< \zc R^2$ coincides in fact with the leading-order result for the mass distribution of jets with a reduced radius $R_\text{sub}$. For somewhat larger values of $v$, $\zc R^2> v > \zc R_\text{sub}^2$, there is a transition to a 
single-logarithmic behaviour as observed for the mMDT and also for pruning in the region $\zc R^2 >v>\zc^2 R^2$. For still larger values, $v > \zc R^2$, as for mMDT and pruning one obtains essentially the result for the plain jet mass distribution for jets with radius $R$, i.e the untrimmed result.

In order to confirm the double logarithmic nature of trimming with \event2 we 
first take our result beyond the soft-collinear limit to incorporate finite $R$ and finite $\zc$ effects. The calculation is straightforward and the details 
are mentioned in appendix \ref{app:trimming}. The result we obtain is
\begin{subequations}
\label{LOtrimmingcomplete}
  \begin{align}
\frac{1}{\sigma}\frac{d \sigma}{d v}^\text{(trimmed, LO, full)}
&=C_F \frac{\alpha_s}{\pi} \frac{1}{v}\left[\ln \left( \frac{4  \tan^2(R_\text{sub}/2)} {v}e^{-\frac{3}{4}}\right)+\ord(v) \right], \, \text{for}\quad v< \zc \Delta_{R_\text{sub}}^2\\
\frac{1}{\sigma}\frac{d \sigma}{d v}^\text{(trimmed, LO, full)}
&=C_F \frac{\alpha_s}{\pi} \frac{1}{v}\left[   \ln \left(\frac{1-\zc}{\zc}e^{ -\frac{3}{4}(1-2\zc) }\right) +\ord(v)  \right],\nonumber \\ & \quad \quad  \quad  \quad \quad \quad  \quad \quad \quad \quad \quad \quad  \quad \quad \quad \quad \quad\text{for} \quad \zc \Delta_{R_\text{sub}}^2 <v< \zc \Delta_R^2,
\end{align}
\end{subequations}
where we chose to focus only on the first two regions of the result i.e $v<\zc \Delta_R^2$, ignoring the less interesting region of largest $v$ values, $v > \zc \Delta_R^2$, where there is a transition to the plain jet mass.  Note also that while the above result correctly accounts for finite $R$, $R_\text{sub}$ and $\zc$ effects in the logarithmic terms, the position of the transition points is still 
approximate since we have ignored the longitudinal recoil of the quark against energetic collinear gluons in the definition of the jet mass, i.e replaced $2x(1-x) \left(1-\cos \theta\right)$ in the jet mass definition by $2x \left(1-\cos \theta \right)$, which is sufficient to obtain the logarithmic structure we 
seek here.

\begin{figure}
\begin{center}
\includegraphics[width=0.49 \textwidth]{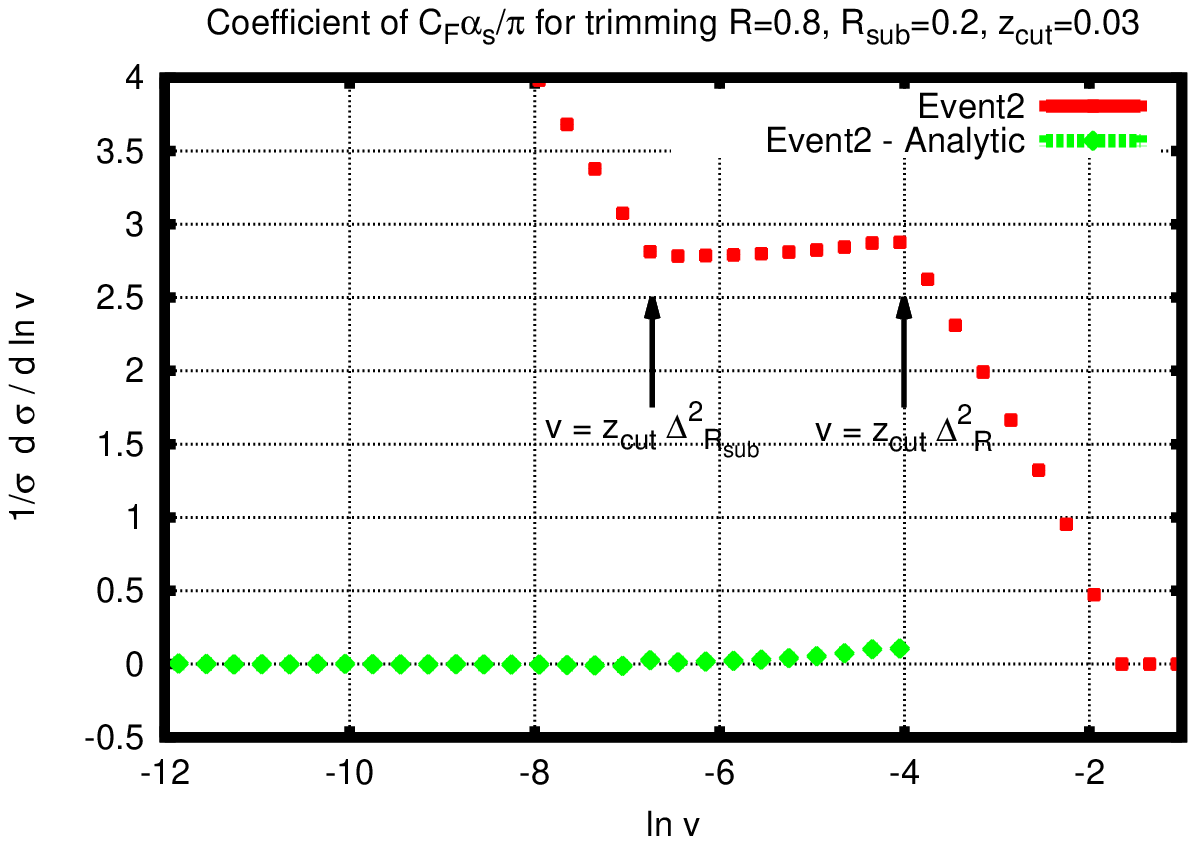}
\includegraphics[width=0.49 \textwidth]{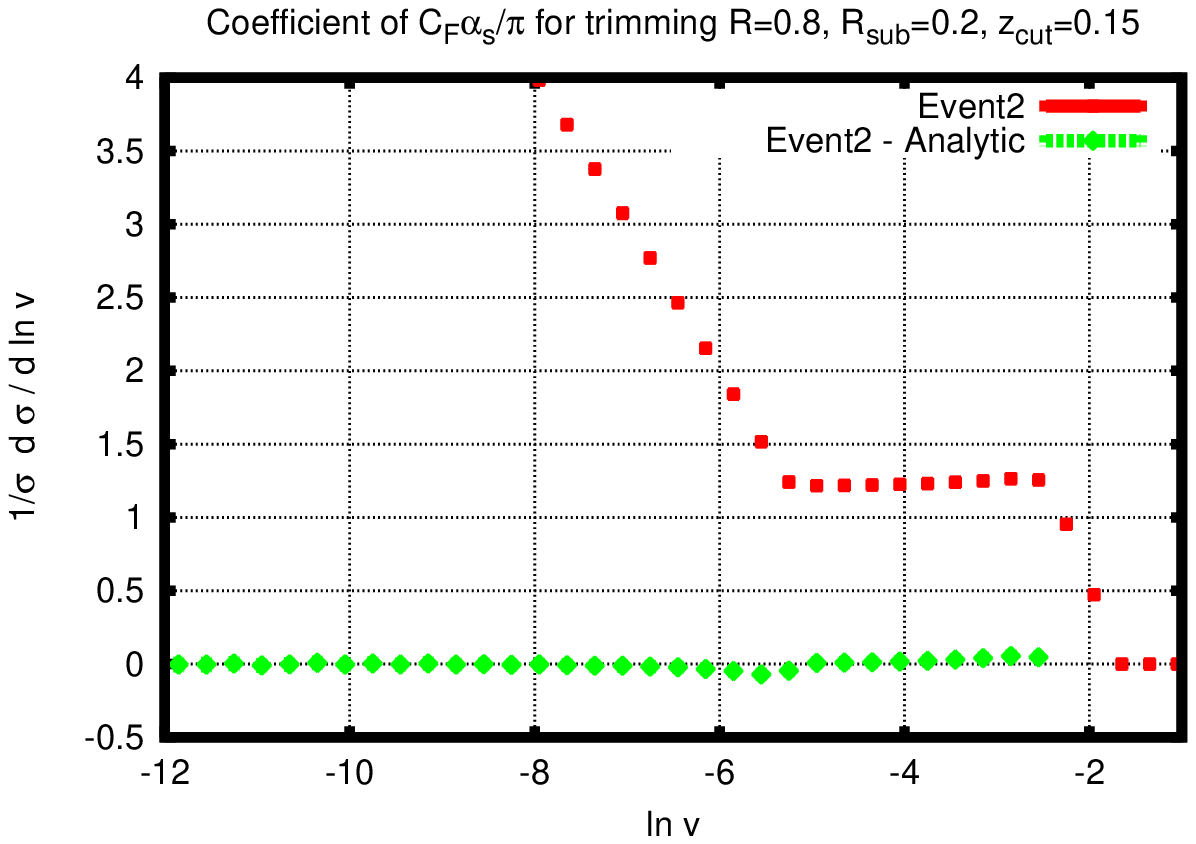}
\caption{Comparison of the analytic calculation Eq.~(\ref{LOtrimmingcomplete}) with \event2 at LO in the region $v< \zc \Delta_R^2$, for different values of $\zc$. The red curve shows the \event2 result alone which exhibits a linear behaviour for the smallest $v$ values, consistent with the double logarithms we find in our calculation. A transition to a flat (single logarithmic) regime is also clearly visible and is as predicted, as is a further transition to another linear (double-logarithmic) regime. The green curve indicates that, after subtracting 
our analytical calculation, the result vanishes at small $v$ as expected.}
\label{fig:trimmingCF}
\end{center}
\end{figure}
The comparison is shown in Fig.~\ref{fig:trimmingCF} for $R=0.8$, $R_\text{sub}=0.2$ and two different values of $\zc$: 0.03 (the value suggested in the original paper~\cite{trimming}) and 0.15. The curves obtained by subtracting Eqs.~(\ref{LOtrimmingcomplete}) from the full LO result shows that 
we have correctly captured the logarithmic behaviour at leading order. Next we 
shall consider the results beyond leading order.

\subsection{Next-to-leading--order results}
\begin{figure}
\begin{center}
\includegraphics[width=0.49\textwidth]{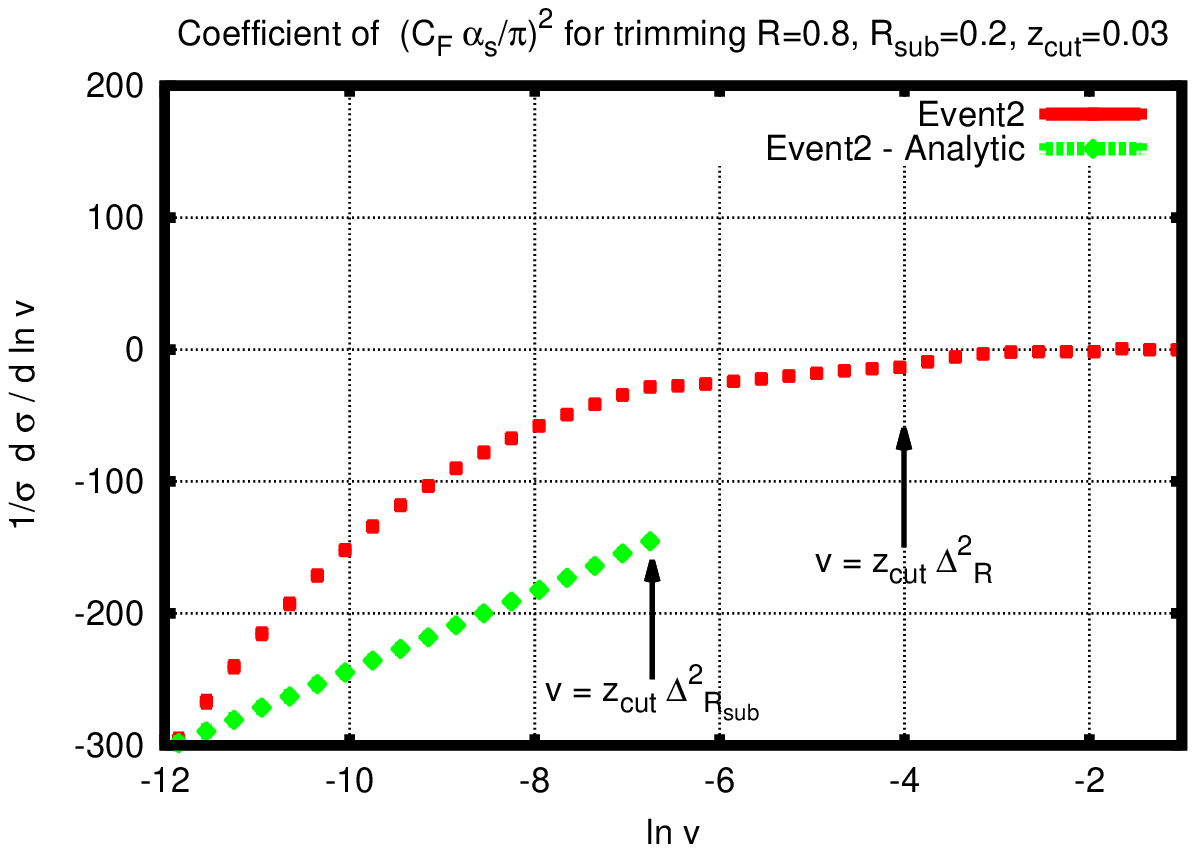}
\includegraphics[width=0.49\textwidth]{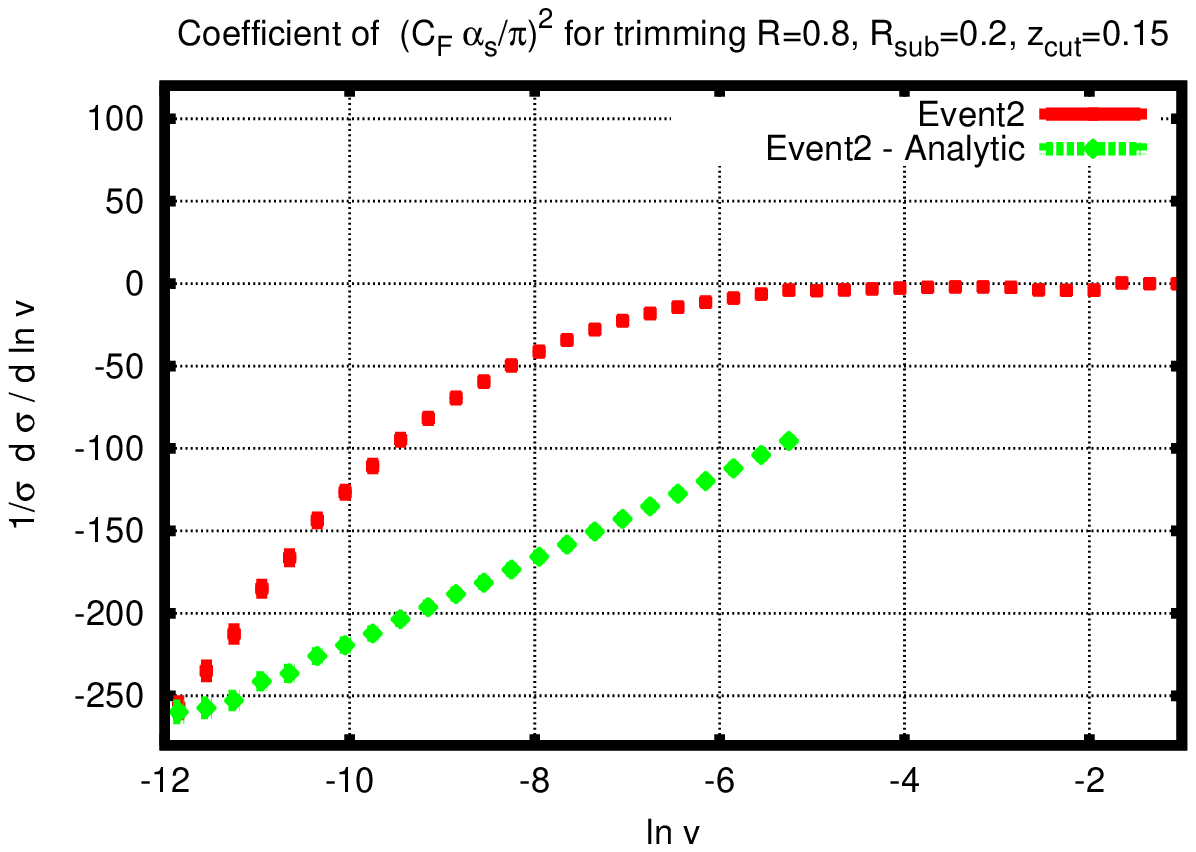}
\includegraphics[width=0.49\textwidth]{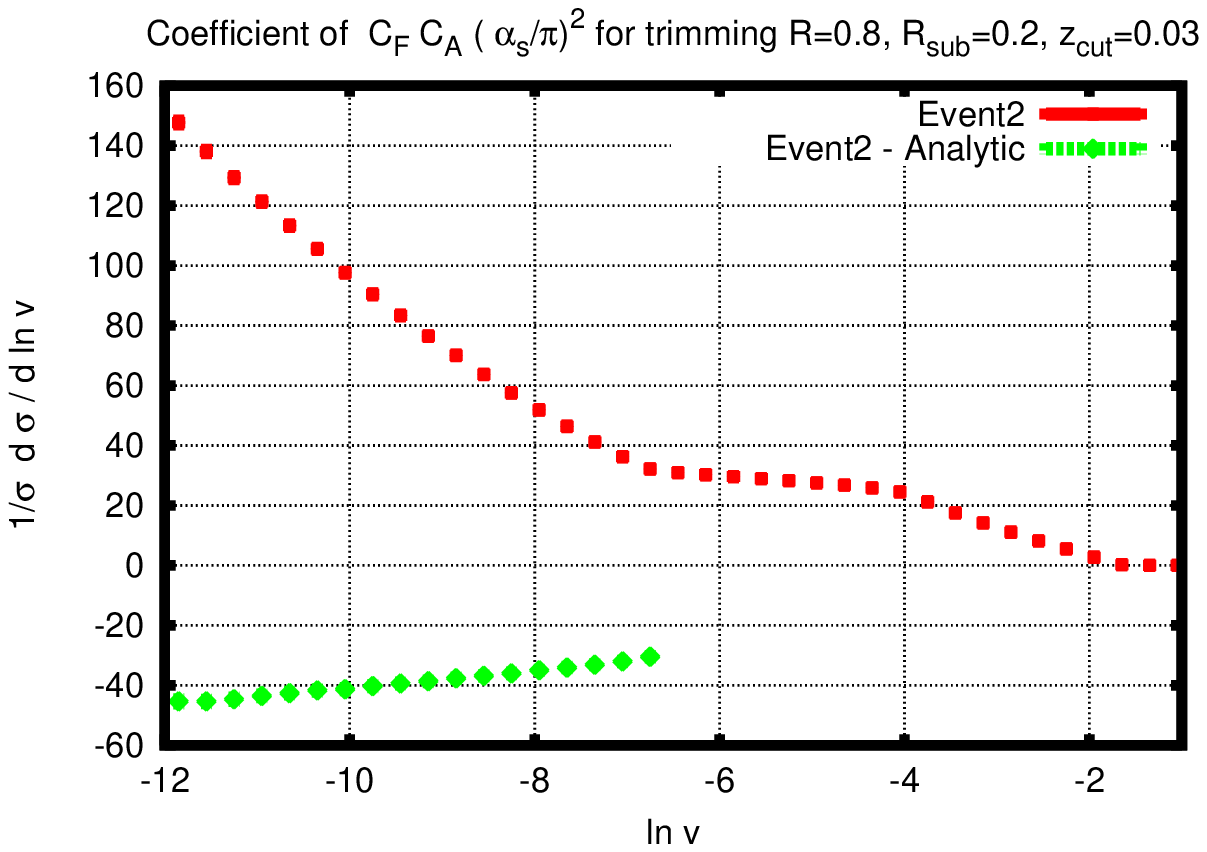}
\includegraphics[width=0.49\textwidth]{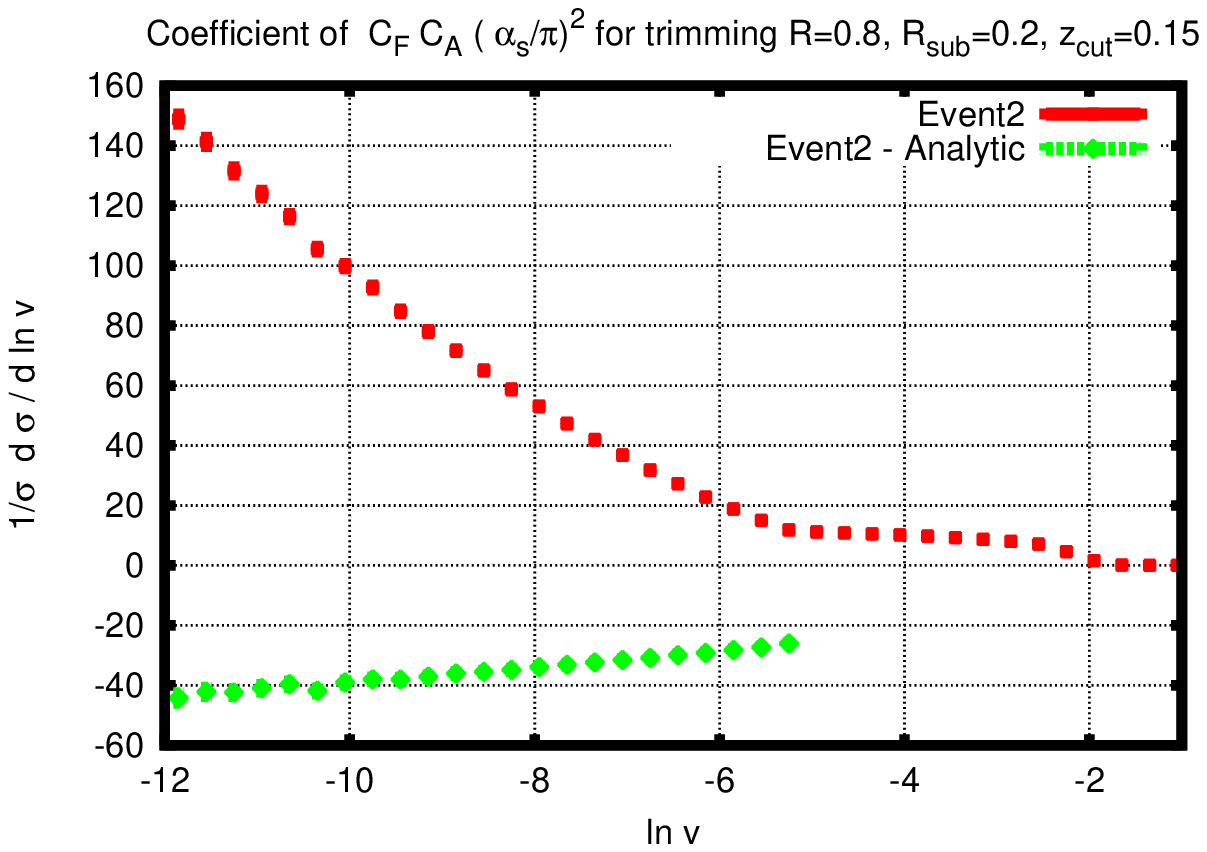}
\includegraphics[width=0.49\textwidth]{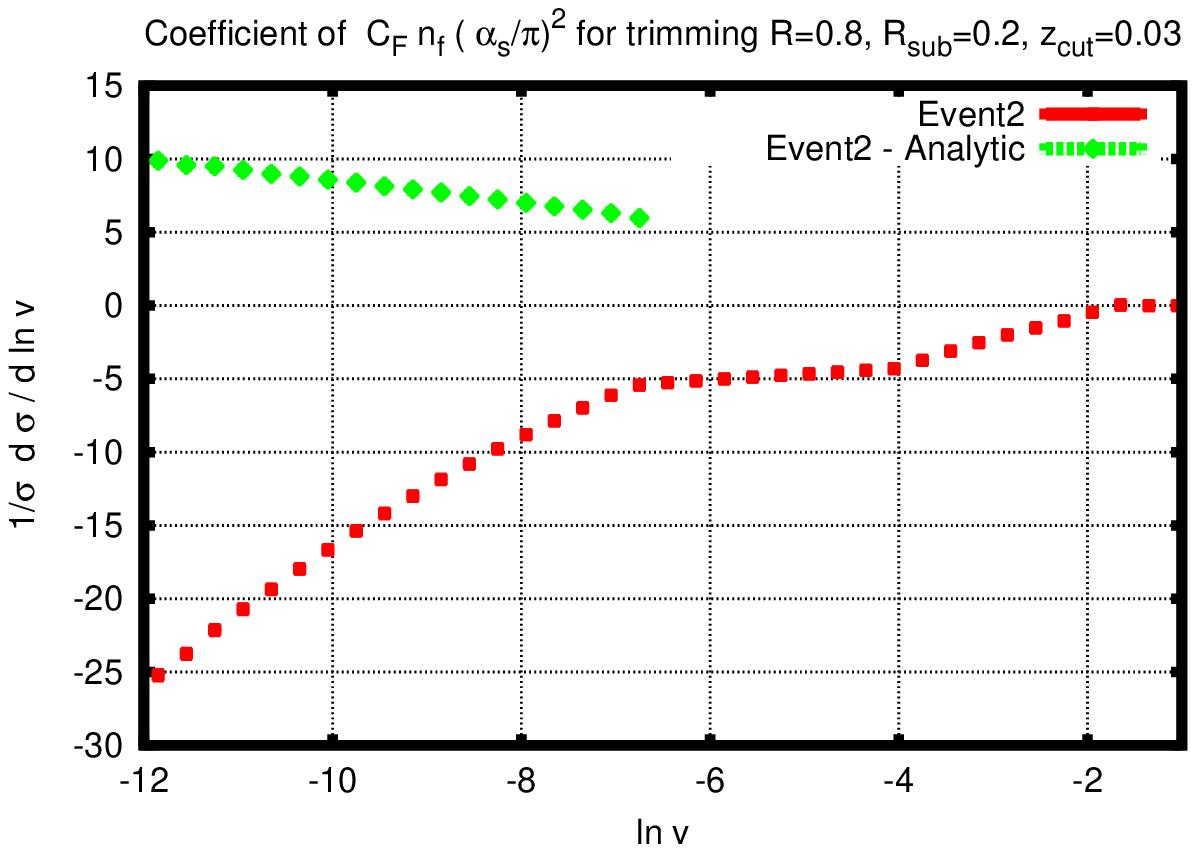}
\includegraphics[width=0.49\textwidth]{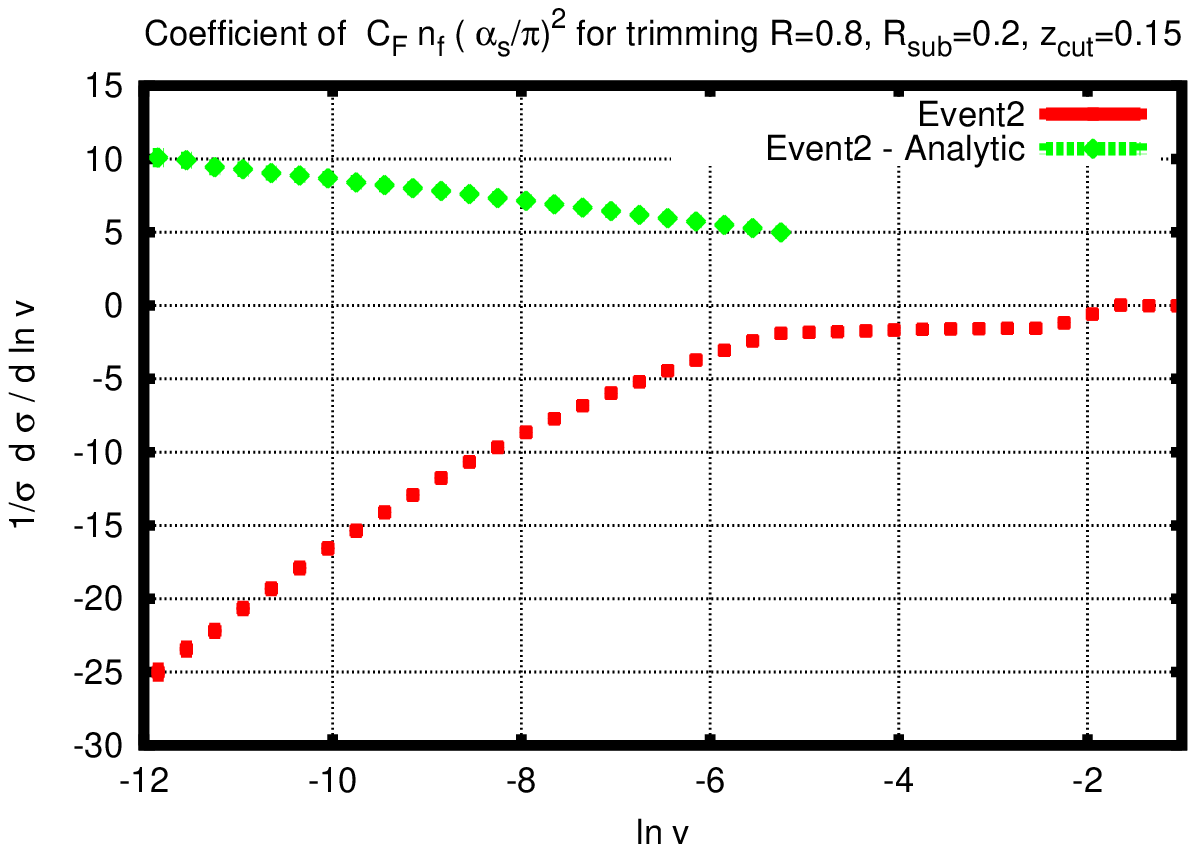}
\caption{
Comparison of the analytic calculation Eq.~(\ref{trimNLOCF2complete}) and Eq.~(\ref{NLOtrimmingRC}) with \event2 at NLO in the region $v< \zc \Delta_{R_\text{sub}}^2$. The green curve indicates that, after subtracting our analytical calculation, the result is a straight line at small $v$ as expected for a single-logarithmic leftover ($\alpha_s^2 \ln^2 1/v$, in the integrated distribution).} 
\label{fig:trimmingNLO}
\end{center}
\end{figure}

 In the previous subsection we have observed that the result for trimming 
contains double logarithms. Physically the origin of the double-logarithmic 
enhancement is rather clear. For emissions that are below $R_\text{sub}$ in angle, there is no cut on the gluon energies and hence the jet mass is trivially the 
usual jet mass with a jet radius corresponding to $R_\text{sub}$. Thus, for the region $v < \zc R_\text{sub}^2$, identified already at leading order, one can 
anticipate (and easily verify) that the NLO result at the accuracy we aim for in this study, i.e.\ $\alpha_s^2 L^4$ and $\alpha_s^2 L^3$ terms in the integrated distribution, will just be the NLO result for jet mass with a jet radius $R_\text{sub}$. We will check this against results from \event2. The $\alpha_s^2 L^2$ terms, like for the plain jet mass originate from a variety of sources including non-global and clustering effects. While these are of course calculable we do not perform explicit calculations for these effects in the present article, where our aim is restricted to verifying the general features and physics of the substructure methods. The results are (note that the terms reported below arise from the region $v< \zc \Delta_{R_\text{sub}}^2$):
\beq \label{trimNLOCF2complete}
\frac{1}{\sigma}\frac{d \sigma}{d v}^\text{(trimmed, $C_F^2$, full)} = -
\left( \frac{\alpha_s C_F}{\pi}\right)^2 \frac{1}{v}  \left[ \frac{1}{2}\ln^3 \frac{1}{v}+\frac{3}{2}\left( 2 \ln \left( 2 \tan\frac{R_\text{sub}}{2}\right)-\frac{3}{4}\right) \ln^2\frac{1}{v} \right] .
\eeq
Moreover, to the accuracy we are working at, all the $C_F C_A$ and $C_F n_f$ 
contributions comes exclusively from the running of the strong coupling:
\beq\label{NLOtrimmingRC}
\frac{1}{\sigma}\frac{d \sigma}{d v}^\text{(trimmed, r.c.)}
= \left( \frac{\alpha_s}{\pi}\right)^2C_F   \frac{11 C_A-2 n_f}{8}  \frac{1}{v} \ln^2\frac{1}{v} 
\eeq
These results are checked against \event2 in Fig.~\ref{fig:trimmingNLO}. One 
notes that the difference between \event2 and the analytic results is 
consistent with a linear behaviour which indicates that in all channel we have a leftover which corresponds to $\alpha_s^2 L^2$ for the integrated distribution.

Also the presence of non-global logarithms for trimming is obvious. This is 
because one can have a soft gluon which makes an angle larger than $R_\text{sub}$ with the hard initiating quark and has energy fraction below $\zc$, emitting a much softer gluon which has an angle less than $R_\text{sub}$ with the hard quark. The first gluon gets removed by trimming 
and hence does not contribute to the trimmed jet mass, while the second much 
softer gluon makes the essential contribution, which corresponds to a 
non-global logarithmic term.\footnote{This is the same configuration as we addressed for non-global logarithms in pruning with the difference arising from the fact that here we have a fixed radius $R_{\text{sub}}$ rather than one chosen according to the fat jet mass.} 
If one ignores clustering effects these non-global logarithms would be the same as for the case of the plain jet mass in the presence of a veto (here the $\zc$ condition), computed for the anti-$k_t$ algorithm in Ref.~\cite{BanDasKKKMar}. However here for the C/A algorithm we employ, jet clustering effects occur which somewhat reduce the non-global component as detailed in~\cite{AppSey1}. Due to the small values of $R_\text{sub}$ used in practice, we expect such clustering effects to be rather small and hence non-global logarithms should be accounted for.
Lastly we note that NLO calculations can be carried out also for the other regions of the jet mass i.e for $v > \zc \Delta_{R_\text{sub}}^2$. We provide the 
details of this calculation, which for economy of presentation we carry 
out in the small $\zc$, $R_\text{sub}$ and small $R$ limit, 
in appendix \ref{app:trimming}. The NLO calculation shows that the feature of 
three distinct regions for trimming, identified at leading order, remains and the result is consistent with an exponentiation of the leading-order result 
for trimming.

\subsection{Summary}
We have noted above that trimming for sufficiently small jet masses gives a result similar to that for the plain jet mass. As for plain jet mass mMDT and pruning, we can summarise the relevant coefficients for trimming which, as before, for brevity we report in the small $R_\text{sub}$ approximation (note that here we define $L = \ln \frac{R_\text{sub}^2}{v}$):
\bea
a_{12}^{\mathrm{trimmed}} &=&  C_F \\ \nonumber
a_{11}^{\mathrm{trimmed}} &=& -\frac{3C_F}{4} \\ \nonumber
a_{24}^{\mathrm{trimmed}} &=& -\frac{C_F^2}{2} \\ \nonumber
a_{23}^{\mathrm{trimmed}} &=& \frac{3}{8} C_F \left(3 C_F + 4  \beta_0 \right),
 \eea
i.e identical to the coefficients for the plain jet-mass for which however the logarithm $L$ was defined in terms of the fat jet radius $R$ instead of $R_{\text{sub}}$.

Finally we note that an all-order result for trimming is also straightforward to obtain. 
The basic form of the resummed integrated distribution, in a fixed-coupling approximation and ignoring finite $\zc$ corrections, is given by the exponentiation of the integrated result for the single-gluon emission Eq.~(\ref{LOtrimming}). 
It therefore follows that the mass 
distribution for trimmed jets, like for the plain jet mass distribution, will have the feature of a Sudakov peak. As shown in Ref.~\cite{DasFregMarSal}, in case of high-$p_t$ jets, the departure of trimming from a single logarithmic mMDT-like behaviour, and the location of Sudakov peak, which happen below $v=\zc \Delta_{R_\text{sub}}^2$, 
can occur in a phenomenologically crucial region, where jet masses are of the 
order of the electroweak scale, and an ideal substructure tool should probably not give 
rise to such structures in the background.

\section{Conclusions and outlook}
In this article we have studied jet masses of QCD jets after the application of boosted-object algorithms, specifically the mass-drop, pruning and trimming 
techniques. A novel feature of our study is that it is analytical, rather than 
one employing Monte Carlo event generators as is the standard practice for most substructure analyses. Here instead we have started to explore the perturbative structure of the substructure methods by using the jet mass as an observable and generating leading and next-to--leading order results in the eikonal approximation extended to treat hard collinear radiation. In the present article, we have explicitly considered the case of jets produced in $e^+e^-$ collisions, 
both in the small-$R$ approximation and for finite $R$. The results in the small-$R$ limit can be also used for (quark) jets in hadron-hadron collisions in 
the same limit. However, going beyond this approximation would require computing contributions from initial-state radiation as well.  These effects have been studied in the case of plain jet mass and they turn out not to be large, mainly affecting the peak region \cite{DasKKKMarSpannow}. One can expect similar effects for our present observables with the exception of the mMDT where due to their pure collinear origin the leading logarithms are independent of $R$. Moreover the Monte Carlo studies of Ref.~\cite{DasFregMarSal} indicate that the finite $R$ correction terms are not critical to an understanding of the main features of taggers. Such terms remain of importance should one wish to carry out accurate phenomenology based on resummed calculations for jet masses and substructure observables.

Our main motivation for this study has been to both understand the features of substructure methods themselves as well as to examine what may be the most accurate methods to calculate such substructure observables. To be more precise, a feature that is common to all the algorithms we have studied, is the fact that they all cut on soft radiation inside a jet, which is necessary to discriminate against 
background QCD jets while having minimal effect on signal jets. A question that one may then ask is whether employing such cuts eliminates the large 
logarithms in $m_j/p_t$ encountered in calculations of the jet mass 
distribution, at least to some degree. If this were the case then there is the possibility that pure fixed-order calculations may be employed to compute such observables most accurately. On the other hand one may also consider that any leftover logarithmic structure may require resummation and examine the feasibility of all-order resummed studies to best describe these substructure observables. 

In the above context we have uncovered several aspects of substructure methods which are both interesting in their own right as well as point the way to future studies that it would be of interest and value to carry out. We started by examining the standard mass-drop tagger~\cite{BDRS} and finding that at leading order the corresponding jet mass distribution had a single logarithmic behaviour (in contrast with double-logarithms obtained for the the plain jet mass). However at NLO the situation changes and one finds an $\alpha_s^2 L^3$ leading term, which arises due to a flaw in the mass-drop procedure (following the more massive rather than harder branch)~\cite{DasFregMarSal}. We then considered the modified mass-drop tagger (mMDT), proposed in the companion article~\cite{DasFregMarSal} which removes the flaw in the mass-drop tagger, referred to above. For mMDT we found a pure single-logarithmic behaviour at both leading order and NLO. We demonstrated that the NLO result is consistent with an exponentiation of the leading order result in the limit of small $\yc$. We confirmed our answers by checking them against exact fixed-order results from the program \event2 and discussed ``flavour changing'' logarithmic effects that are needed to go beyond the small $\yc$ approximation. We also demonstrated the absence of non-global logarithms and emphasised that the pure single logarithmic results devoid of non-global logarithms made the mMDT jet mass distribution unique amongst single jet observables at hadron colliders. A more complete treatment of the resummation for jet masses with mMDT can be found in Ref.~\cite{DasFregMarSal}. In future work we intend to investigate 
phenomenologically the accuracy of both fixed-order as well as matched resummed results for the mMDT, by direct comparisons to LHC data.

Next we turned our attention to pruning. Again we found that at leading-order 
the result was single logarithmic along the same lines as for the (m)MDT. At 
NLO however the situation changed and one encountered double logarithms $\alpha_s^2 L^4$ which are as singular as those for the plain jet mass. Additionally 
we also pointed out the presence of non-global and clustering logarithms for pruning in contrast to the mMDT case. We noticed via the NLO calculation, that pruning has {\it{multiple transition points}}. In particular for $\zc R^2> v > \zc^2 R^2$ we observed that the double-logarithmic behaviour gives way to a pure single-logarithmic result which suggests that the mMDT and pruning are identical (up to subleading corrections we do not treat here) in this regime. However for small jet masses $v < \zc R^2$ it is clear from our calculations that one can expect the presence of Sudakov-peak like structures in all-order results owing to the double logarithmic resummation required. We confirmed our calculations with \event2 and also explained in detail the physical origin of the double logarithms. These arise from configurations in which the emission that dominates the fat-jet mass, and hence sets the pruning radius, is removed by pruning, leaving arbitrarily soft and collinear emissions in the core of the jet, which are never tested for pruning.
We next considered a variant of pruning ($\sanepruning$~\cite{DasFregMarSal}) and demonstrated that this eliminated the one-prong configurations and double logarithms so that at 
$n^{\mathrm{th}}$ order the leading logarithms are $\alpha_s ^n L^{2n-1}$, rather than double logarithms. We pointed out that both pruning and $\sanepruning$ are resummable though the resummation for pruning is significantly more complicated, on a technical level, than that for the plain jet mass. Detailed formulae encapsulating the resummation for pruning and $\sanepruning$ can be found in Ref.~\cite{DasFregMarSal}. Additionally the presence of double logarithms and consequent Sudakov peaks as well as transition points for pruning give the QCD background an uncomfortably rich structure in terms of phenomenological studies aiming to discover signal peaks associated with the presence of new particles. 

Lastly, we considered jet mass distributions obtained after employing the method of trimming. Here in contrast to the (m)MDT and pruning we found a double 
logarithmic term, $\alpha_s L^2$ at leading order, which persists at NLO via an $\alpha_s^2 L^4$ term and at all subsequent orders. There are also transition points for trimming similar to the pruning case. In fact for $\zc R^2 > v> \zc R_{\text{sub}}^2$ one observes a single logarithmic behaviour as is the case 
for mMDT in the entire range of $v$ and for pruning at intermediate values of $v$. A basic resummed result for jet masses with trimming essentially amounts to an exponentiation of the leading order result which then needs to be 
corrected for various single logarithmic effects such as non-global and clustering logarithms as well as the effect of multiple emissions. In practice, this resummation is similar to that required for the plain jet mass, which was treated for anti-$k_t$ jets in Refs.~\cite{BanDasKKKMar,DasKKKMarSpannow}. As for pruning, potentially unwelcome Sudakov peaks in the background also exist for the 
trimming case and more details for this, in the context of phenomenology, can be found in Ref.~\cite{DasFregMarSal}. 

Our results can be exploited in different ways. Having analytical formulae at hand helps us to understand the dependence of observables such as jet masses on the parameters of jet finding, on those involved in substructure algorithms and the interplay between them. This information is valuable while making choices of parameters for phenomenological studies, especially in a discovery context. In order for such work to be considered more complete, analytical studies for the impact of the substructure methods on the signal should also be carried out in the future and the performance of the substructure methods should be considered and compared also in the context of signal to background ratios. Knowing about features of substructure methods such as transition points and Sudakov peaks identified by our present studies is also crucial when it comes to data driven background estimates and this is another phenomenological aspect where we expect our studies to be of value. 
Lastly, knowledge 
of the perturbative structure and specifically about the presence of large 
logarithms alerts us to whether one can use fixed-order tools to compute tagged jet masses (which may be possible for the mMDT where there are no double logarithms), or whether one can carry out a resummed calculation with fixed-order matching. Also the analytical estimates can 
be used for direct comparison to event generator tools, a task we have embarked on in Ref.~\cite{DasFregMarSal}. All of the above should help us to make 
better estimates of the true uncertainty involved in theory predictions for 
jet mass and other distributions for jet-substructure observables,  which in turn is 
important for future LHC phenomenology with boosted objects. 

We conclude by stressing that a full analytical understanding of substructure methods and consequently the development of optimal techniques can still be considered in its infancy and there is scope for substantial progress to be made in the near future, guided at least in part by our current findings.

\section*{Acknowledgements}
We would like to thank Gavin Salam for several useful comments and discussions throughout the course of this work. One of us (MD) would like to thank the IPPP Durham for hospitality during part of this work. 
AF acknowledges useful discussions with Mike Seymour on \event2 and related topics. AF is supported by an EPSRC studentship.
This work is supported by the UK's STFC.

\appendix

\section{MDT}
\subsection{Leading-order result beyond small $\yc,R$}
\label{mdtlofull}
The full result in the soft-approximation, for the MDT jet mass distribution, 
may be obtained by considering the emission pattern of a soft gluon from a $q \bar{q}$ pair. In the soft limit one can neglect recoil and assume the $q$ and $\bar{q}$ to be back-to--back and hence write the kinematics as:

\bea
p &=& E_q (1,0,0,1), \\ \nonumber
\bar{p} &=& E_{\bar{q}}(1,0,0,-1), \\ \nonumber
k &=& E_g \left(1,0,\sin \theta, \cos \theta \right).
\eea
In the eikonal approximation the squared matrix-element for gluon emission from the $q \bar{q}$ dipole is given by 
\begin{equation}
\label{prob}
W (k) = C_F \frac{\alpha_s}{\pi} \frac{\left(p_1.p_2 \right)}{\left(p_1.k \right) \left(p_2.k \right)} =  \frac{2}{E_g^2} C_F \frac{\alpha_s}{\pi} \frac{1}{\left(1-\cos^2 \theta \right)}.
\end{equation}
Using the above emission probability and integrating over the gluon emission 
phase-space gives, for the jet-mass $v$, after applying MDT cuts:

\bea \label{LOmdtsetup}
\frac{1}{\sigma}\frac{d \sigma}{d v}^\text{(MDT, LO, full)}&=& \frac{2\as C_F}{\pi}\int \sin \theta \frac{d \theta}{1-\cos^2 \theta}\int \frac{d x}{x}\Theta\left(\Delta_R^2 -\Delta_\theta^2 \right)\Theta\left(x-\frac{\yc}{1+\yc} \right) \nonumber\\
&&\Theta \left (\frac{1}{1+\yc}-x \right) \delta \left(v- 2 x (1-\cos \theta) \right) \nonumber \\
&=& \frac{\as C_F}{\pi}\int \sin \theta \, d \theta \left( \frac{1}{1-\cos \theta}+\frac{1}{1+\cos \theta}\right)\int \frac{d x}{x}\Theta\left(\Delta_R^2 -\Delta_\theta^2 \right)
\nonumber\\
&&\Theta\left(x-\frac{\yc}{1+\yc} \right) \Theta \left (\frac{1}{1+\yc}-x \right) \delta \left(v- 2 x (1-\cos \theta) \right),\nonumber \\
\eea
where we have separated the singular behaviour in the limit where the gluon is emitted collinear to the measured jet, i.e. $\theta\to 0$. In order to capture single logarithms arising from hard-collinear emission, the integral in this region must be performed with the full splitting function $p_{gq}(x)$ rather than its soft approximation $1/x$ , where 
\begin{equation}
p_{gq}(x) = \frac{1+(1-x)^2}{2x}. 
\end{equation}
 Note that in the definition of jet-mass we have gone beyond the small-angle approximation. We can continue to neglect at our accuracy the effects of energy-loss of the quark in the definition of the jet mass i.e treat $x(1-x)  \approx x$, since retaining the full result only changes our answer at the level of non-singular terms.

Carrying out the required integrals then produces the result quoted in the main text and used for comparisons to \event2.

\section{Pruning}
\subsection{Leading-order calculation}
The leading order result can be obtained by considering a soft gluon emitted off a $q \bar{q}$ dipole as for the MDT described above. We consider that the gluon is recombined with say the quark and use precisely the same kinematic pattern as for the MDT calculation performed in the preceding section. One then gets:
\begin{multline} 
\label{LOpruningsetup}
\frac{1}{\sigma}\frac{d \sigma}{d v}^\text{(pruned, LO)} = \frac{2\as C_F}{\pi}\int d \theta \frac{\sin \theta}{1-\cos^2 \theta}\int_{\zc}^{1-\zc}\frac{d x}{x}\Theta\left(\Delta_R^2 -\Delta_\theta^2 \right) \delta \left(v- 2 x (1-x) (1-\cos \theta) \right).
\end{multline}

Once again as for MDT we can split the integral in $\theta$ so as to separate out the collinear singular $1/(1-\cos \theta)$ piece and in this region we replace the divergence $1/x$ by the full splitting function $p_{gq}(x)$. Performing the integral then produces the result Eq.~(\ref{LOpruningfull}) quoted in the main text.

\subsection{Next-to--leading order, independent emission}
\label{prune:nlo}
Here we report the calculations for the integrals $I_1$ and $I_2$, Eqs.~(\ref{eq:intsc1}) and (\ref{eq:intsc2}) in the main text, beyond the soft {\it{and}} collinear approximation. First we lift the requirement of collinearity and use the full emission pattern for each soft gluon by a $q \bar{q}$ antenna. We are still working in the soft limit and can neglect the recoil of the $q \bar{q}$ pair which are back-to--back. Thus the integrals $I_1$ and $I_2$ generalise to 
\begin{multline}
I_1 =  \left(\frac{2 \alpha_s C_F}{\pi} \right)^2 \int  \frac{d x_1}{x_1}\frac{d x_2}{x_2} \frac{d \cos \theta_1}{1-\cos^2 \theta_1} \frac{d \cos \theta_2}{1-\cos^2 \theta_2} \delta \left(v-2 x_2 \left(1-\cos \theta_2 \right) \right)\\ 
\Theta(\zc-x_1) \Theta(2(1-\cos \theta_1)-R_\text{prune}^2) \Theta(R_\text{prune}^2-2(1-\cos \theta_2)) 
\end{multline}
and 
\begin{multline}
I_2 =-  \left(\frac{2 \alpha_s C_F}{\pi} \right)^2 \int \frac{d x_1}{x_1}\frac{d x_2}{x_2} \frac{d \cos \theta_1}{1-\cos^2 \theta_1} \frac{d \cos \theta_2}{1-\cos^2 \theta_2}\delta \left(v-2 x_2 \left(1-\cos \theta_2 \right) \right) \\ 
\Theta(x_2-\zc)\Theta(1-\zc-x_2) \Theta(2(1-\cos \theta_1)-R_\text{prune}^2) \Theta(R_\text{prune}^2-2(1-\cos \theta_2)) 
\end{multline}

We have $R_\text{prune}^2 = 2 x_1 \left(1-\cos \theta_1 \right)+2 x_2 \left(1-\cos \theta_2 \right)$, where to accommodate large angles we do not make any collinear approximation in the definition of $R_\text{prune}$, while continuing to use the soft limit $x_{1,2} \ll 1$. One can anticipate that the leading singular contributions will arise from $I_1$ since unlike $I_2$ there is no infrared 
cut-off $\zc$ on the energy fractions.
It also proves convenient to perform a decomposition
\begin{equation}
\frac{1}{1-\cos^2 \theta_i} =\frac{1}{2} \left ( \frac{1}{1-\cos \theta_i}+\frac{1}{1+\cos \theta_i} \right)
\end{equation}
for each gluon $i$, to separate the leading soft and collinear contributions 
from the less singular soft wide-angle terms. 
We thus generate four terms for each of $I_1$ and $I_2$. Of these we find that only the two terms containing the collinear singularity for gluon $k_2$, give rise to relevant large logarithmic terms. For convenience we change variables to $t_i= 2 (1-\cos\theta_i)$ and we first consider the most singular term involving a collinear singularity for each gluon in $I_1$. We also make the substitution $\frac{1}{x_i}\to p_{gq}(x_i)$ in order to capture hard-collinear radiation, for the terms where gluon $i$ goes collinear to the quark. We then define the collinear term:
\begin{eqnarray} \label{I1c}
I_1^c &=&  C_F^2 \left(\frac{\alpha_s}{\pi} \right)^2 \frac{1}{v} \int dx_1 dx_2 p_{gq}(x_1) p_{gq}(x_2) \frac{dt_1}{t_1}  \Theta \left(t_1-v \frac{(1-x_2)}{x_1 x_2} \right) \Theta(1-x_1-x_2) \\ \nonumber  &\times& \Theta(\zc-x_1)\Theta (\Delta_R^2-t_1) 
\end{eqnarray}
where $c$ denotes that both gluons can go collinear i.e the $1/(1-\cos \theta_i)$ terms only. The other contribution to $I_1$ is the soft term $I_1^s$ where emission $1$ does not have a collinear enhancement but is soft. The integral to evaluate is the same as in Eq.~(\ref{I1c}), with a different matrix element obtained by replacing $\frac{1}{t_1} \to \frac{1}{4-t_1}$, which has a finite behaviour as $t_1 \to 0$. Hence we do not employ the splitting function $p_{gq}(x_1)$ here unlike in the case of $I_1^c$ but just work with the soft $1/x_1$ pole. For gluon 2 on the other hand, which has a collinear enhancement, we continue to employ the full splitting function. The results are (recall 
$\Delta_R^2=2\left(1-\cos R \right)$):
\begin{equation} \label{pruningNLOCF2_1}
I^c_{1} =  \left(\frac{\alpha_s C_F}{\pi} \right)^2\frac{1}{v} \left[ \frac{1}{6} \ln^3 \frac{1}{v}+\left( \frac{1}{2} \ln \Delta_R^2 -\frac{1}{2} \left(\zc-\frac{\zc^2}{4}-\ln \zc+\frac{3}{4}  \right) \right)  \ln^2 \frac{1}{v} \right].
\end{equation}

\beq  \label{pruningNLOCF2_3}
I^s_{1} =  -\left(\frac{\alpha_s C_F}{\pi} \right)^2\frac{1}{2v}    \ln \left( 1- \frac{\Delta_R^2}{4}\right)\ln^2 \frac{1}{v}.
\eeq

A similar treatment for $I_2$ can be carried out. Since this integral is less singular, only the term where both gluons contribute a collinear singularity 
matters at our accuracy. The result is 

\begin{equation}  \label{pruningNLOCF2_2}
I^c_{2} = - \left(\frac{\alpha_s C_F}{\pi} \right)^2 \frac{1}{2v} \left( \ln \frac{1-\zc}{\zc}+\frac{3}{2}\zc-\frac{3}{4}  \right)  \ln^2 \frac{1}{v}.
\end{equation}

The final result quoted in the main text Eq.~(\ref{pruningNLOCF2_final}) 
corresponds to $I_1^c+I_1^s+I_2^c$.

\subsection{The region $\zc^2 R^2 < v < \zc R^2$}
\label{sec:prunemdt}
In this appendix we compute the behaviour of pruning in the intermediate mass region $\zc^2 R^2 < v < \zc R^2$. Because our checks with \event2 are confined to the small-$v$ region, we are not concerned with this here. As a consequence, we decide to work in the small-$\zc$ approximation, which simplifies the algebra and it is enough to highlight the point we wish to make.

We shall need to consider the independent emission of two soft gluons where the region of integration considered is $\theta_{1}^2,\theta_{2}^2 > 
R_{\mathrm{prune}}^2$. 
Once again one considers double-real and one-real 
one-virtual contributions together. Specifically, in the double real term we can have both $x_1,x_2 > \zc$ or only one of them greater than $\zc$. When both emissions have energy fractions below $\zc$, they are both removed and there is no contribution. Thus we have 
\begin{equation}
I_3 = C_F^2 \left(\frac{\alpha_s}{\pi}\right)^2 \frac{1}{2!} \int \frac{d x_1}{x_1} \frac{d x_2}{x_2} \frac{d \theta_1^2}{\theta_1^2} \frac{d\theta_2^2}{\theta_2^2} \Theta \left(\theta_1^2 -R_{\text{prune}}^2 \right) \Theta \left (\theta_2^2-R_{\text{prune}}^2 \right) \Delta [\Theta],
\end{equation}
 where one has 
\begin{align}
\Delta[\Theta] = &\Theta \left ( x_1-\zc \right) \Theta \left( x_2 -\zc \right) 
\delta \left ( v -x_1 \theta_1^2 -x_2 \theta_2^2 \right) \nonumber \\
+& \Theta \left ( x_1-\zc \right) \Theta \left( \zc-x_2 \right ) \delta \left (v-x_1 \theta_1^2 \right) \nonumber \\ 
+& \Theta \left( x_2-\zc \right) \Theta \left( \zc-x_1 \right ) \delta \left ( v-x_2 \theta_2^2 \right)\nonumber \\
-&\Theta \left (x_2-\zc\right) \delta \left (v -x_2 \theta_2^2 \right) - \Theta \left(x_1-\zc \right) \delta \left(v-x_1 \theta_1^2 \right).
\end{align}

The first term in $\Delta[\Theta]$ above refers to the contribution when both 
real emissions have energy fractions above $\zc$ while the next two terms 
correspond to having either $x_1$ or $x_2$ below $\zc$, which results in the 
emission being pruned away. The final two terms (with negative signs) are the 
contributions when $k_1$ or $k_2$ is virtual with the other emission being real. For such contributions one always needs a cut on the energy fraction of the 
real emission, to obtain a finite jet mass. 

One can further combine real and virtual terms to obtain
\begin{equation}
\Delta [\Theta]= \Theta \left (x_1-\zc \right) \Theta \left(x_2 -\zc \right)\left 
[\delta \left (v-x_1 \theta_1^2-x_2 \theta_2^2\right) -\delta \left ( v -x_1 \theta_1^2\right ) -\delta \left ( v-x_2 \theta_2^2\right) \right ]
\end{equation}

Using the fact that $R_{\text{prune}}^2 = x_1 \theta_1^2 +x_2 \theta_2^2$, one can evaluate the integral $I_3$. No logarithmically enhanced terms 
are found for the region $v < \zc^2 R^2$ and hence the results we obtained from the integrals $I_1$ and $I_2$ , reported in the main text, correspond to the full answer for pruning. However in the region $v > \zc^2 R^2$ one finds a 
double logarithmic behaviour that cancels the contribution from $I_1$ in the main text:

\begin{equation}
I _3 = C_F^2 \left(\frac{\alpha_s}{\pi}\right)^2 \frac{1}{v}\left[-\frac{1}{6} \ln^3 \frac{R^2 \zc}{v}+\mathcal{O} \left (\ln^2 \frac{1}{v} \right) \right] \Theta \left(v-\zc^2 R^2 \right) \left(\zc R^2-v \right).
\end{equation}

Combining all contributions in the region $\zc^2 R^2 < v < \zc R^2$ one gets
\bea 
\label{pruningNLOCF2_mmdt}
I_1+I_2+I_3 = -\left(\frac{C_F \alpha_s}{\pi} \right)^2 \frac{1}{v} \ln^2 \frac{1}{\zc} \ln \frac{R^2}{v}, \, \, \zc^2 R^2 < v < \zc R^2,
\eea
the result quoted in the main text.

\section{Trimming} \label{app:trimming}
\subsection{Leading-order calculation}
The leading-order result can be obtained by considering a soft gluon emitted off a $q \bar{q}$ dipole as for the MDT and pruning cases described above. Considering the gluon to be recombined with the quark, one gets:
\bea \label{LOtrimmingsetupcomplete}
\frac{1}{v}\frac{d \sigma}{d v}^\text{(trimmed, LO)}&=& \frac{2\as C_F}{\pi}\int \frac{\sin \theta d \theta}{1-\cos^2 \theta}\int \frac{d x}{x}\Theta\left(\Delta_R^2 -\Delta_\theta^2 \right)\left[\Theta\left(\Delta_{R_\text{sub}}^2-\Delta_\theta^2 \right)\right. \nonumber \\ 
&+& \left. \Theta\left(\Delta_\theta^2-\Delta_{R_\text{sub}}^2 \right)\Theta(x-\zc)\Theta(1-\zc-x)\right] \delta \left(v- 2 x (1-x) (1-\cos \theta) \right). \nonumber \\
\eea

As in previous cases we can split the integral in $\theta$ so as to separate out the collinear singular $1/(1-\cos \theta)$ piece and in this region we 
replace the divergence $1/x$ by the full splitting function $p_{gq}(x)$. 
Performing the integral then produces the result Eq.~(\ref{LOtrimmingcomplete}) quoted in the main text, up to power corrections of $\ord(v)$. 

\subsection{Next-to--leading order calculation}
\label{trimmingnlo}
Here we carry out the NLO calculation for trimming in the soft and collinear approximation relevant to generating results for small $\zc$, $R_{\text{sub}}$ 
and $R$. The extension of our methods to obtain results beyond these limits is completely straightforward and can be carried out along the lines of the corresponding leading-order calculation.

Since we wish to highlight the simple relationship of the NLO result to that 
obtained at leading order, let us briefly revisit the leading order calculation. We note that Eq.~(\ref{LOtrimmingsetup}) can be expressed as 
\begin{equation}
\frac{1}{\sigma}\frac{d \sigma}{d v}^\text{(trimmed, LO)} = - \frac{d}{dv} (I^\mathrm{in}+I^\mathrm{out})
\end{equation}

with 

\bea 
\label{LOtrimminginout}
I^{\mathrm{in}} &=& \frac{\as C_F}{\pi}\int \frac{d \theta^2}{\theta^2}\int \frac{d x}{x}
\Theta\left(R_\text{sub}^2-\theta^2 \right)\Theta \left(x \theta^2 -v \right), \\ \nonumber
I^{\mathrm{out}} &=&  \frac{\as C_F}{\pi}\int \frac{d \theta^2}{\theta^2}\int \frac{d x}{x}
  \Theta\left(R^2-\theta^2 \right)\Theta\left(\theta^2-R_\text{sub}^2 \right)\Theta(x-\zc) \Theta \left(x \theta^2 -v \right),
\eea
where the integrals $I^{\mathrm{in}}$ and $I^{\mathrm{out}}$ correspond to 
evaluating the integrated cross-section at leading order, in the angular region where the gluon emission is inside and outside $R_\text{sub}$, respectively.

Moving to NLO let us consider the independent emission of two gluons $k_1$ and $k_2$ as for the case of the other substructure techniques. Here one can write the differential distribution, by extension of the leading order notation, and by considering the region in angle where both emissions are outside $R_{\text{sub}}$, inside $R_{\text{sub}}$ or two identical contributions from the region where a given gluon is out and the other is in and vice-versa:
\begin{equation}
\frac{1}{\sigma} \frac{d\sigma}{dv}^{(\mathrm{trimmed}, \,C_F^2)} =\frac{1}{\sigma} \frac{d\sigma}{dv}^{\mathrm{out,out}}+\frac{1}{\sigma} \frac{d\sigma}{dv}^{\mathrm{in,in}}+2 \times \frac{1}{\sigma} \frac{d\sigma}{dv}^{\mathrm{in,out}},
\end{equation}
which can be expressed in terms of the integrals entering the integrated cross-section as before,
\begin{equation}
\frac{1}{\sigma} \frac{d\sigma}{dv}^{(\mathrm{trimmed}, \,C_F^2)}= \frac{d}{dv} \left(I^{\mathrm{out,out}}+I^{\mathrm{in,in}}+2 \times I^{\mathrm{in,out}} \right).
\end{equation}

We address first the region of angle where $\theta_1,\theta_2 > R_{\text{sub}}$ and consider both double real and one-real one-virtual terms in this angular region. Here real gluons only contribute to the jet mass distribution if they have energy fraction $x_i > \zc$. Taking this into 
account one can write (we avoid explicitly writing the condition $R^2 > \theta_{1,2}^2$, which should be understood from now on):

\bea
\label{trimout2}
I^{\mathrm{out},\mathrm{out}} = C_F^2 \left(\frac{\alpha_s}{\pi}\right)^2 \frac{1}{2!} \int \frac{d x_1}{x_1} \frac{d x_2}{x_2} \frac{d \theta_1^2}{\theta_1^2} \frac{d\theta_2^2}{\theta_2^2} \Theta \left(\theta_1^2 -R_{\text{sub}}^2 \right) \Theta \left (\theta_2^2-R_{\text{sub}}^2 \right) \Delta [\Theta],
\eea
 
where we have that 
\begin{align}
\Delta[\Theta] = &\Theta \left ( x_1-\zc \right) \Theta \left( x_2 -\zc \right) 
\Theta \left ( v-x_1 \theta_1^2 -x_2 \theta_2^2 \right) \nonumber \\
+& \Theta \left ( x_1-\zc \right) \Theta \left( \zc-x_2 \right ) \Theta \left (v-x_1 \theta_1^2 \right) \nonumber \\ 
+& \Theta \left( x_2-\zc \right) \Theta \left( \zc-x_1 \right ) \Theta \left( v-x_2 \theta_2^2 \right)\nonumber \\
-&\Theta \left (x_2-\zc\right) \Theta \left (v -x_2 \theta_2^2 \right) - \Theta \left(x_1-\zc \right) \Theta\left(v-x_1 \theta_1^2 \right).
\end{align}
where, as for the case of pruning, the first three terms on the RHS of the 
above arise from double real emission while the last two terms, with a minus sign, are the contributions from one-real and one-virtual emission. 

To leading logarithmic accuracy (i.e. up to single-logarithmic correction terms which we do not attempt to treat here), one can make the replacement $$\Theta \left(v-x_1 \theta_1^2-x_2 \theta_2^2 \right) \to \Theta \left(v-x_1 \theta_1^2\right) \Theta \left(v-x_2 \theta_2^2 \right)$$ and then combine terms to obtain 

\begin{equation}
\Delta [\Theta] = \Theta \left(x_1-\zc \right) \Theta \left(x_2-\zc \right) \left [\Theta \left(v-x_1 \theta_1^2 \right) \Theta \left(v-x_2 \theta_2^2 \right) -\Theta \left(v-x_1 \theta_1^2 \right)-\Theta \left(v-x_2 \theta_2^2 \right) \right],
\end{equation}
which implies that 
\begin{equation}
\frac{d}{dv} \Delta [\Theta] = \frac{d}{dv} \left(\Theta \left(x_1-\zc \right) \Theta \left(x_2-\zc \right) \Theta \left(x_1 \theta_1^2 -v\right)\Theta \left(x_2 \theta_2^2 -v \right)\right).
\end{equation}

Using this result in Eq.~(\ref{trimout2}) we obtain 
\begin{multline}
\frac{d}{dv} I^{\mathrm{out,out}} = \frac{d}{dv} \, \frac{1}{2!}  C_F^2 \left(\frac{\alpha_s}{\pi}\right)^2 \int \frac{d x_1}{x_1} \frac{d \theta_1^2}{\theta_1^2}   \Theta \left(x_1-\zc \right) \Theta \left(\theta_1^2 -R_{\text{sub}}^2 \right) \Theta \left(x_1 \theta_1^2 -v\right) \\ \nonumber
\times \int\frac{d x_2}{x_2} \frac{d\theta_2^2}{\theta_2^2} \Theta \left(x_2-z\right) \Theta \left (\theta_2^2-R_{\text{sub}}^2 \right) \Theta \left(x_2 \theta_2^2 -v\right)
\end{multline}
which is just $\frac{d}{dv}\frac{1}{2!} \left(I^{\mathrm{out}}\right)^2$.

We now consider the region where both real gluons are within an angle $R_{\text{sub}}$ and the one-real one-virtual corrections in this angular region. Here the gluons contribute for all values of energy fractions and one can write
\bea
\label{trimint}
I^{\mathrm{in},\mathrm{in}} = C_F^2 \left(\frac{\alpha_s}{\pi}\right)^2 \frac{1}{2!} \int \frac{d x_1}{x_1} \frac{d x_2}{x_2} \frac{d \theta_1^2}{\theta_1^2} \frac{d\theta_2^2}{\theta_2^2} \Theta \left(R_{\text{sub}}^2 -\theta_1^2\right) \Theta \left (R_{\text{sub}}^2 -\theta_2^2\right) \Delta [\Theta],
\eea
 
where we have that 
\begin{equation}
\Delta[\Theta] = \Theta \left (v-x_1 \theta_1^2 -x_2\theta_2^2 \right) - 
\Theta \left (v -x_2 \theta_2^2 \right) - \Theta\left(v-x_1 \theta_1^2 \right).
\end{equation}

Following the same steps as before (i.e.\ factorising the constraint involving the sum of contributions from both emissions) it is easy to see that this contribution leads to 
\begin{equation}
\frac{d}{dv} I^{\mathrm{in},\mathrm{in}}= \frac{d}{dv}\frac{1}{2!} 
\left(I^{\mathrm{in}}\right)^2.
\end{equation}

Finally we need to consider the contribution with one gluon (say $k_1$) with 
$\theta_1 < R_{\text{sub}}$ and the other with $\theta_2 > R_{\text{sub}}$ (and an equal contribution with $k_1$ and $k_2$ exchanged), which applying identical methods to those above, can be expressed as:
\begin{equation}
2 \frac{d}{dv} I^{\mathrm{in},\mathrm{out}} = \frac{d}{dv}\frac{1}{2!} \left( 2 I^{\mathrm{in}}I^{\mathrm{out}} \right).
\end{equation}

Combining terms one observes that 
\begin{equation}
\frac{1}{\sigma} \frac{d\sigma}{dv}^{(\mathrm{trimmed}, \,C_F^2)} = \frac{d}{dv} \frac{(I^{\mathrm{in}}+I^{\mathrm{out}})^2}{2!},
\end{equation}
consistent with a simple exponentiation of the leading-order result for the integrated cross-section for trimming. Our arguments here can easily be extended to all orders to verify the exponentiation.

\end{document}